\documentclass[10pt,twocolumn]{article}
\usepackage[a4paper,hmargin=1.8cm,top=2.2cm,bottom=2.4cm,columnsep=0.7cm,headsep=10pt]{geometry}
\usepackage[T1]{fontenc}
\usepackage{amsmath,mathtools}
\usepackage[colorlinks,citecolor=blue,urlcolor=blue,linkcolor=blue]{hyperref}
\usepackage{iftex}
\ifPDFTeX
  \usepackage{newtxtext,newtxmath}
  \usepackage[fontset=none]{ctex}

\else
  \IfFileExists{newtxmath.sty}{\usepackage{newtxmath}}{\usepackage{txfonts}}
  \usepackage{ctex}
  \IfFontExistsTF{Noto Serif CJK SC Light}{\setCJKmainfont{Noto Serif CJK SC Light}}{}
\fi
\usepackage{bm}
\usepackage{etoolbox}
\usepackage{multicol}
\usepackage{array}
\usepackage{indentfirst}
\usepackage{graphicx}
\usepackage{adjustbox}
\usepackage{caption}
\usepackage{tikz}
\usetikzlibrary{patterns,decorations.markings,decorations.pathmorphing,arrows.meta,calc}
\tikzset{
  vtxq/.style={circle,fill,inner sep=1.5pt},
  vtx/.style={circle,fill,inner sep=2.6pt},
  extin/.style={thick,postaction={decorate},decoration={markings,
    mark=at position 0.55 with {\arrow{Stealth[length=5pt,width=4pt]}}}},
  extout/.style={thick,postaction={decorate},decoration={markings,
    mark=at position 0.55 with {\arrow{Stealth[length=5pt,width=4pt]}}}},
  ext/.style={thick},
  int/.style={thick},
  lab/.style={font=\footnotesize},
  cut/.style={dashed,red!70!black,thick}
}
\newcommand{\fpanel}[2]{%
 \begin{tikzpicture}[scale=0.75,baseline=(current bounding box.north)]
  \useasboundingbox (-2.9,-2.4) rectangle (2.9,1.45);
  #1
  \node[lab,align=center,anchor=north] at (0,-1.7) {#2};
 \end{tikzpicture}}

\renewcommand{\refname}{References}
\renewcommand{\appendixname}{Appendix}
\ctexset{figurename=Fig.,tablename=Table,contentsname=Contents}
\newcommand{\dd}{\mathrm d}
\newcommand{\ii}{\mathrm i}
\newcommand{\Lag}{\mathcal L}
\newcommand{\M}{\mathcal M}
\newcommand{\MlowII}{\M_{\substack{(\partial\phi)^4,\,\rm low\\ \rm tree,\,\mathit{s,t,u}\ bubble}}}
\newcommand{\MeikII}{\M_{\substack{(\partial\phi)^4,\,\rm eikonal\\ \mathit{t,u}\ \rm bubble}}}
\newcommand{\MlowIII}{\M_{\substack{(\partial\phi)^6,\,\rm low\\ \mathit{s,t,u}\ \rm banana}}}
\newcommand{\MeikIII}{\M_{\substack{(\partial\phi)^6,\,\rm eikonal\\ \mathit{t,u}\ \rm banana}}}
\newcommand{\MlowIIs}{\M_{\substack{(\partial\phi)^4,\,\rm low\\ \rm tree,\,\mathit{s}\ bubble}}}
\newcommand{\MlowIItu}{\M_{\substack{(\partial\phi)^4,\,\rm low\\ \mathit{t,u}\ \rm bubble}}}
\newcommand{\MlowIIIs}{\M_{\substack{(\partial\phi)^6,\,\rm low\\ \mathit{s}\ \rm banana}}}
\newcommand{\MlowIIItu}{\M_{\substack{(\partial\phi)^6,\,\rm low\\ \mathit{t,u}\ \rm banana}}}
\newcommand{\PhiPS}{\mathrm d\Phi}
\usepackage{titlesec}
\titleformat{\section}[block]{\normalfont\large\bfseries\filcenter}{\thesection.}{0.5em}{\MakeUppercase}
\titleformat{\subsection}{\normalfont\normalsize\bfseries\filcenter}{\thesubsection.}{0.5em}{}
\titlespacing*{\section}{0pt}{2.2ex plus 0.6ex minus 0.3ex}{1.2ex plus 0.3ex}
\titlespacing*{\subsection}{0pt}{1.6ex plus 0.5ex minus 0.2ex}{0.8ex plus 0.2ex}
\usepackage{fancyhdr}
\fancypagestyle{cpcrule}{\fancyhf{}\fancyhead[L]{\small Ji Li}\fancyfoot[C]{\thepage}%
 }
\fancypagestyle{plain}{\fancyhf{}\fancyhead[L]{\small Ji Li}\fancyfoot[C]{\thepage}%
 }
\renewcommand{\footnoterule}{\kern 4pt\hrule width \textwidth\kern 7pt}
\usepackage{orcidlink}
\renewcommand{\thesection}{\Roman{section}}
\renewcommand{\thesubsection}{\Alph{subsection}}
\makeatletter\renewcommand{\p@subsection}{\thesection\,}\makeatother

\begin{document}
\twocolumn[{%
\begin{center}
{\Large\bfseries Regge bounds and eikonal estimates\\[0.2em] in scalar EFT}

\vspace{1.0em}
{\large Ji Li (李骥)$^{1,2,3}$\,\orcidlink{0009-0007-1666-7474}}

\vspace{0.6em}
{\small
$^1$School of Fundamental Physics and Mathematical Sciences, Hangzhou Institute for Advanced Study,\\
University of Chinese Academy of Sciences, Hangzhou 310024, China\\
$^2$Institute of Theoretical Physics, Chinese Academy of Sciences, Beijing 100190, China\\
$^3$University of Chinese Academy of Sciences, Beijing 100049, China\par}
\end{center}
\vspace{0.8em}
\begin{list}{}{\leftmargin=2.2em\rightmargin=2.2em}\item[]
\normalsize
\noindent\textbf{Abstract:} Positivity bounds are derived from twice-subtracted dispersion relations,
which hold only if the amplitude grows more slowly than $s^2$.  For the massless
shift-symmetric scalar with $g_4(\partial\phi)^4/\Lambda^4$ and
$g_6(\partial\phi)^6/\Lambda^8$, we derive the dispersion relation, whose cuts
start at zero energy, and study the large-$s$ growth at fixed $t<0$ of the
tree, the one-loop bubble and the three-loop banana.  The partial-wave sum
is split at a transition spin: low spins are bounded by unitarity, channel by
channel, and high spins are evaluated with the eikonal phase of the
crossed-channel logarithms by stationary phase.  Matching the two regions gives
the phase-model growth $s^{8/7}$ for the quartic sector and $s^{16/15}$ for the
sextic sector, while a transition scale fixed by the cutoff gives the conditional
bound $s^{7/4}$; in the forward limit the same construction gives only the
marginal growth $s^2$, so that Regge boundedness remains an assumption on the
completion.  Under this assumption the \(2\to2\) and \(2\to4\) absorptive parts
below the cutoff restrict the couplings to $0<g_4\le80\pi^2/7$ and
$-7.2\times10^5\lesssim g_6\lesssim6.9\times10^5$.

\vspace{0.5em}
\end{list}
\vspace{1.2em}
}]
{\renewcommand{\thefootnote}{}%
 \footnotetext{\hspace*{-1.8em}\phantom{Received}\par
 \vspace*{0.45em}\noindent\phantom{Supported by}\par
 \vspace*{0.45em}\noindent\phantom{Corresponding author}\par
 \vspace*{0.3em}}}

\section{Introduction}
\label{sec:intro}

We consider a real massless shift-symmetric scalar field with the derivative
expansion
\begin{equation}
\begin{aligned}\Lag={}&-\frac12\partial_\mu\phi\partial^\mu\phi
  +\frac{g_4}{\Lambda^4}(\partial_\mu\phi\partial^\mu\phi)^2\\
  &+\frac{g_6}{\Lambda^8}(\partial_\mu\phi\partial^\mu\phi)^3+\cdots ,\end{aligned}\label{eq:lagrangian}
\end{equation}
where \(\Lambda\) is the cutoff of the EFT and \(g_4\), \(g_6\) are
dimensionless~\cite{Weinberg}.  With
\(\eta_{\mu\nu}=\operatorname{diag}(-1,+1,+1,+1)\), all momenta incoming and
\(s_{ij}=-(p_i+p_j)^2\), the variables \(s=s_{12}\), \(t=s_{13}\),
\(u=s_{14}\) satisfy \(s+t+u=0\), and every crossing-symmetric contact term is
a polynomial in the invariants~\cite{Basile,PengShaoTokarevaXu}
\begin{equation}
 \sigma_2=-\tfrac12\bigl(s^2+t^2+u^2\bigr)=st+tu+us,\qquad
 \sigma_3=stu .\label{eq:sigma-def}
\end{equation}
With
\(\partial_\mu\to\ii p_\mu\) and \(p_i\cdot p_j=-s_{ij}/2\), the quartic
operator gives
\begin{equation}
 \M^{\rm tree}_{(\partial\phi)^4}(s,t)=\frac{2g_4}{\Lambda^4}\left(s^2+t^2+u^2\right)
 =-\frac{4g_4}{\Lambda^4}\,\sigma_2 ,\label{eq:general-tree-ansatz}
\end{equation}
and we write $c\equiv2g_4/\Lambda^4$ for the coefficient of $s^2+t^2+u^2$.

\begin{figure*}[!htbp]
\centering
\begin{tikzpicture}[scale=0.85,>=Stealth]
 \draw[->,gray] (-3.1,0)--(3.2,0) node[right,black]{\small$\operatorname{Re}s$};
 \draw[->,gray] (0,-2.9)--(0,2.9) node[above,black]{\small$\operatorname{Im}s$};
 \draw[very thick,red!70!black,decorate,decoration={zigzag,amplitude=1.3pt,segment length=4pt}] (0,0)--(3.0,0);
 \draw[very thick,red!70!black,decorate,decoration={zigzag,amplitude=1.3pt,segment length=4pt}] (-3.0,0)--(0,0);
 \draw[blue,thick,postaction={decorate},decoration={markings,mark=at position 0.13 with {\arrow{Stealth[length=5pt]}},mark=at position 0.63 with {\arrow{Stealth[length=5pt]}}}] (0,0) circle (2.4);
 \draw[blue,thick,postaction={decorate},decoration={markings,mark=at position 0.3 with {\arrow{Stealth[length=4pt]}}}] (0,0) circle (0.45);
 \draw[blue,thick,->] (0.55,0.16)--(1.6,0.16); \draw[blue,thick] (1.6,0.16)--(2.35,0.16);
 \draw[blue,thick,->] (2.35,-0.16)--(1.4,-0.16); \draw[blue,thick] (1.4,-0.16)--(0.55,-0.16);
 \draw[blue,thick,->] (-2.35,0.16)--(-1.4,0.16); \draw[blue,thick] (-1.4,0.16)--(-0.55,0.16);
 \draw[blue,thick,->] (-0.55,-0.16)--(-1.6,-0.16); \draw[blue,thick] (-1.6,-0.16)--(-2.35,-0.16);
 \node[blue] at (0.32,0.62) {\small$\varepsilon$};
 \node[blue] at (1.95,2.0) {\small$|s|=R$};
 \node at (0,-3.25) {\small(a) $t=0$};
\end{tikzpicture}\hspace{1cm}
\begin{tikzpicture}[scale=0.85,>=Stealth]
 \draw[->,gray] (-3.1,0)--(3.2,0) node[right,black]{\small$\operatorname{Re}s$};
 \draw[->,gray] (0,-2.9)--(0,2.9) node[above,black]{\small$\operatorname{Im}s$};
 \draw[very thick,red!70!black,decorate,decoration={zigzag,amplitude=1.3pt,segment length=4pt}] (0,0.05)--(3.0,0.05);
 \draw[very thick,orange!80!black,decorate,decoration={zigzag,amplitude=1.3pt,segment length=4pt}] (-3.0,-0.05)--(0.45,-0.05);
 \fill (0.45,0) circle (1.2pt); \node[font=\scriptsize] at (0.45,0.35) {$-t$};
 \draw[blue,thick,postaction={decorate},decoration={markings,mark=at position 0.13 with {\arrow{Stealth[length=5pt]}},mark=at position 0.63 with {\arrow{Stealth[length=5pt]}}}] (0,0) circle (2.4);
 \draw[green!50!black,thick,postaction={decorate},decoration={markings,mark=at position 0.3 with {\arrow{Stealth[length=5pt]}}}] (0,0) circle (1.0);
 \draw[blue,thick,->] (1.1,0.2)--(1.8,0.2); \draw[blue,thick] (1.8,0.2)--(2.35,0.2);
 \draw[blue,thick,->] (2.35,-0.2)--(1.6,-0.2); \draw[blue,thick] (1.6,-0.2)--(1.1,-0.2);
 \draw[blue,thick,->] (-2.35,0.2)--(-1.6,0.2); \draw[blue,thick] (-1.6,0.2)--(-1.1,0.2);
 \draw[blue,thick,->] (-1.1,-0.2)--(-1.8,-0.2); \draw[blue,thick] (-1.8,-0.2)--(-2.35,-0.2);
 \node[green!50!black,font=\small] at (0.55,1.25) {$|s|=r$};
 \node[green!50!black,font=\scriptsize] at (-0.05,-0.62) {EFT};
 \node[blue] at (1.95,2.0) {\small$|s|=R$};
 \node at (0,-3.25) {\small(b) $-t<r<\Lambda^2$};
\end{tikzpicture}
\caption{(color online) Contour in the complex \(s\) plane at fixed \(t\).  The \(s\)-channel cut
runs along \(s\ge0\) and the \(u\)-channel cut along \(u=-s-t\ge0\), i.e.\
\(s\le-t\), both starting at zero energy.  (a) \(t=0\): the cuts meet at the
origin, around which a loop of radius \(\varepsilon\) is drawn.  (b)
\(t<0\): the cuts overlap on \(0\le s\le-t\), and the region \(|s|<r\) is
described by the EFT.}
\label{fig:contour}
\end{figure*}
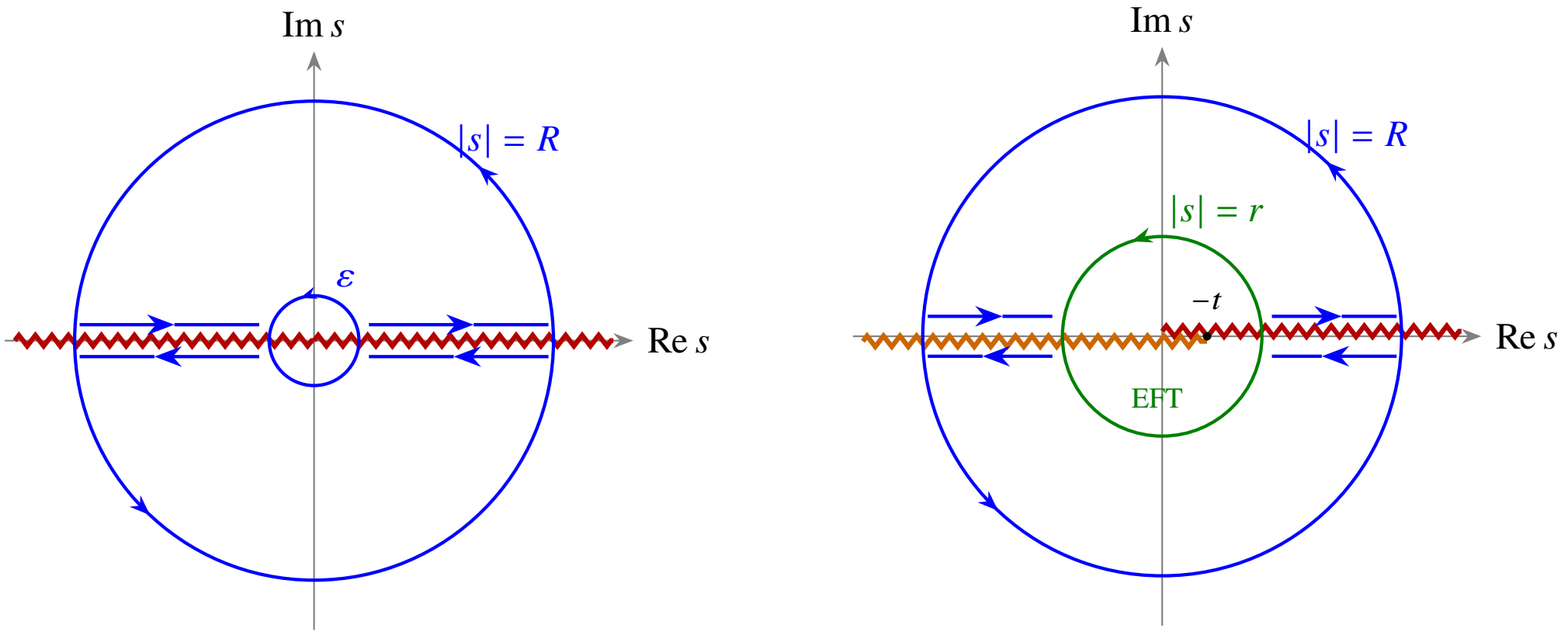

The motivation is the dispersive bootstrap of Wilson
coefficients~\cite{PhamTruong,Adams,DeRham,BellazziniMoments,Tolley,EFThedron,CaronHuotVanDuong,CaronHuotSwampland,Chiang,Deproject,NRTpositivity}.
Analyticity, crossing and unitarity of the complete
amplitude~\cite{Mandelstam,Lehmann,BEG} give, for the
coefficient of $s^2$ in the forward limit,
\begin{equation}
\begin{aligned}\frac{2g_4}{\Lambda^4} &=\frac1\pi\int_{s_{\rm th}}^\infty\dd s\,
 \frac{\operatorname{Im}\M(s,0)}{s^3} \\
 &\quad+\lim_{R\to\infty}\frac{1}{4\pi\ii}\oint_{|s'|=R}
 \frac{\M(s',0)}{s'^3}\,\dd s' .\end{aligned}\label{eq:intro-dispersion}
\end{equation}
where \(s_{\rm th}\) is the start of the cuts, \(s_{\rm th}=0\) in the
massless theory (Sec.~\ref{sec:dispersion}).  The last term of Eq.~\eqref{eq:intro-dispersion} is the contribution of
the large contour \(|s'|=R\) in Fig.~\ref{fig:contour}(a), the two arcs that
close the cuts at large \(|s'|\), and is referred to below as the arc.  It
vanishes as \(R\to\infty\) if \(|\M(s,0)|/|s|^2\to0\) for \(|s|\to\infty\), the
condition of Regge boundedness, which is an assumption on the completion of
the massless theory~\cite{Adams,TokarevaXu}; only then does
$\operatorname{Im}\M\ge0$ imply $g_4>0$~\cite{Adams,Distler}.  Bounds on the Regge growth of a completion, with an exponent not exceeding two at fixed \(t<0\), have been derived from the bound on chaos~\cite{ChandorkarRegge} and, for gravitational scattering, from unitarity and analyticity in impact parameter~\cite{HaringZhiboedov}.  In a massless theory the EFT loops
also produce $\operatorname{Im}\M(s,0)\propto g_4^2s^4/\Lambda^8$ at low energy, a
calculable part that sharpens the bounds~\cite{BellazziniRiembauRiva,YeHeGu,TokarevaXu}.

\enlargethispage{-74pt}
The input of Eq.~\eqref{eq:intro-dispersion} is the growth of the amplitude at
large $s$, and this paper determines the growth implied by the EFT itself.  In the
massless theory the cuts start at zero energy, the origin is a branch point,
and the coefficient $2c$ of $s^2$ in $\M(s,0)$ is defined by the limit of a small loop around the branch point (Sec.~\ref{sec:contour}).  We derive the dispersion relation for this case and
find that the truncated EFT does not satisfy it on its own: at one loop its
arc grows as $R^2$, Eq.~\eqref{eq:eft-arc}, so that the completion must transfer $c$ to cuts above the cutoff, Eq.~\eqref{eq:uv-transfer}.  We then estimate the growth of the EFT contributions at large $s$ and fixed $t<0$ (Secs.~\ref{sec:split}--\ref{sec:growth}).  At fixed order the quartic tree grows as $s^2$, the
one-loop bubble as $s^4\log s$ and the three-loop banana as $s^8\log s$, Eqs.~\eqref{eq:general-tree-ansatz}, \eqref{eq:x2-channel-decomposition} and~\eqref{eq:x3-elastic-full}.  The
partial-wave sum is split at a transition spin $\ell_*=b_*\sqrt s/2$ (Fig.~\ref{fig:overview}): below
it only $|S_\ell|\le1$ is used, and above it the EFT enters through the phase
shift generated by the crossed-channel logarithms, among them
$C(s)\,t^2\log(-t)$ with $C(s)\propto g_4^2s^2/\Lambda^8$, Eq.~\eqref{eq:x2-regge}.  The resulting powers bound the EFT contributions, whereas a Regge bound concerns the exponent of the completion~\cite{Regge,Collins}.

The main results are the following.  Within the EFT the \(s\)-channel contributions are bounded by the constant \(192\pi\), Eqs.~\eqref{eq:x2-pwu-explicit-bound} and~\eqref{eq:x3-pwu-explicit-bound}; the phase models of the crossed logarithms give \(s^{8/7}\) for the quartic sector and \(s^{16/15}\) for the high-spin part of the sextic sector, Eqs.~\eqref{eq:x2-low-high-matched} and~\eqref{eq:x3-high-scaling}; and a transition at the boundary of the region controlled by the EFT gives \(s^{7/4}<s^2\) at fixed \(q\neq0\), Eq.~\eqref{eq:eft-controlled-bound}, and the marginal \(s^2\) in the forward limit, Eq.~\eqref{eq:forward-eft-controlled}.  The EFT contributions are thus compatible with a Regge-bounded completion at fixed \(t<0\), while Regge boundedness itself, Eq.~\eqref{eq:regge-bounded-condition}, remains an assumption on the completion.  Assuming it, the \(2\to2\) and the complete \(2\to4\) absorptive parts give \(g_4>0\) and the region~\eqref{eq:allowed-region}, centred on \(g_6=-c\,g_4^2\).  Three questions remain open (Sec.~\ref{sec:outlook}): the exponents for operators \((\partial\phi)^{2k}\) with \(k\ge4\) rest on crossed-logarithm coefficients that have not been computed, Table~\ref{tab:xk}; the bounds use the leading absorptive parts, and the two-particle cuts at two and three loops, Eq.~\eqref{eq:x6-power-counting}, together with an infrared treatment adapted to loop-level positivity~\cite{BellazziniRiembauRiva,AlberteQED,HerreroValea}, would sharpen them; and Regge boundedness of the completion is assumed rather than derived~\cite{ChandorkarRegge,HaringZhiboedov}.

The paper is organized as follows.  Section~\ref{sec:pw-chapter} collects the properties of the amplitude used
throughout: the shift symmetry and crossing, Eq.~\eqref{eq:analytic}, the
analytic structure with cuts starting at the origin, and the partial waves,
the unitarity disk and the phase shifts,
Eqs.~\eqref{eq:pwu-convention}--\eqref{eq:pwu-one-loop-pw}, which supply the
two ingredients used later, the bound \(|S_\ell|\le1\) and the phase.
Section~\ref{sec:dispersion} derives the dispersion relation of the massless
theory, in which the coefficient \(c\) is defined by the small loop around the
branch point, Eqs.~\eqref{eq:forward-soft} and
\eqref{eq:cauchy-cut}--\eqref{eq:uv-transfer}; it shows that the one-loop arc of
the truncated EFT grows as \(R^2\), Eq.~\eqref{eq:eft-arc}, and relates Regge
boundedness to the exponent \(a(t)\) of the completion,
Eqs.~\eqref{eq:regge-exponent}--\eqref{eq:t-channel-spin}, and thereby fixes
the condition that the rest of the paper examines.  Section~\ref{sec:split} splits the partial-wave sum at
\(\ell_*=b_*\sqrt s/2\), bounds the low-spin part by unitarity,
Eq.~\eqref{eq:method-low-spin-bound}, and evaluates the high-spin part by
stationary phase, Eqs.~\eqref{eq:general-eikonal}--\eqref{eq:general-saddle-amplitude};
this method does not depend on the operator (Fig.~\ref{fig:overview}).
Sections~\ref{sec:amplitudes} and~\ref{sec:sextic} compute the tree and loop
amplitudes and the complete partial waves of the \((\partial\phi)^4\) and
\((\partial\phi)^6\) sectors, Eqs.~\eqref{eq:x2-tree-cyclic}--\eqref{eq:x3-t-pw-main},
which form the input of the method: the imaginary parts are confined to \(\ell=0,2\), and the
high-spin parts come from the logarithms non-analytic in the angle.

Sections~\ref{sec:x2-low} and~\ref{sec:x3-low-eikonal} apply the method to the
two sectors at fixed \(t<0\), Sec.~\ref{sec:growth} combines the two phases,
fixes the transition scale by the cutoff and takes the forward limit, and
Sec.~\ref{sec:bounds} derives the allowed region of \(g_4\) and \(g_6\).
Section~\ref{sec:conclusions} summarizes the results, and
Sec.~\ref{sec:outlook} collects conjectures for operators of higher degree that
are not used elsewhere.  Appendix~\ref{app:pw}
gives the partial-wave projections, Appendix~\ref{app:integrals} lists the
integrals, Appendix~\ref{app:dispersion-regge} derives the dispersion relation
with \(n\) subtractions and its relation to the Regge exponent,
Appendix~\ref{app:impact} derives the impact-parameter form of the high-spin
sum, Appendix~\ref{app:P4} renormalizes the one-loop amplitude, and
Appendix~\ref{app:rambo} checks the four-particle cut by Monte Carlo
integration.

\section{Symmetry, analyticity and unitarity}
\label{sec:pw-chapter}

This section collects the three properties of the amplitude on which the
dispersion relation and the growth estimates rest.

\subsection{Symmetry}
\label{sec:symmetry}

The Lagrangian~\eqref{eq:lagrangian} is invariant under the shift
\(\phi\to\phi+{\rm const}\), so \(\phi\) enters only through its
derivatives \(\partial_\mu\phi\).  This symmetry forbids a potential
\(V(\phi)\), and with it a mass term, so that the scalar is massless, and
every vertex carries at least one power of momentum on each external
leg~\cite{Adams}.  The Lagrangian is also invariant under
\(\phi\to-\phi\), so there is no cubic vertex.  The particles are identical and the amplitude
is symmetric under permutations of the external legs, which at fixed \(t\)
reduces to \(s\leftrightarrow u\) crossing~\cite{Mandelstam}; together with
real analyticity~\cite{Collins},
\begin{equation}
\begin{gathered}\M(s^*,t)=\M(s,t)^*, \\
\M(s,t)=\M(u,t),\quad u=-s-t .\end{gathered}\label{eq:analytic}
\end{equation}
Bose symmetry restricts the partial waves of the \(s\) channel to even
spins, Sec.~\ref{sec:pwu-method}.

\subsection{Analyticity}
\label{sec:analytic}

At fixed \(t\le0\) the amplitude \(\M(s,t)\) is assumed to be an analytic
function of \(s\) whose singularities are those of its absorptive
parts~\cite{Mandelstam,BEG}.  Below the cutoff the scalar is massless and
has no cubic vertex (Sec.~\ref{sec:symmetry}), so there is no single-particle
pole and the only singularities are the cuts of the multi-particle states;
heavy states of the completion enter only at \(s\gtrsim\Lambda^2\), where they
contribute to \(\operatorname{Im}\M\ge0\).  The \(s\)-channel cut starts at the
two-particle threshold \(s=0\) and runs along \(s\ge0\), and the \(u\)-channel
cut runs along \(s\le-t\), where \(u=-s-t\ge0\)~\cite{Mandelstam,Cutkosky}.  Since the threshold is at zero, the two cuts meet at the origin for \(t=0\) and
overlap on \(0\le s\le-t\) for \(t<0\) (Fig.~\ref{fig:contour}).

\subsection{Partial waves and unitarity}
\label{sec:pwu-method}

For identical massless scalars
\begin{equation}
\begin{gathered}\M(s,t)
 =16\pi\sum_{\ell\ {\rm even}}
 (2\ell+1)\,f_\ell(s)\,P_\ell(x), \\
f_\ell(s)=\frac{1}{32\pi}\int_{-1}^{1}\dd x\,
 P_\ell(x)\,\M(s,t(x)),\end{gathered}\label{eq:pwu-convention}
\end{equation}
with $x=1+2t/s$ and $S_\ell=1+\ii f_\ell$; the factor $1/32\pi$ includes the identical-particle $1/2!$~\cite{Collins,CaronHuotVanDuong}.
A partial wave is labelled by the diagram as a superscript and by the
sector as a subscript, and $f_\ell$ without labels is the complete partial wave; where a
single channel is meant, as in the low-spin analysis of
Secs.~\ref{sec:x2-low} and~\ref{sec:x3-low} and in Appendix~\ref{app:pw},
the channel is added to the subscript, as in
$f^{\rm bubble}_{\ell,s,(\partial\phi)^4}$ and $f^{\rm bubble}_{\ell,t+u,(\partial\phi)^4}$.
Probability conservation, $|S_\ell|\le1$,
confines $f_\ell$ to the disk $|f_\ell-\ii|\le1$~\cite{Collins},
\begin{equation}
\begin{gathered}\operatorname{Im}f_\ell\ \ge\ \tfrac12|f_\ell|^2, \\
\left|\operatorname{Re}f_\ell\right|\le1, \\
|f_\ell|\le2, \\
1-|S_\ell|^2=2\operatorname{Im}f_\ell-|f_\ell|^2\le1,\end{gathered}\label{eq:pwu-unitarity}
\end{equation}
with equality in the first when the scattering is completely elastic.
These conditions hold for the complete partial wave whether or not other
final states can be produced; in the sextic sector, where
$\phi\phi\to\phi\phi\phi\phi$ is open, $|S_\ell|<1$ and $1-|S_\ell|^2$ is the
absorption~\cite{Collins}.

\subsection{Cuts, phase shifts and absorption}
\label{sec:phase-shifts}

The imaginary parts of the loop diagrams follow from the cutting relation~\cite{Cutkosky}
\begin{equation}
\begin{aligned}
 &2\operatorname{Im}\M(12\to34)\\
 &\qquad=\sum_n\int\dd\Pi_n\,\M(12\to n)\,\M^*(34\to n),
\end{aligned}\label{eq:pwu-cut}
\end{equation}
summed over physically accessible states, so that a diagram is complex only where
its internal lines can be put on shell together.  For $s>0$, $t,u<0$ only
the $s$-channel loop is complex, $\log\frac{-s-\ii0}{\mu^2}=\log(s/\mu^2)-\ii\pi$; the
$t$- and $u$-channel logarithms are real~\cite{Cutkosky}.  For elastic
scattering the unitarity relation is solved by a real phase,
\begin{equation}
\begin{gathered}S_\ell=e^{2\ii\delta_\ell}, \\
f_\ell=\frac{e^{2\ii\delta_\ell}-1}{\ii}
 =2\delta_\ell+2\ii\delta_\ell^2-\frac43\delta_\ell^3+\cdots, \\
\operatorname{Im}f_\ell=\tfrac12|f_\ell|^2 ,\end{gathered}\label{eq:pwu-phase-promotion}
\end{equation}
and for the general partial wave $S_\ell=\eta_\ell e^{2\ii\delta_\ell}$,
$0\le\eta_\ell\le1$,
\begin{equation}
 \operatorname{Im}f_\ell=\tfrac12|f_\ell|^2+\tfrac12\bigl(1-\eta_\ell^2\bigr),\label{eq:pwu-absorption}
\end{equation}
an elastic part and an absorption term, the second requiring a complex
phase $\delta_\ell+\ii\chi_\ell$~\cite{Collins}; the $s$-channel banana is of this kind at $\ell=0,2$, Eq.~\eqref{eq:x3-complex-phase}.  The phase shift serves only to express unitarity; the eikonal of Sec.~\ref{sec:eikonal-stationary} is a transverse Fourier transform followed by
a resummation, and its phase coincides with $\delta_\ell$ only at large
$\ell$, Eq.~\eqref{eq:pwu-to-eikonal}.

\subsection{Tree-level violation and one-loop unitarity}
\label{sec:pw-computed}

The quartic tree, $\M^{\rm tree}_{(\partial\phi)^4}=c(s^2+t^2+u^2)
=c\,s^2\bigl[\tfrac53P_0(x)+\tfrac13P_2(x)\bigr]$, gives
\begin{equation}
\begin{gathered}f_{0,(\partial\phi)^4}^{\rm tree}=\frac{5g_4}{24\pi}\,\frac{s^2}{\Lambda^4}, \\
f_{2,(\partial\phi)^4}^{\rm tree}=\frac{g_4}{120\pi}\,\frac{s^2}{\Lambda^4}, \\
f_{\ell\ge4,(\partial\phi)^4}^{\rm tree}=0 ,\end{gathered}\label{eq:x2-tree-pw-main}
\end{equation}
real, confined to $\ell\le2$ and growing as $s^2$; a contact term of degree $n$ is a polynomial of degree $n$ in $x$ and therefore contributes to only $\ell\le n$ by the orthogonality of the Legendre polynomials, Eq.~\eqref{eq:int-orth}~\cite{Szego}.
A real fixed-order partial wave violates probability conservation,
\begin{equation}
 |S_\ell^{\rm tree}|^2=\bigl|1+\ii f^{\rm tree}_{\ell,(\partial\phi)^4}\bigr|^2
 =1+\bigl(f_{\ell,(\partial\phi)^4}^{\rm tree}\bigr)^2>1 ,\label{eq:pwu-tree-violation}
\end{equation}
and at one loop the $s$-channel bubble supplies the missing imaginary
part.  Expanding the phase of Eq.~\eqref{eq:pwu-phase-promotion} to second
order, or evaluating the two-particle cut~\eqref{eq:pwu-cut}, gives
\begin{equation}
 \operatorname{Im}f_{\ell,(\partial\phi)^4}^{\rm bubble}=\tfrac12\bigl(f_{\ell,(\partial\phi)^4}^{\rm tree}\bigr)^2,
 \qquad
 f_{\ell\ {\rm odd},(\partial\phi)^4}^{\rm bubble}=0,\label{eq:pwu-one-loop-pw}
\end{equation}
the imaginary part being confined to $\ell=0,2$ and the real part
contributing to every even $\ell$ (Sec.~\ref{sec:x2-s-channel}), with the
counterterm contributing for $\ell\le4$.

\section{Dispersion relation and Regge boundedness}
\label{sec:regge-method}
\label{sec:dispersion}

\subsection{Contour and dispersion relation}
\label{sec:contour}

The dispersion relation is written for the forward amplitude, whose behaviour near the origin fixes the definition of the coefficient \(c\).  In the forward limit, where the \(t\)-channel bubble vanishes as
\(t^2\log(-t)\), Eqs.~\eqref{eq:general-tree-ansatz}
and~\eqref{eq:x2-channel-decomposition} give at one loop
\begin{equation}
\adjustbox{max width=207.8pt}{$\displaystyle
\begin{aligned}\M(s,0) &=2c\,s^2-\frac{7g_4^2}{20\pi^2\Lambda^8}\,s^4
 \left[\log\frac{-s-\ii0}{\mu^2}+\log\frac{s}{\mu^2}\right] \\
 &\quad+P_4(s,0),\end{aligned}
$}\label{eq:forward-one-loop}
\end{equation}
with \(P_4(s,0)\propto s^4\).  The first logarithm, from the \(s\)-channel
bubble, is complex for \(s>0\), the second, from the \(u\)-channel bubble, for \(s<0\), as follows from Eq.~\eqref{eq:x2-channel-decomposition} at \(t=0\).  There is no \(s^0\) or \(s^1\) term, by the shift symmetry and by crossing at \(t=0\), Sec.~\ref{sec:symmetry} and Eq.~\eqref{eq:analytic},
so that near the origin
\begin{equation}
 \M(s,0)=\M(-s,0)=2cs^2+O(s^4\log s),\label{eq:forward-soft}
\end{equation}
Although \(\M(s,0)/s^3\to2c/s\) near the origin by Eq.~\eqref{eq:forward-soft}, the
origin is a branch point and not an isolated singularity, so
\(\M(s,0)/s^3\) has no residue there, and the residue theorem is replaced by
Cauchy's theorem on the plane cut along the real axis between the loops
\(|s|=\varepsilon\) and \(|s|=R\) of Fig.~\ref{fig:contour}(a).  Cauchy's theorem~\cite{Titchmarsh} requires only analyticity in \(s\) at fixed \(t\), which
is assumed for every \(t\le0\) in Sec.~\ref{sec:analytic}, so the relation is
stated for \(\M(s,t)\); the forward amplitude of Eq.~\eqref{eq:forward-soft} is
its limit \(t\to0\), Eq.~\eqref{eq:forward-one-loop}, while for \(t<0\) the two cuts
overlap on \(0\le s\le-t\), Fig.~\ref{fig:contour}(a), and the small loop is
treated at the end of this section.  With both loops counter-clockwise,
\begin{equation}
\begin{aligned}
 &\frac{1}{2\pi\ii}\oint_{|s|=\varepsilon}\frac{\M(s,t)}{s^3}\,\dd s
 =\frac{1}{2\pi\ii}\oint_{|s|=R}\frac{\M(s,t)}{s^3}\,\dd s\\
 &\qquad+\frac{1}{2\pi\ii}\int_{\varepsilon<|s'|<R}\dd s'\,
 \frac{\operatorname{Disc}\M(s',t)}{s'^3},
\end{aligned}\label{eq:cauchy-cut}
\end{equation}
in which the integral of the discontinuity across the cuts takes the place
of the sum of residues; the derivation, for an arbitrary number of
subtractions, is given in Appendix~\ref{app:dispersion-regge}, where the small loop is kept at finite radius, since in the massless theory the coefficient of \(s^n\) at the origin is infrared divergent for \(n\ge4\), Eq.~\eqref{eq:app-soft}.  At \(t=0\)
real analyticity and crossing, Eq.~\eqref{eq:analytic}, make the two cuts
contribute equally,
\begin{equation}
\begin{aligned}
 &\frac{1}{2\pi\ii}\int_{\varepsilon<|s'|<R}\dd s'\,
 \frac{\operatorname{Disc}\M(s',0)}{s'^3}\\
 &\qquad=\frac2\pi\int_\varepsilon^R\dd s\,\frac{\operatorname{Im}\M(s,0)}{s^3},
\end{aligned}\label{eq:cuts}
\end{equation}
with a limit \(\varepsilon\to0\) that exists because the absorptive part starts
at \(s^4\).  On the large loop
\begin{equation}
 \Bigl|\oint_{|s|=R}\frac{\M(s,0)}{s^3}\,\dd s\Bigr|
 \le\frac{2\pi\max_{|s|=R}|\M(s,0)|}{R^2},\label{eq:large-circle}
\end{equation}
which vanishes as \(R\to\infty\) if the amplitude is Regge bounded,
Eq.~\eqref{eq:regge-bounded-condition}; Eqs.~\eqref{eq:cauchy-cut}--\eqref{eq:cuts}
then give Eq.~\eqref{eq:intro-dispersion} without the arc.

The truncated EFT does not satisfy this relation on its own, as Eqs.~\eqref{eq:eft-cut} and~\eqref{eq:eft-arc} show.  Its
one-loop absorptive part, Eq.~\eqref{eq:x2-optical}, and the cut integral are
\begin{equation}
\begin{gathered}\operatorname{Im}\M(s,0)=\frac{7g_4^2s^4}{20\pi\Lambda^8}, \\
\frac2\pi\int_0^R\dd s\,\frac{\operatorname{Im}\M(s,0)}{s^3}
 =\frac{7g_4^2}{20\pi^2\Lambda^8}\,R^2,\end{gathered}\label{eq:eft-cut}
\end{equation}
and Eq.~\eqref{eq:cauchy-cut} gives for the one-loop
amplitude~\eqref{eq:forward-one-loop}
\begin{equation}
 \frac{1}{2\pi\ii}\oint_{|s|=R}\frac{\M_{\rm EFT}(s,0)}{s^3}\,\dd s
 =2c-\frac{7g_4^2}{20\pi^2\Lambda^8}\,R^2 ,\label{eq:eft-arc}
\end{equation}
which grows as \(R^2\).
The coefficient \(c\) is given by the arc in the EFT, and a Regge-bounded
completion must transfer it to cuts above the cutoff,
\begin{equation}
\begin{aligned}&c-\frac1\pi\int_0^{\Lambda^2}\dd s\,\frac{\operatorname{Im}\M_{\rm EFT}(s,0)}{s^3}\\
 &\quad=\frac1\pi\int_{\Lambda^2}^\infty\dd s\,\frac{\operatorname{Im}\M(s,0)}{s^3}\ge0 ,\end{aligned}\label{eq:uv-transfer}
\end{equation}
the relation used in Sec.~\ref{sec:bounds}.  For \(t<0\) the overlapping cuts
prevent an expansion about the origin, and the small loop is replaced by
a loop \(|s|=r\), \(-t<r<\Lambda^2\), on which the amplitude is computed in
the EFT (Fig.~\ref{fig:contour}(b)); the same steps give
\begin{equation}
\adjustbox{max width=202.8pt}{$\displaystyle
\begin{aligned}\frac{1}{2\pi\ii}\oint_{|s|=r}\frac{\M(s,t)}{s^3}\,\dd s &=\frac1\pi\int_r^\infty\dd s\,\frac{\operatorname{Im}\M(s,t)}{s^3} \\
 &\quad+\frac1\pi\int_{r-t}^\infty\dd u\,\frac{\operatorname{Im}\M(u,t)}{(u+t)^3},\end{aligned}
$}\label{eq:dispersion-fixed-t}
\end{equation}
the left-hand side being fixed by the EFT and the right-hand side by the
absorptive parts of the completion, expanded in partial waves in
Sec.~\ref{sec:pwu-method}.  From the ultraviolet side, unitarity and Regge boundedness of
the completion constrain the infrared coefficients, which gives the
positivity of \(g_4\) and the correlation of \(g_4\) and \(g_6\) in
Sec.~\ref{sec:bounds}.  From the infrared side, the EFT fixes the left-hand
side, which any completion without new light states must reproduce; since
\(C(s)\,t^2\log(-t)\) is non-analytic at \(t=0\) while every
\(P_\ell(1+2t/s)\) is a polynomial in \(t\), it is reproduced only by partial
waves of arbitrarily high spin, at large impact parameter~\cite{TokarevaXu},
and enters the completion as low-energy matching data.

\subsection{Regge exponent and Regge boundedness}

\label{sec:regge}

The Regge limit~\cite{Regge,ChewFrautschi,Collins} is $s\to\infty$ at fixed
$t=-q^2$, in which the ultraviolet-complete amplitude behaves as
\begin{equation}
\begin{gathered}\M_{\rm UV}(s,t)\simeq\beta_{\rm R}(t)\left(\frac{s}{s_0}\right)^{a(t)}, \\
s\to\infty,\quad t<0\ \text{fixed}.\end{gathered}\label{eq:regge-exponent}
\end{equation}
The arc in Eq.~\eqref{eq:intro-dispersion} vanishes only if
\begin{equation}
 \lim_{s\to\infty}\frac{\M(s,t)}{s^2}=0 ,\label{eq:regge-bounded-condition}
\end{equation}
and an upper bound on $a(t)$, a Regge bound, is a property of the completion; the one-loop $s^2t^2\log(-t)$ is a fixed-order term with no direct relation to $a(t)$, and the powers obtained below from the EFT phase are
scalings of a phase model.

We first select the crossed $t$-channel contribution from the complete
crossing-symmetric amplitude, and only then substitute $u=-s-t$ and expand at large $s$, as in Appendix~\ref{app:pw}.  In either sector this channel has the form
\begin{equation}
 \M_t(s,t)=\mathcal P_t(s,t)\log\frac{-t-\ii0}{\mu^2}
 +\mathcal Q_t(s,t),\label{eq:regge-channel-template}
\end{equation}
with $\mathcal Q_t$ analytic in $t$, hence a contact term at $b=0$,
whereas the logarithm generates the power-law falloff at $b>0$, Eq.~\eqref{eq:log-transform}.  We take
$\mu^2=s$.  A partial wave of spin $\ell$ corresponds to $b=2\ell/\sqrt s$,
so the Regge limit is the limit of large $\ell$ at fixed $b$, while a
partial wave of fixed $\ell$ corresponds to $b\to0$, where contact terms reside~\cite{LevySucher,DiVecchiaEikonal}.

With \(s^{-n-1}\) in place of \(s^{-3}\) and, for \(n\ge4\), the small loop at finite radius, Eq.~\eqref{eq:app-identity}, Eq.~\eqref{eq:cauchy-cut} is an
identity for every \(R\), satisfied also by the truncated EFT,
Eq.~\eqref{eq:eft-arc}, and becomes a dispersion relation only when the arc
vanishes.  For \(\M\simeq\beta_{\rm R}(t)s^{a(t)}\) the arc is bounded by
\(\max_{|s|=R}|\M|/R^n\sim R^{\,a(t)-n}\), Eq.~\eqref{eq:app-arc-bound},
so that \(n\) subtractions require \(a(t)<n\) (Appendix~\ref{app:dispersion-regge}).
If \(a(0)\ge2\), the arc in Eq.~\eqref{eq:intro-dispersion} survives with a
sign set by the asymptotics, Eq.~\eqref{eq:app-arc-power},
\(\operatorname{Im}\M\ge0\) no longer implies \(g_4>0\), and the bounds of
Sec.~\ref{sec:bounds} are lost; the dispersion relation itself holds for any
\(a\) as an identity and therefore cannot establish \(a(0)<2\).

The \(s^2\) growth originates in the \(t\)-channel partial-wave expansion,
\begin{equation}
\begin{gathered}\M(s,t)=16\pi\sum_J(2J+1)f^{(t)}_J(t)P_J(z),\\
 z=1+\frac{2s}{t},\end{gathered}\label{eq:t-channel-pw}
\end{equation}
in which, since the leading coefficient of \(P_J(z)\) is \((2J)!/[2^J(J!)^2]\)~\cite{AbramowitzStegun} and \(z\simeq2s/t\) at fixed \(t\), a partial wave of spin \(J\) grows as
\begin{equation}
 P_J(z)\simeq\frac{(2J)!}{2^J(J!)^2}\Bigl(\frac{2s}{t}\Bigr)^J\propto s^J,
 \qquad s\to\infty .\label{eq:t-channel-spin}
\end{equation}
The contact tree contains only \(J=0,2\) (Sec.~\ref{sec:x2-tree}) and grows as
\(s^2\), at the boundary of Eq.~\eqref{eq:regge-bounded-condition}, and any finite
set of \(t\)-channel spins grows as \(s^{J_{\max}}\).  A Regge-bounded
completion must therefore replace the spin-two contact interaction by an
infinite set of spins whose sum behaves as \(s^{a(t)}\) with \(a(t)<2\),
Eq.~\eqref{eq:app-regge-pole}.

In the massless theory the partial waves fall only as a power of \(\ell\),
Eq.~\eqref{eq:x2-crossed-pw}, so that unitarity, \(|f_\ell|\le2\), does not cut
the partial-wave sum off at any finite spin, and
Eq.~\eqref{eq:regge-bounded-condition} is an assumption on the completion.
Section~\ref{sec:forward-regge} compares the EFT estimate with this assumption.

\section{Low-spin bound and high-spin eikonal bound on the amplitude}
\label{sec:split}
\label{sec:eikonal-method}

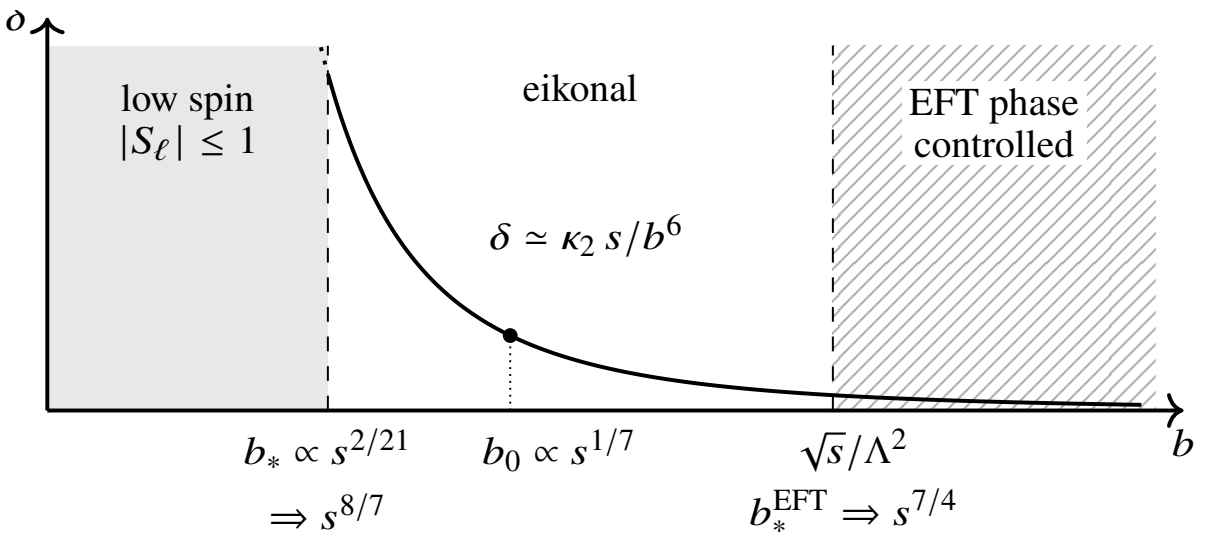
\begin{figure}[!tbp]
\centering
\adjustbox{max width=\linewidth}{%
\begin{tikzpicture}[x=1cm,y=1cm,font=\footnotesize]
 \fill[gray!18] (0,0) rectangle (2.0,2.6);
 \clip (-0.6,-1.1) rectangle (8.4,2.85);
 \fill[pattern=north east lines,pattern color=gray!60] (5.6,0) rectangle (7.9,2.6);
 \draw[->,thick] (0,0) -- (8.1,0) node[below]{$b$};
 \draw[->,thick] (0,0) -- (0,2.8) node[left]{$\delta$};
 \draw[thick,dotted,domain=1.95:2.0,samples=8] plot (\x,{2.4*pow(2/\x,3)});
 \draw[thick,domain=2.0:7.8,samples=100,smooth] plot (\x,{2.4*pow(2/\x,3)});
 \node[anchor=west] at (3.0,1.25) {$\delta\simeq\kappa_2\,s/b^6$};
 \draw[dashed] (2.0,0) -- (2.0,2.6);
 \draw[dashed] (5.6,0) -- (5.6,2.6);
 \draw[densely dotted] (3.3,0) -- (3.3,{2.4*pow(2/3.3,3)});
 \fill (3.3,{2.4*pow(2/3.3,3)}) circle (1.5pt);
 \node[below] at (2.0,0) {$b_*\propto s^{2/21}$};
 \node[below] at (3.65,0) {$b_0\propto s^{1/7}$};
 \node[below] at (5.75,0) {$\sqrt s/\Lambda^2$};
 \node[align=center] at (1.0,2.05) {low spin\\[-1pt]$|S_\ell|\le1$};
 \node[align=center] at (3.8,2.3) {eikonal};
 \node[align=center,fill=white,inner sep=1.5pt] at (6.75,2.05) {EFT phase\\[-1pt]controlled};
 \node[align=center] at (2.0,-0.72) {$\Rightarrow s^{8/7}$};
 \node[align=center] at (5.75,-0.72) {$b_*^{\rm EFT}\Rightarrow s^{7/4}$};
\end{tikzpicture}}
\caption{Impact-parameter picture of the \((\partial\phi)^4\) estimate at fixed
\(s\) and \(t=-q^2\), with \(\ell=b\sqrt s/2\).  Partial waves with \(b<b_*\) are
bounded by unitarity, Eq.~\eqref{eq:method-low-spin-bound}; for \(b>b_*\) the
phase of Eq.~\eqref{eq:x2-impact-phase} is exponentiated, its stationary point
\(b_0\) follows from Eq.~\eqref{eq:x2-saddle-scaling}, and matching at
\(b_*\propto s^{2/21}\) gives \(s^{8/7}\), Eq.~\eqref{eq:x2-low-high-matched}.  The
EFT controls the phase only for \(b\gg\sqrt s/\Lambda^2\),
Eq.~\eqref{eq:higher-degree-phase}; a transition there gives \(s^{7/4}\),
Eq.~\eqref{eq:eft-controlled-bound}.}
\label{fig:overview}
\end{figure}

The resummed amplitude
\begin{equation}
\adjustbox{max width=202.8pt}{$\displaystyle
\M_{\rm PWU}(s,t)
 =16\pi\sum_{\substack{\ell=0\\ \ell\ {\rm even}}}^{\infty}(2\ell+1)\,
 f_\ell^{\rm PWU}(s)\,
 P_\ell(x)
$}\label{eq:pwu-resummed-amplitude}
\end{equation}
is split at \(\ell_*=b_*\sqrt s/2\) into a low-spin and a high-spin part, Eq.~\eqref{eq:logic} and Fig.~\ref{fig:overview}.  We
constrain the low-spin contribution from partial-wave unitarity, and the
high-spin part from the eikonal resummation of the exponentiated partial waves~\cite{LevySucher,tHooft,ACV,KabatOrtiz,DiVecchiaEikonal}.  The two parts of Eq.~\eqref{eq:pwu-resummed-amplitude} are denoted
\(\M_{\rm low}\) and \(\M_{\rm eikonal}\).
Where the fixed-order partial waves lie outside the unitarity disk the phase
loses its meaning, Eq.~\eqref{eq:pwu-tree-violation}, and there they are replaced by the unitarity bound; at high spin the phase is real and generated entirely by the logarithms non-analytic in the angle, Eqs.~\eqref{eq:x2-delta-ell} and~\eqref{eq:x3-delta-ell}.  The EFT partial waves are thus used in two ways,
\begin{equation}
f_\ell\ \longrightarrow\
 \begin{cases}
  \ell<\ell_*: & \text{exact unitarity bound}\\
               & |S_\ell|\le1,\\[0.3em]
  \ell>\ell_*: & \text{phase matching and}\\
               & \text{eikonal analysis},
 \end{cases}
\label{eq:logic}
\end{equation}
where phase matching identifies, at large \(\ell\), the phase shift with the eikonal phase \(\delta(s,b)\) of Sec.~\ref{sec:eikonal-stationary},
\begin{equation}
 \delta_\ell(s)=\delta(s,b)\big|_{b=2\ell/\sqrt s}.\label{eq:pwu-to-eikonal}
\end{equation}
At low spin only \(|S_\ell|\le1\) is used, which holds for the complete
partial wave, including absorption and whatever the ultraviolet
completion, so it requires neither accurate EFT partial waves nor the phase
promotion~\eqref{eq:pwu-phase-promotion}.    The prescription that fixes which phase is exponentiated, and that it is taken real, defines the phase model; the high-spin results hold within this model and are reliable only where that phase is controlled.

The three ingredients therefore enter at different impact parameters.  The low-spin part, \(b<b_*\), uses no EFT input: its bound~\eqref{eq:method-low-spin-bound} follows from \(|S_\ell|\le1\) of the complete amplitude, Eq.~\eqref{eq:pwu-unitarity}, and grows only with the number of partial waves.  The high-spin part, \(b>b_*\), uses the EFT through one quantity, the leading phase of the crossed-channel logarithms, Eqs.~\eqref{eq:x2-delta-ell} and~\eqref{eq:x3-delta-ell}, resummed by the eikonal~\cite{LevySucher,KabatOrtiz,DiVecchiaEikonal}; since the EFT is an expansion in \(s/\Lambda^4b^2\), Eq.~\eqref{eq:higher-degree-phase}, this phase is controlled only for \(b\gg\sqrt s/\Lambda^2\).  With \(b_*=b_*^{\rm EFT}\), Eq.~\eqref{eq:bstar-eft}, every partial wave is treated either by unitarity or by the controlled phase, giving the conditional bound~\eqref{eq:eft-controlled-bound}; with \(b_*\) fixed by matching, Eqs.~\eqref{eq:x2-low-high-matched} and~\eqref{eq:x3-self-matching}, the phase is extrapolated into \(b<\sqrt s/\Lambda^2\), and the exponents are those of the phase models (Sec.~\ref{sec:eft-controlled}).

\subsection{The low-spin amplitude \texorpdfstring{$\M_{\rm low}$}{M low}: partial-wave unitarity bound}
\label{sec:low-spin-bound}

For \(\ell<\ell_*\) we use only unitarity, which bounds every partial wave
by \(|f_\ell|\le2\), Eq.~\eqref{eq:pwu-unitarity}, whatever its fixed-order
growth.  We take
\begin{equation}
 |f_\ell(s)|\lesssim1,
 \qquad
 2\ell+1\simeq2\ell,\label{eq:method-pwu-degeneracy}
\end{equation}
the overall \(16\pi\) and the factor two of \(|f_\ell|\le2\) being absorbed
into \(\lesssim\) below.  The Legendre polynomials obey Bernstein's
inequality~\cite{Szego}, \(|P_\ell(\cos\theta)|<[2/(\pi\ell\sin\theta)]^{1/2}\)
for \(\ell\ge1\) and \(0<\theta<\pi\); in the centre-of-mass frame
\(\cos\theta=1-2q^2/s\), and at small angle \(\sin\theta\simeq2q/\sqrt s\), so
that
\begin{equation}
 \left|P_\ell\!\left(1-\frac{2q^2}{s}\right)\right|
 <\left(\frac{\sqrt s}{\pi\ell q}\right)^{1/2}
 \lesssim
 \left(\frac{q\ell}{\sqrt{s}}\right)^{-1/2},\label{eq:method-gegenbauer-bound}
\end{equation}
valid for every \(\ell\ge1\) and hence over the entire low-spin region,
unlike the Bessel asymptotics~\eqref{eq:gegenbauer-bessel} used below only
for \(\ell>\ell_*\).  The low-spin part of
Eq.~\eqref{eq:pwu-resummed-amplitude} is therefore bounded term by term,
\begin{equation}
\begin{aligned}|\M_{\rm low}(s,-q^2)| &=\Bigl|16\pi\sum_{\ell<\ell_*}(2\ell+1)f_\ell(s)P_\ell(x)\Bigr| \\
 &\lesssim\sum_{\ell<\ell_*}2\ell\left(\frac{q\ell}{\sqrt s}\right)^{-1/2} \\
 &=2q^{-1/2}s^{1/4}\sum_{\ell<\ell_*}\ell^{1/2}.\end{aligned}\label{eq:low-spin-sum}
\end{equation}
The \(\ell=0\) term is bounded by \(|P_0|=1\) and, like the approximation
\(2\ell+1\simeq2\ell\), does not affect the result, since the sum is
dominated by \(\ell\approx\ell_*\).  With
\(\sum_{\ell<\ell_*}\ell^{1/2}\simeq\tfrac23\ell_*^{3/2}\) and
\(\ell_*=b_*\sqrt s/2\),
\begin{equation}
 |\M_{\rm low}(s,-q^2)|
 \lesssim
 \frac{4}{3\sqrt q}\,s^{1/4}\ell_*^{3/2}
 =\frac{\sqrt2}{3}\,\frac{s\,b_*^{3/2}}{\sqrt q},\label{eq:method-low-spin-bound}
\end{equation}
and for \(b_*=Bs^\gamma\) the bound grows as \(s^{1+3\gamma/2}\), Eq.~\eqref{eq:method-low-spin-bound}.  The growth arises from the number of partial waves below \(\ell_*\), each of which is bounded by unitarity, Eq.~\eqref{eq:pwu-unitarity}.

In the scalar EFT, \(b_*\) is fixed by matching the low-spin and eikonal
descriptions of each sector, with an alternative fixed by the EFT cutoff in
Sec.~\ref{sec:eft-controlled}.

\subsection{The high-spin amplitude \texorpdfstring{$\M_{\rm eikonal}$}{M eikonal}: stationary phase}
\label{sec:eikonal-stationary}

In the Regge limit the partial waves with \(\ell>\ell_*\) have
\(\ell\to\infty\) at fixed \(b=2\ell/\sqrt s\), so the sum becomes an
integral over \(b\) and the partial-wave phases become the eikonal
phase~\cite{LevySucher,tHooft,ACV,KabatOrtiz,Giddings,DiVecchiaEikonal}.
With \(b\) fixed as \(\ell\to\infty\), Hilb's formula for the Legendre
polynomials~\cite{Szego} gives the Bessel asymptotics and the change of
variable from \(\ell\) to \(b\), derived with their corrections in
Appendix~\ref{app:impact},
\begin{equation}
\begin{gathered}P_\ell\!\left(1-\frac{q^2b^2}{2\ell^2}\right)
 \simeq J_0(bq), \\
2\ell+1\simeq\sqrt s\,b, \\
\dd\ell=\frac{\sqrt{s}}2\,\dd b .\end{gathered}\label{eq:gegenbauer-bessel}
\end{equation}
The high-spin part of Eq.~\eqref{eq:pwu-resummed-amplitude} is the sum over
\(\ell>\ell_*\) in the first line of Eq.~\eqref{eq:general-eikonal}.  With
\(f_\ell=\ii(1-S_\ell)\), \(S_\ell=e^{2\ii\delta_\ell}\),
Eq.~\eqref{eq:pwu-convention}, the phase \(\delta_\ell=\delta(s,b)\) of
Eq.~\eqref{eq:pwu-to-eikonal}, and the sum over even \(\ell\) replaced by
\(\tfrac12\int\dd\ell\), Eq.~\eqref{eq:gegenbauer-bessel} turns it into an
integral over \(b\), with relative corrections of order \(q/\sqrt s\),
\begin{equation}
\adjustbox{max width=202.8pt}{$\displaystyle
\begin{aligned}
 \M_{\rm eikonal}(s,-q^2)
 &=16\pi\!\!\sum_{\substack{\ell>\ell_*\\ \ell\ {\rm even}}}\!\!(2\ell+1)f_\ell(s)P_\ell\Bigl(1-\frac{2q^2}{s}\Bigr)\\
 &\simeq16\pi\cdot\frac12\int_{b_*}^\infty\frac{\sqrt s}{2}\,\dd b\\
 &\qquad\times\sqrt s\,b\;\ii\bigl[1-e^{2\ii\delta(s,b)}\bigr]J_0(qb)\\
 &=4\pi\ii s\int_{b_*}^{\infty}\dd b\,bJ_0(bq)
 \left[1-e^{2\ii\delta(s,b)}\right],
\end{aligned}
$}\label{eq:general-eikonal}
\end{equation}
with the phase at impact parameter \(b=2\ell/\sqrt s\) given, at leading order, by the transverse Fourier transform of the \(t\)-channel amplitude~\cite{LevySucher,KabatOrtiz},
\begin{equation}
 \delta(s,b)=\frac{1}{4s}\int\frac{\dd^2\bm q}{(2\pi)^2}
 e^{\ii\bm q\cdot\bm b}\M_t(s,-q^2).\label{eq:eikonal-definition}
\end{equation}
We evaluate the high-spin part assuming that elastic scattering dominates,
i.e.\ that the real phase saturates the unitarity condition at
\(\ell>\ell_*\).  The integrand of Eq.~\eqref{eq:general-eikonal} is
oscillatory, and for a function \(g\) slowly varying compared with the
phase \(f\) the stationary-phase approximation~\cite{BenderOrszag} gives
\begin{equation}
\begin{aligned}
 &\int\dd x\,g(x)\,e^{\ii f(x)}
 \simeq g(x_0)\,e^{\ii f(x_0)+\ii\frac\pi4\operatorname{sign}f''(x_0)}\\
 &\qquad\times\left(\frac{2\pi}{|f''(x_0)|}\right)^{1/2},\qquad f'(x_0)=0 .
\end{aligned}\label{eq:stationary-phase}
\end{equation}
For large \(qb\) the Bessel function is replaced by its asymptotic form~\cite{AbramowitzStegun}, the sum of an outgoing and an incoming cylindrical wave,
\begin{equation}
 J_0(qb)\simeq
 \sqrt{\frac{2}{\pi qb}}\,
 \frac{e^{\ii(qb-\pi/4)}+e^{-\ii(qb-\pi/4)}}{2},\label{eq:j0-large-argument}
\end{equation}
so that the exponential in Eq.~\eqref{eq:general-eikonal} combines with
one of the two terms into \(e^{\ii\Phi(b)}\), and the stationary point
follows from
\begin{equation}
 \frac{\partial}{\partial b}\Bigl(\pm qb+2\delta(s,b)\Bigr)=0 .\label{eq:saddle-condition}
\end{equation}
The loop-generated phases of this paper fall as a power of the impact
parameter,
\begin{equation}
 \delta(s,b)=\kappa\,\frac{s}{b^{n}},
 \qquad
 \kappa=\text{const},\label{eq:power-phase}
\end{equation}
with the power \(n\) fixed by the diagram.  On the
branch with a positive real solution, \(\Phi(b)=qb+2\kappa s/b^n\),
Eq.~\eqref{eq:saddle-condition} reads
\(q-2n\kappa s\,b^{-n-1}=0\), and the main contribution comes from the
neighbourhood of
\begin{equation}
\begin{gathered}b_0=\left(\frac{2n\kappa s}{q}\right)^{\frac1{n+1}}\propto s^{\frac1{n+1}}, \\
\Phi''(b_0)=\frac{2n(n+1)\kappa s}{b_0^{\,n+2}}=\frac{(n+1)q}{b_0},\end{gathered}\label{eq:general-saddle}
\end{equation}
where the classical deflection \(-2\,\partial\delta/\partial b\) equals the
momentum transfer \(q\).  Taking the integral around this point with
Eq.~\eqref{eq:stationary-phase}, the prefactor
\(4\pi s\,b_0\,(2/\pi qb_0)^{1/2}/2\) and the Gaussian factor
\([2\pi/\Phi''(b_0)]^{1/2}=[2\pi b_0/((n+1)q)]^{1/2}\) give
\begin{equation}
\adjustbox{max width=202.8pt}{$\displaystyle
|\M_{\rm eikonal}|
 \simeq\frac{4\pi}{\sqrt{n+1}}\,\frac{s\,b_0}{q}
 \ \propto\ \kappa^{\frac1{n+1}}\,s^{\frac{n+2}{n+1}}\,q^{-\frac{n+2}{n+1}} .
$}\label{eq:general-saddle-amplitude}
\end{equation}
For every \(n\ge1\) the power lies between \(s\) and \(s^2\), so the high-spin part is
bounded by \(s^2\).  The result is independent of \(b_*\) as long as the
saddle lies in the eikonal region, \(b_0>b_*\), which for
\(b_*=Bs^\gamma\) requires \(\gamma<1/(n+1)\) at large \(s\), by Eq.~\eqref{eq:general-saddle}.

The coefficient \(\mathcal P_t\) of the crossed logarithm is computed in the
low-energy EFT, and its use at \(s\gtrsim\Lambda^2\) is not justified by the EFT~\cite{Weinberg}; the high-spin results therefore hold within the phase model
defined above, in which absorption is not included at high spin.

\section{Tree and loop amplitudes and partial waves of the \texorpdfstring{$(\partial\phi)^4$}{(∂φ)⁴} sector}
\label{sec:loops}
\label{sec:amplitudes}

The tree and one-loop amplitudes of the quartic operator and their partial
waves, to which the estimates of Sec.~\ref{sec:split} are applied, are
collected here; the sextic operator is treated in the same way in
Sec.~\ref{sec:sextic}.  Every amplitude is
denoted by \(\M\): a subscript \((\partial\phi)^4\) or \((\partial\phi)^6\) marks the
sector and \(s,t,u\) the channel, a superscript tree, bubble or banana the
fixed-order diagram, as in \(\M^{\rm bubble}_{t,(\partial\phi)^4}\) and
\(\M^{\rm banana}_{t,(\partial\phi)^6}\), and \(\M_{\rm UV}\) is the complete amplitude.

\subsection{Tree and one-loop amplitudes}
\label{sec:x2-amplitude}
\label{sec:quartic}
\label{sec:x2-tree}
\label{sec:x2-loop}

A single insertion of \(g_4(\partial\phi)^4/\Lambda^4\) gives the local tree of
Eq.~\eqref{eq:general-tree-ansatz}, with the partial waves of
Eq.~\eqref{eq:x2-tree-pw-main} (Fig.~\ref{fig:x2-tree}).  In the crossed
process, with \(z=1+2s/t\) the cosine of the angle,
\begin{equation}
 \M^{\rm tree}_{(\partial\phi)^4}=c\,t^2\Bigl[\tfrac53P_0(z)+\tfrac13P_2(z)\Bigr],\label{eq:x2-tree-cyclic}
\end{equation}
it contains only the spins \(J=0,2\) (Appendix~\ref{app:pw}), and since
\(P_J(z)\propto z^J\) at large \(z\), the spin-two component grows as \(s^2\)
at fixed \(t\), Eq.~\eqref{eq:t-channel-spin}.  Two insertions give the one-loop
amplitude
\begin{equation}
\adjustbox{max width=202.8pt}{$\displaystyle
\begin{gathered}
 \begin{aligned}
 \M^{\rm bubble}_{(\partial\phi)^4}={}&-\frac{g_4^2}{60\pi^2\Lambda^8}\sum_{(X;Y,Z)}
 X^2\left(21X^2-YZ\right)\\
 &\times\log\frac{-X-\ii0}{\mu^2}+P_4,
 \end{aligned}\\
 (X;Y,Z)\in\{(s;t,u),(t;u,s),(u;s,t)\},
\end{gathered}
$}\label{eq:x2-channel-decomposition}
\end{equation}
the sum of the three diagrams of Figs.~\ref{fig:x2-loop-s}--\ref{fig:x2-loop-u}.
The counterterm is a crossing-symmetric polynomial of degree four in the
Mandelstam variables; since \(\sigma_2\) and \(\sigma_3\) of
Eq.~\eqref{eq:sigma-def} are of degree two and three, it is
\begin{equation}
 P_4=\frac{g_4^2}{\Lambda^8}\,c_P(\mu)\,\sigma_2^{\,2},\label{eq:x2-P4}
\end{equation}
with one scheme-dependent constant \(c_P(\mu)\) (Appendix~\ref{app:P4}); it
contributes to \(\ell\le4\) and is supported at \(b=0\).  For \(s+t+u=0\) the power sums \(p_k=s^k+t^k+u^k\) obey Newton's
identities~\cite{Macdonald},
\begin{equation}
\begin{gathered}
 p_k=-\sigma_2\,p_{k-2}+\sigma_3\,p_{k-3},\\
 p_0=3,\quad p_1=0,\quad p_2=-2\sigma_2,\quad p_3=3\sigma_3,\\
 p_4=2\sigma_2^{\,2},\qquad p_5=-5\sigma_2\sigma_3,\\
 p_6=-2\sigma_2^{\,3}+3\sigma_3^{\,2},\qquad
 p_8=2\sigma_2^{\,4}-8\sigma_2\sigma_3^{\,2},
\end{gathered}\label{eq:newton}
\end{equation}
and with \(YZ=\sigma_3/X\) the coefficients of the three logarithms of
Eq.~\eqref{eq:x2-channel-decomposition} add up to
\begin{equation}
\begin{aligned}
 \sum_{X=s,t,u}X^2\bigl(21X^2-YZ\bigr)&=21\,p_4-\sigma_3\,p_1\\
 &=21\cdot2\sigma_2^{\,2}-\sigma_3\cdot0=42\,\sigma_2^{\,2} .
\end{aligned}\label{eq:x2-sigma-sum}
\end{equation}
the coefficient of the pole that \(P_4\) cancels (Appendix~\ref{app:P4}).  The prefactor is fixed by unitarity, and its partial waves enter the analysis
complete or, at low spin, separated into the \(s\)-channel and the \(t\)-
and \(u\)-channel parts (Sec.~\ref{sec:x2-low}).

\begin{figure*}[!htbp]
\centering
\begin{minipage}[t]{0.235\textwidth}
\centering\captionsetup{justification=centering,font=footnotesize}
\resizebox{\linewidth}{!}{\fpanel{\node[vtxq] (v) at (0,0) {};\draw[extin] (-1.25,0.95) node[lab,left]{$p_1$} -- (v);\draw[extin] (-1.25,-0.95) node[lab,left]{$p_2$} -- (v);\draw[extout] (v) -- (1.25,0.95) node[lab,right]{$p_3$};\draw[extout] (v) -- (1.25,-0.95) node[lab,right]{$p_4$};}{}}
\caption{Quartic contact tree. Arrows indicate the physical process \(12\to34\).}
\label{fig:x2-tree}
\end{minipage}\hfill
\begin{minipage}[t]{0.235\textwidth}
\centering\captionsetup{justification=centering,font=footnotesize}
\resizebox{\linewidth}{!}{\fpanel{\node[vtxq] (a) at (-0.75,0) {}; \node[vtxq] (b) at (0.75,0) {};\draw[extin] (-1.8,0.95) node[lab,left]{$p_1$} -- (a);\draw[extin] (-1.8,-0.95) node[lab,left]{$p_2$} -- (a);\draw[extout] (b) -- (1.8,0.95) node[lab,right]{$p_3$};\draw[extout] (b) -- (1.8,-0.95) node[lab,right]{$p_4$};\draw[int] (a) to[out=60,in=120] (b); \draw[int] (a) to[out=-60,in=-120] (b);}{(12$|$34)}}
\caption{\(s\)-channel bubble.}
\label{fig:x2-loop-s}
\end{minipage}\hfill
\begin{minipage}[t]{0.235\textwidth}
\centering\captionsetup{justification=centering,font=footnotesize}
\resizebox{\linewidth}{!}{\fpanel{\node[vtxq] (a) at (-0.75,0) {}; \node[vtxq] (b) at (0.75,0) {};\draw[extin] (-1.8,0.95) node[lab,left]{$p_1$} -- (a);\draw[extout] (a) -- (-1.8,-0.95) node[lab,left]{$p_3$};\draw[extin] (1.8,0.95) node[lab,right]{$p_2$} -- (b);\draw[extout] (b) -- (1.8,-0.95) node[lab,right]{$p_4$};\draw[int] (a) to[out=60,in=120] (b); \draw[int] (a) to[out=-60,in=-120] (b);}{(13$|$24)}}
\caption{\(t\)-channel bubble.}
\label{fig:x2-loop-t}
\end{minipage}\hfill
\begin{minipage}[t]{0.235\textwidth}
\centering\captionsetup{justification=centering,font=footnotesize}
\resizebox{\linewidth}{!}{\fpanel{\node[vtxq] (a) at (-0.75,0) {}; \node[vtxq] (b) at (0.75,0) {};\draw[extin] (-1.8,0.95) node[lab,left]{$p_1$} -- (a);\draw[extout] (a) -- (-1.8,-0.95) node[lab,left]{$p_4$};\draw[extin] (1.8,0.95) node[lab,right]{$p_2$} -- (b);\draw[extout] (b) -- (1.8,-0.95) node[lab,right]{$p_3$};\draw[int] (a) to[out=60,in=120] (b); \draw[int] (a) to[out=-60,in=-120] (b);}{(14$|$23)}}
\caption{\(u\)-channel bubble.}
\label{fig:x2-loop-u}
\end{minipage}
\end{figure*}

\subsection{Partial waves}
\label{sec:x2-s-channel}
\label{sec:x2-t-channel}
\label{sec:coefficient}
\label{sec:x2-u-channel}

In the angle \(x\) at fixed \(s\) the three logarithms of
Eq.~\eqref{eq:x2-channel-decomposition} are
\begin{equation}
\begin{gathered}\log\frac{-s-\ii0}{\mu^2}=\log(s/\mu^2)-\ii\pi, \\
\log\frac{-t}{\mu^2}=\log(s/\mu^2)+\log\frac{1-x}{2}, \\
\log\frac{-u}{\mu^2}=\log(s/\mu^2)+\log\frac{1+x}{2},\end{gathered}\label{eq:x2-logs-angle}
\end{equation}
so that the projection~\eqref{eq:pwu-convention} separates three terms: \(\log(s/\mu^2)\)
multiplies a polynomial of degree four in \(x\) and contributes to \(\ell\le4\); the
imaginary part \(-\ii\pi\) multiplies \(\tfrac{s^4}{4}(83+x^2)\), of degree two,
and contributes to only \(\ell=0,2\); and \(\log(1\mp x)\), non-analytic at
\(x=\pm1\), contributes to every \(\ell\), the two endpoints giving equal even and
opposite odd partial waves, so that the odd ones cancel.  The complete
one-loop partial waves are, in units of \(g_4^2s^4/(\pi^3\Lambda^8)\) and up to
the counterterm,
\begin{equation}
\adjustbox{max width=202.8pt}{$\displaystyle
\begin{aligned}
 f_{0,(\partial\phi)^4}^{\rm bubble}
 &=-\frac{49}{1600}\log(s/\mu^2)+\frac{3199}{1\,728\,000}+\frac{25\pi}{1152}\,\ii,\\
 f_{2,(\partial\phi)^4}^{\rm bubble}
 &=-\frac{1}{400}\log(s/\mu^2)-\frac{2561}{12\,096\,000}+\frac{\pi}{28\,800}\,\ii,\\
 f_{4,(\partial\phi)^4}^{\rm bubble}
 &=-\frac{1}{14\,400}\log(s/\mu^2)-\frac{6487}{72\,576\,000},\\
 f_{\ell\ {\rm odd},(\partial\phi)^4}^{\rm bubble}&=0,
\end{aligned}
$}\label{eq:x2-full-pw-main}
\end{equation}
the coefficients of \(\log(s/\mu^2)\) being the projections of the
term in \(\log(s/\mu^2)\) of Eq.~\eqref{eq:x2-channel-decomposition}, whose
coefficient is Eq.~\eqref{eq:x2-sigma-sum}.  With
\(\log[(-X-\ii0)/\mu^2]=\log(s/\mu^2)+\log[(-X-\ii0)/s]\), the bubbles
separate as
\begin{equation}
\adjustbox{max width=207.8pt}{$\displaystyle
\begin{aligned}
 \M^{\rm bubble}_{(\partial\phi)^4}
 &=-\frac{7g_4^2}{10\pi^2\Lambda^8}\,\sigma_2^2\log\frac{s}{\mu^2}\\
 &\quad-\frac{g_4^2}{60\pi^2\Lambda^8}\sum_{X=s,t,u}X^2(21X^2-YZ)\log\frac{-X-\ii0}{s}+P_4 ,
\end{aligned}
$}\label{eq:x2-L-part}
\end{equation}
in which only the first term depends on \(\mu\), the dependence that the
running of \(P_4\) compensates (Appendix~\ref{app:P4}).  In the angle \(x\),
with \(t=-\tfrac s2(1-x)\) and \(u=-\tfrac s2(1+x)\), Eq.~\eqref{eq:pwu-convention},
the invariants are
\begin{equation}
 \sigma_2=-\frac{s^2}{4}\,(3+x^2),\qquad
 \sigma_3=\frac{s^3}{4}\,(1-x^2),\label{eq:sigma-angle}
\end{equation}
so that \(\sigma_2^2\) is a polynomial of degree four in \(x\), and the first
term of Eq.~\eqref{eq:x2-L-part}, like \(P_4\), contributes only to
\(\ell\le4\), Eq.~\eqref{eq:x2-full-pw-main}.
For even \(\ell\ge6\), with (Appendix~\ref{app:pw})
\begin{equation}
 F_k(\ell)\equiv\frac{(\ell-k-1)!}{(\ell+k+1)!},\label{eq:Fk-def}
\end{equation}
the \(t\)-channel projection, Eq.~\eqref{eq:app-t-high}, doubled by the \(u\) channel, Eq.~\eqref{eq:app-u-bubble-pw}, gives the partial waves
\begin{equation}
\adjustbox{max width=202.8pt}{$\displaystyle
\begin{aligned}f_{\ell,(\partial\phi)^4}^{\rm bubble} &=\frac{g_4^2s^4}{120\pi^3\Lambda^8}
 \bigl[F_2(\ell)+9F_3(\ell)+3024F_4(\ell)\bigr] \\
 &\simeq\frac{g_4^2s^4}{120\pi^3\Lambda^8\,\ell^6},\end{aligned}
$}\label{eq:x2-crossed-pw}
\end{equation}
real and fixed entirely by the endpoint non-analyticity.  The imaginary part,
present only at \(\ell=0,2\), is that of the amplitude,
\begin{equation}
\begin{gathered}\operatorname{Im}\M^{\rm bubble}_{(\partial\phi)^4}(s,t)=\frac{c^2s^4}{960\pi}\left(83+x^2\right), \\
\operatorname{Im}\M^{\rm bubble}_{(\partial\phi)^4}(s,0)=\frac{7c^2s^4}{80\pi}=s\,\sigma_{2\to2},\end{gathered}\label{eq:x2-optical}
\end{equation}
the forward limit being the optical theorem~\cite{Collins}.  By elastic unitarity,
Eq.~\eqref{eq:pwu-one-loop-pw}, it is fixed by the tree:
\(\tfrac12\bigl(f_{0,(\partial\phi)^4}^{\rm tree}\bigr)^2\) and
\(\tfrac12\bigl(f_{2,(\partial\phi)^4}^{\rm tree}\bigr)^2\) reproduce the
imaginary parts of Eq.~\eqref{eq:x2-full-pw-main}, and
\(\operatorname{Im}f_{\ell\ge4,(\partial\phi)^4}^{\rm bubble}=0\).  The termination at
\(\ell=2\) reflects the conservation of \(\ell\) through the two-particle
intermediate state~\cite{Collins}, to which the quartic vertex couples only for \(\ell\le2\), Eq.~\eqref{eq:x2-tree-pw-main}.

At fixed \(t\) and large \(s\) the only term non-analytic at \(t=0\) is the
logarithm of \(-t\), which arises only from the term \(X=t\) of
Eq.~\eqref{eq:x2-channel-decomposition}, the \(t\)-channel bubble.  With
\(us=-s^2-st\), so that \(t^2(21t^2-us)=s^2t^2+st^3+21t^4\),
\begin{equation}
\adjustbox{max width=202.8pt}{$\displaystyle
\begin{aligned}
 \M^{\rm bubble}_{t,(\partial\phi)^4}(s,t)
 &=-\frac{g_4^2}{60\pi^2\Lambda^8}
 t^2(21t^2-us)\log\frac{-t}{\mu^2}\\
 &=C(s)\,t^2\Bigl(1+\frac ts+\frac{21t^2}{s^2}\Bigr)\log\frac{-t}{\mu^2},
\end{aligned}
$}\label{eq:x2-regge}
\end{equation}
with \(C(s)=-g_4^2s^2/(60\pi^2\Lambda^8)\) and \(\mu^2=s\), the terms \(t/s\) and \(21t^2/s^2\) being suppressed at large \(s\); the coefficient agrees with that of
\(s^2t^2\log(-t)\) obtained in Ref.~\cite{TokarevaXu}, where this term is traced
to the high-spin part of the dispersive integral.  The other logarithms are analytic at \(t=0\), Eq.~\eqref{eq:forward-one-loop}.  The power \(s^2\) is the
spin-two component of the tree in the crossed process continued to large
\(z\) (Appendix~\ref{app:pw}, Eq.~\eqref{eq:app-Cn}); its projection gives the
high-spin partial waves~\eqref{eq:x2-crossed-pw}.  With the transverse Fourier transform
\begin{equation}
 \int\frac{\dd^2\bm q}{(2\pi)^2}e^{-\ii\bm q\cdot\bm b}
 (q^2)^n\log q^2
 =(-1)^{n+1}\frac{4^n(n!)^2}{\pi b^{2n+2}} ,\label{eq:log-transform}
\end{equation}
up to contact terms, Eq.~\eqref{eq:eikonal-definition} gives the phase
\begin{equation}
\adjustbox{max width=202.8pt}{$\displaystyle
\begin{gathered}\delta_{(\partial\phi)^4}(s,b)=\frac{g_4^2}{\pi^3\Lambda^8}
 \left(\frac{4s}{15b^6}+\frac{48}{5b^8}
 +\frac{64512}{5sb^{10}}\right)
 \simeq\kappa_2\frac{s}{b^6}, \\
\kappa_2=\frac{4g_4^2}{15\pi^3\Lambda^8},\end{gathered}
$}\label{eq:x2-impact-phase}
\end{equation}
equal to \(\tfrac12f_{\ell,(\partial\phi)^4}^{\rm bubble}\) of
Eq.~\eqref{eq:x2-crossed-pw} at \(\ell=b\sqrt s/2\).  

\section{Tree and loop amplitudes and partial waves of the \texorpdfstring{$(\partial\phi)^6$}{(∂φ)⁶} sector}
\label{sec:sextic}

The sextic operator is treated as the quartic one in Sec.~\ref{sec:amplitudes}:
its tree and three-loop amplitudes are followed by their partial waves,
which enter the estimates of Sec.~\ref{sec:split}.

\subsection{Tree and three-loop amplitudes}
\label{sec:x3-amplitude}
\label{sec:x3}
\label{sec:x3-tree}
\label{sec:x3-amplitudes}
\label{sec:x3-banana}

\begin{figure*}[!htbp]
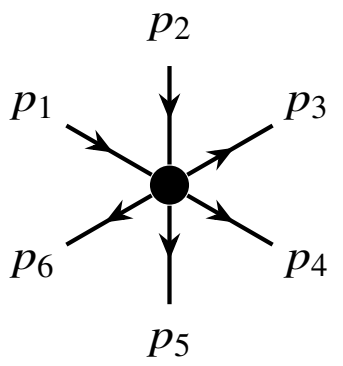

\centering
\begin{minipage}[t]{0.40\textwidth}
\centering\captionsetup{justification=centering,font=footnotesize}
\fpanel{\node[vtx] (v) at (0,0) {};\draw[extin] (150:1.05) node[lab,shift={(150:0.26)}]{$p_1$} -- (v);\draw[extin] (90:1.05) node[lab,shift={(90:0.26)}]{$p_2$} -- (v);\draw[extout] (v) -- (30:1.05) node[lab,shift={(30:0.26)}]{$p_3$};\draw[extout] (v) -- (-30:1.05) node[lab,shift={(-30:0.26)}]{$p_4$};\draw[extout] (v) -- (-90:1.05) node[lab,shift={(-90:0.26)}]{$p_5$};\draw[extout] (v) -- (-150:1.05) node[lab,shift={(-150:0.26)}]{$p_6$};}{}
\caption{Sextic contact tree. Arrows indicate the process \(12\to3456\).}
\label{fig:x3-tree}
\end{minipage}
\end{figure*}

A single \(g_6(\partial\phi)^6/\Lambda^8\) insertion gives the six-point tree.
A derivative acting on an incoming leg gives \(\ii p_\mu\), and each of the
\(15\) pairings of the six legs arises \(3!\,2^3=48\) times among the \(6!\)
contractions of \((\partial\phi\cdot\partial\phi)^3\); with
\(p_i\cdot p_j=-s_{ij}/2\) for massless legs,
\begin{equation}
\begin{gathered}
\begin{aligned}
 \M^{\rm tree}_{6,(\partial\phi)^6}
 &=-48\,\frac{g_6}{\Lambda^8}\sum_{\text{pairings}}
 (p_i\!\cdot p_j)(p_k\!\cdot p_l)(p_m\!\cdot p_n)\\
 &=6\,\frac{g_6}{\Lambda^8}\sum_{\text{15 pairings}}s_{ij}s_{kl}s_{mn},
\end{aligned}\\
\left.\M^{\rm tree}_{(\partial\phi)^6}\right|_{g_6}=0,
\end{gathered}\label{eq:x3-tree-permutation}
\end{equation}
symmetric under all permutations of the six legs (Fig.~\ref{fig:x3-tree}).
The same counting, with three pairings each arising \(8\) times,
reproduces Eq.~\eqref{eq:general-tree-ansatz}.  With no internal line the
vertex gives no elastic four-point tree: at first order in \(g_6\) the two
remaining legs close into the tadpole \(\int\dd^d\ell/\ell^2\), and the
\(g_4g_6\) two-loop graphs depend only on a single external momentum \(p\)
with \(p^2=0\); both integrals are scaleless and vanish in dimensional regularization~\cite{tHooftVeltman}.  The elastic partial waves therefore vanish at this order
and there is no lowest-order real phase, Eqs.~\eqref{eq:x3-tree-permutation} and~\eqref{eq:x3-s-pw-main}.  The operator enters \(2\to2\)
scattering only through the four-particle cut (Fig.~\ref{fig:x3-cut}), whose
imaginary part is the four-particle production rate: quadratically at order
\(g_6^2\), and linearly at order \(g_4^2g_6\) through the interference with the
exchange of one line between two quartic vertices, Sec.~\ref{sec:bounds}.

The elastic \(g_6^2\) amplitude is the three-loop banana of two sextic
vertices joined by four lines
(Figs.~\ref{fig:x3-loop-s}--\ref{fig:x3-loop-u}), homogeneous of degree
\(16\) in the momenta.  Its discontinuity is the four-particle cut of the
sextic tree (Fig.~\ref{fig:x3-cut}, Eq.~\eqref{eq:banana-cut}), and with
crossing
\begin{equation}
\begin{aligned}
 &\M^{\rm banana}_{(\partial\phi)^6}
 =-\frac{g_6^2}{512\,10!\,\pi^6\Lambda^{16}}\sum_{(X;Y,Z)}X^6\\
 &\qquad\times\bigl(1210X^2-31YZ\bigr)\log\frac{-X-\ii0}{\mu^2}+P,
\end{aligned}\label{eq:x3-elastic-full}
\end{equation}
with \((X;Y,Z)\) as in Eq.~\eqref{eq:x2-channel-decomposition}.  The
coefficients of the three logarithms add up to a crossing-symmetric
polynomial in the invariants~\eqref{eq:sigma-def}: with \(YZ=\sigma_3/X\) and
the power sums~\eqref{eq:newton},
\begin{equation}
\begin{aligned}
&\sum_{X=s,t,u}X^6\bigl(1210X^2-31YZ\bigr)=1210\,p_8-31\,\sigma_3\,p_5\\
 &\quad=2420\,\sigma_2^{\,4}-9680\,\sigma_2\sigma_3^{\,2}+155\,\sigma_2\sigma_3^{\,2}\\
 &\quad=2420\,\sigma_2^{\,4}-9525\,\sigma_2\sigma_3^{\,2},
\end{aligned}\label{eq:x3-sigma-sum}
\end{equation}
and the counterterm is a crossing-symmetric polynomial of degree eight in the
Mandelstam variables; since \(\sigma_2\) and \(\sigma_3\) of
Eq.~\eqref{eq:sigma-def} are of degree two and three, the only such
structures are \(\sigma_2^{\,4}\) and \(\sigma_2\sigma_3^{\,2}\).  Since
the amplitude~\eqref{eq:x3-elastic-full} does not depend on \(\mu\), a change
\(\mu_0\to\mu\) of the logarithms, multiplied by the
sum~\eqref{eq:x3-sigma-sum}, is compensated by the counterterm, and
{\allowdisplaybreaks
\begin{align}
 P&=\frac{g_6^2}{\Lambda^{16}}\bigl[c_a(\mu)\,\sigma_2^{\,4}+c_b(\mu)\,\sigma_2\sigma_3^{\,2}\bigr]\nonumber\\
 &=\frac{g_6^2}{\Lambda^{16}}\bigl[c_a(\mu)\,(st+tu+us)^4\nonumber\\
 &\qquad+c_b(\mu)\,(st+tu+us)\,s^2t^2u^2\bigr],\label{eq:x3-P}\\
 c_a(\mu)&=c_a(\mu_0)-\frac{2420}{512\,10!\,\pi^6}\log\frac{\mu^2}{\mu_0^2},\nonumber\\
 c_b(\mu)&=c_b(\mu_0)+\frac{9525}{512\,10!\,\pi^6}\log\frac{\mu^2}{\mu_0^2},\nonumber
\end{align}}
where the values \(c_a(\mu_0)\) and \(c_b(\mu_0)\) at a reference scale are
scheme dependent and fixed by matching to the completion~\cite{Weinberg}, as
\(c_P\) of Eq.~\eqref{eq:x2-P4} (Appendix~\ref{app:P4}); \(P\) contributes to
\(\ell\le8\) and is supported at \(b=0\).  At fixed \(t=-q^2\) the only term non-analytic at \(t=0\) is the
logarithm of \(-t\), which arises only from the term \(X=t\) of
Eq.~\eqref{eq:x3-elastic-full}, the \(t\)-channel banana.  With \(u=-s+q^2\),
so that \(1210t^2-31su=31s^2-31sq^2+1210q^4\),
\begin{equation}
\begin{aligned}
 &\M^{\rm banana}_{t,(\partial\phi)^6}(s,-q^2)
 =-\frac{g_6^2}{512\,10!\,\pi^6\Lambda^{16}}\,q^{12}\\
 &\qquad\times\bigl(31s^2-31sq^2+1210q^4\bigr)\log\frac{q^2}{\mu^2},
\end{aligned}\label{eq:x3-t-channel-exact}
\end{equation}
which grows as \(s^2\), like its quartic counterpart~\eqref{eq:x2-regge}, but
with \(t^6\); the other two logarithms are analytic at \(t=0\) and grow as
\(s^8\log s\), contained in a fixed number of partial waves.

\begin{figure*}[!htbp]
\centering
\begin{minipage}[t]{0.235\textwidth}
\centering\captionsetup{justification=centering,font=footnotesize}
\resizebox{\linewidth}{!}{\fpanel{\node[vtx] (a) at (-0.95,0) {}; \node[vtx] (b) at (0.95,0) {};\draw[extin] (-1.9,0.95) node[lab,left]{$p_1$} -- (a);\draw[extin] (-1.9,-0.95) node[lab,left]{$p_2$} -- (a);\draw[extout] (b) -- (1.9,0.95) node[lab,right]{$p_3$};\draw[extout] (b) -- (1.9,-0.95) node[lab,right]{$p_4$};\foreach \o in {72,26,-26,-72} \draw[int] (a) to[out=\o,in=180-\o] (b);}{(12$|$34)}}
\caption{\(s\)-channel banana.}
\label{fig:x3-loop-s}
\end{minipage}\hfill
\begin{minipage}[t]{0.235\textwidth}
\centering\captionsetup{justification=centering,font=footnotesize}
\resizebox{\linewidth}{!}{\fpanel{\node[vtx] (a) at (-0.95,0) {}; \node[vtx] (b) at (0.95,0) {};\draw[extin] (-1.9,0.95) node[lab,left]{$p_1$} -- (a);\draw[extout] (a) -- (-1.9,-0.95) node[lab,left]{$p_3$};\draw[extin] (1.9,0.95) node[lab,right]{$p_2$} -- (b);\draw[extout] (b) -- (1.9,-0.95) node[lab,right]{$p_4$};\foreach \o in {72,26,-26,-72} \draw[int] (a) to[out=\o,in=180-\o] (b);}{(13$|$24)}}
\caption{\(t\)-channel banana.}
\label{fig:x3-loop-t}
\end{minipage}\hfill
\begin{minipage}[t]{0.235\textwidth}
\centering\captionsetup{justification=centering,font=footnotesize}
\resizebox{\linewidth}{!}{\fpanel{\node[vtx] (a) at (-0.95,0) {}; \node[vtx] (b) at (0.95,0) {};\draw[extin] (-1.9,0.95) node[lab,left]{$p_1$} -- (a);\draw[extout] (a) -- (-1.9,-0.95) node[lab,left]{$p_4$};\draw[extin] (1.9,0.95) node[lab,right]{$p_2$} -- (b);\draw[extout] (b) -- (1.9,-0.95) node[lab,right]{$p_3$};\foreach \o in {72,26,-26,-72} \draw[int] (a) to[out=\o,in=180-\o] (b);}{(14$|$23)}}
\caption{\(u\)-channel banana.}
\label{fig:x3-loop-u}
\end{minipage}\hfill
\begin{minipage}[t]{0.235\textwidth}
\centering\captionsetup{justification=centering,font=footnotesize}
\resizebox{\linewidth}{!}{\fpanel{\node[vtx] (a) at (-0.95,0) {}; \node[vtx] (b) at (0.95,0) {};\draw[extin] (-1.9,0.95) node[lab,left]{$p_1$} -- (a);\draw[extin] (-1.9,-0.95) node[lab,left]{$p_2$} -- (a);\draw[extout] (b) -- (1.9,0.95) node[lab,right]{$p_3$};\draw[extout] (b) -- (1.9,-0.95) node[lab,right]{$p_4$};\foreach \o in {72,26,-26,-72} \draw[int] (a) to[out=\o,in=180-\o] (b);\draw[cut] (0,1.3) -- (0,-1.3);}{(12$|$34)}}
\caption{Four-particle cut of the \(s\)-channel banana.}
\label{fig:x3-cut}
\end{minipage}
\end{figure*}

\subsection{Partial waves}\label{sec:x3-pw}
\label{sec:x3-s-channel}
\label{sec:x3-t-channel}
\label{sec:x3-u-channel}

With the logarithms~\eqref{eq:x2-logs-angle}, the projection of
Eq.~\eqref{eq:x3-elastic-full} again separates three terms.  The coefficient of
\(\log(s/\mu^2)\), Eq.~\eqref{eq:x3-sigma-sum}, and the coefficient of
\(-\ii\pi\), the term \(X=s\), follow from Eq.~\eqref{eq:sigma-angle}, with
\(tu=\sigma_3/s\) and \(x^2=\tfrac13\bigl[P_0(x)+2P_2(x)\bigr]\), as
\begin{equation}
\adjustbox{max width=197.8pt}{$\displaystyle
\begin{aligned}
 2420\,\sigma_2^{\,4}-9525\,\sigma_2\sigma_3^{\,2}
 &=\frac{s^8}{64}\Bigl[605\,(3+x^2)^4\\
 &\qquad+9525\,(3+x^2)(1-x^2)^2\Bigr],\\
 s^6\bigl(1210s^2-31tu\bigr)
 &=\frac{s^8}{4}\,\bigl(4809+31x^2\bigr)\\
 &=s^8\Bigl[\tfrac{7229}{6}P_0(x)+\tfrac{31}{6}P_2(x)\Bigr],
\end{aligned}
$}\label{eq:x3-coeff-angle}
\end{equation}
polynomials of degree eight and two in \(x\): \(\log(s/\mu^2)\) contributes to
\(\ell\le8\), \(-\ii\pi\) to only \(\ell=0,2\), and \(\log(1\mp x)\) to every even
\(\ell\).  The imaginary part of the amplitude,
\begin{equation}
\begin{aligned}
 &\operatorname{Im}\M^{\rm banana}_{(\partial\phi)^6}
 =\frac{g_6^2}{1\,857\,945\,600\,\pi^5\Lambda^{16}}\\
 &\qquad\times s^6\bigl(1210s^2-31tu\bigr),\\
 &\operatorname{Im}\M^{\rm banana}_{(\partial\phi)^6}(s,0)=\frac{121\,g_6^2s^8}{185\,794\,560\,\pi^5\Lambda^{16}},
\end{aligned}\label{eq:banana-imaginary}
\end{equation}
is reproduced by a RAMBO~\cite{RAMBO,RAMBOdiet} evaluation of the cut to
\(2\times10^{-4}\) (Appendix~\ref{app:rambo}, Table~\ref{tab:rambo}).  The
complete partial waves with \(\ell\le8\) are listed in
Table~\ref{tab:x3-pw} of Appendix~\ref{app:pw}; their imaginary parts,
\begin{equation}
\begin{gathered}\operatorname{Im}f_{0,(\partial\phi)^6}^{\rm banana}
 =\frac{7229\,g_6^2s^8}{178\,362\,777\,600\,\pi^6\Lambda^{16}}, \\
\operatorname{Im}f_{2,(\partial\phi)^6}^{\rm banana}
 =\frac{31\,g_6^2s^8}{891\,813\,888\,000\,\pi^6\Lambda^{16}}, \\
1-\eta_\ell^2=2\operatorname{Im}f_{\ell,(\partial\phi)^6}^{\rm banana},\end{gathered}\label{eq:x3-s-pw-main}
\end{equation}
are the four-particle production rate.  They terminate at \(\ell=2\) for the same reason as in the quartic
sector, the sextic vertex, linear in every external momentum, coupling only to \(\ell\le2\), Eq.~\eqref{eq:app-x3-s}.  For even \(\ell\ge10\) (Appendix~\ref{app:pw})
\begin{equation}
\begin{aligned}
 f_{\ell,(\partial\phi)^6}^{\rm banana}
 &=\frac{g_6^2s^8}{\pi^7\Lambda^{16}}
 \Bigl[\frac{31}{28\,672}F_6(\ell)+\frac{217}{4096}F_7(\ell)\\
 &\qquad+\frac{4235}{32}F_8(\ell)\Bigr]\\
 &\simeq\frac{31\,g_6^2s^8}{28\,672\,\pi^7\Lambda^{16}\,\ell^{14}},
\end{aligned}\label{eq:x3-t-pw-main}
\end{equation}
real and falling as \(\ell^{-14}\), the profile being set by \(t^6\) in
Eq.~\eqref{eq:x3-t-channel-exact}; odd partial waves vanish.

\section{Eikonal estimate of the Regge growth: the \texorpdfstring{$(\partial\phi)^4$}{(∂φ)⁴} sector}
\label{sec:x2-low}
\label{sec:x2-low-pwu}

The partial-wave sum of the sector is split at \(\ell_*=b_*\sqrt s/2\),
\begin{equation}
\begin{aligned}&\M^{\rm tree}_{(\partial\phi)^4}+\M^{\rm bubble}_{(\partial\phi)^4}\\
 &\qquad=\MlowII+\MeikII ,\end{aligned}\label{eq:x2-low-eikonal-split}
\end{equation}
where the left-hand side is the amplitude truncated at one loop and the two
parts are its partial-wave sums over \(\ell<\ell_*\) and
\(\ell\ge\ell_*\); the subscripts name the diagrams of the truncation, not
separate pieces of the complete amplitude \(\M_{\rm UV}\).  The truncation is
unitary only order by order: with
\(S_\ell=1+\ii\bigl(f^{\rm tree}_{\ell,(\partial\phi)^4}+f^{\rm bubble}_{\ell,(\partial\phi)^4}\bigr)\)
and Eq.~\eqref{eq:pwu-one-loop-pw},
\begin{equation}
\begin{aligned}|S_\ell|^2-1 &=2f^{\rm tree}_{\ell,(\partial\phi)^4}\operatorname{Re}f^{\rm bubble}_{\ell,(\partial\phi)^4}\\
 &\quad+\bigl|f^{\rm bubble}_{\ell,(\partial\phi)^4}\bigr|^2=O(g_4^3),\end{aligned}\label{eq:x2-truncated-unitarity}
\end{equation}
the order \(g_4^2\) of Eq.~\eqref{eq:pwu-tree-violation} being cancelled and a
remainder of order \(g_4^3s^6\log s/\Lambda^{12}\), which grows without bound,
being left.  The bounds on the low-spin part
below are therefore bounds on the partial waves \(f_\ell\) of \(\M_{\rm UV}\),
which obey Eq.~\eqref{eq:pwu-unitarity} exactly, the truncated partial waves
being their first two terms while \(f^{\rm tree}_{0,(\partial\phi)^4}\ll1\),
\(s\ll\Lambda^2(24\pi/5g_4)^{1/2}\), Eq.~\eqref{eq:x2-tree-pw-main}; the
high-spin part enters only through the leading term of the phase,
Eq.~\eqref{eq:x2-delta-ell}, and \(S_\ell=e^{2\ii\delta_\ell}\) is unitary for every real
\(\delta_\ell\).  The growth of the sector is to be compared with the \(s^2\)
required by Eq.~\eqref{eq:regge-bounded-condition}.

\subsection{Tree and \texorpdfstring{$s$}{s}-channel bubble: low spin}
\label{sec:x2-low-s}

The partial waves
\(\ell\le4\) of Eq.~\eqref{eq:x2-full-pw-main} and of the tree grow as \(s^2\)
and \(s^4\log s\) at fixed order; the tree alone violates probability conservation, Eq.~\eqref{eq:pwu-tree-violation}, and at sufficiently large \(s\) its \(s\) wave lies outside the unitarity disk.  The one-loop imaginary parts,
Eq.~\eqref{eq:x2-full-pw-main}, cancel the excess of the tree, and summed
over \(\ell\) with \(P_\ell(1)=1\) they give the forward absorptive part of
Eq.~\eqref{eq:x2-optical},
\(8\pi\bigl[(f_{0,(\partial\phi)^4}^{\rm tree})^2+5(f_{2,(\partial\phi)^4}^{\rm tree})^2\bigr]=7g_4^2s^4/(20\pi\Lambda^8)\),
the low-energy input of the dispersion relation in Sec.~\ref{sec:bounds}.  The low-spin part contains
two kinds of contribution.  The tree and the \(s\)-channel bubble depend on the angle only
through \(s^2+t^2+u^2\) and \(tu=\tfrac14s^2(1-x^2)\), polynomials of degree two
in \(x\), and contribute to only \(\ell=0,2\), Eqs.~\eqref{eq:int-orth} and~\eqref{eq:app-s-bubble-pw}.  With the \(s\)-channel logarithm
\(\log(s/\mu^2)-\ii\pi\) of Eq.~\eqref{eq:x2-logs-angle}, the \(s\)-channel bubble projects
onto
\begin{equation}
\begin{gathered}f_{\ell,s,(\partial\phi)^4}^{\rm bubble}
 =\frac{\ii\pi-\log(s/\mu^2)}{\pi}\operatorname{Im}f_{\ell,(\partial\phi)^4}^{\rm bubble}, \\
\ell=0,2, \\
f_{\ell\ge4,s,(\partial\phi)^4}^{\rm bubble}=0,\end{gathered}\label{eq:x2-s-channel-pw}
\end{equation}
with the imaginary parts of Eq.~\eqref{eq:x2-full-pw-main}, so that
\begin{equation}
\begin{aligned}\MlowIIs &=16\pi\sum_{\ell=0,2}(2\ell+1)
 \Bigl[f_{\ell,(\partial\phi)^4}^{\rm tree} \\
 &\quad+f_{\ell,s,(\partial\phi)^4}^{\rm bubble}\Bigr]P_\ell(x)\end{aligned}\label{eq:x2-low-s-part}
\end{equation}
contains the tree and the entire imaginary part.  At fixed order it is
largest in the forward direction, where the tree gives
\begin{equation}
\adjustbox{max width=202.8pt}{$\displaystyle
\bigl|\M^{\rm tree}_{(\partial\phi)^4}\bigr|\le\frac{4g_4}{\Lambda^4}\,s^2
 =16\pi\Bigl(f_{0,(\partial\phi)^4}^{\rm tree}+5f_{2,(\partial\phi)^4}^{\rm tree}\Bigr),
$}\label{eq:x2-low-fixed-order}
\end{equation}
while with \(|P_\ell(x)|\le1\)~\cite{Szego}, sharp at fixed \(\ell\) as \(x\to1\), exact unitarity of the partial waves of \(\M_{\rm UV}\), of which Eq.~\eqref{eq:x2-low-s-part} is the truncation,
applied to these two partial waves gives
\begin{equation}
 \bigl|\MlowIIs\bigr|
 \le16\pi\,(1+5)\cdot2=192\pi :\label{eq:x2-pwu-explicit-bound}
\end{equation}
the growth \(s^2\) and \(s^4\log s\) of the tree and the \(s\)-channel bubble is
replaced by a constant.

\subsection{\texorpdfstring{$t$}{t}- and \texorpdfstring{$u$}{u}-channel bubbles: low spin}
\label{sec:x2-low-spin}

The \(t\)- and \(u\)-channel bubbles contain \(\log(1\mp x)\) and contribute to every even \(\ell\), the low spins included, Eqs.~\eqref{eq:x2-logs-angle} and~\eqref{eq:x2-crossed-pw}, and are treated separately.
Subtracting Eq.~\eqref{eq:x2-s-channel-pw} from
Eq.~\eqref{eq:x2-full-pw-main}, their partial waves at \(\ell\le4\) are, in
units of \(g_4^2s^4/(\pi^3\Lambda^8)\) and up to the counterterm,
\begin{equation}
\begin{gathered}f_{0,t+u,(\partial\phi)^4}^{\rm bubble}=-\tfrac{257}{28\,800}\log(s/\mu^2)+\tfrac{3199}{1\,728\,000}, \\
f_{2,t+u,(\partial\phi)^4}^{\rm bubble}=-\tfrac{71}{28\,800}\log(s/\mu^2)-\tfrac{2561}{12\,096\,000}, \\
f_{4,t+u,(\partial\phi)^4}^{\rm bubble}=f_{4,(\partial\phi)^4}^{\rm bubble},\end{gathered}\label{eq:x2-tu-low-pw}
\end{equation}
all real, and for even \(\ell\ge6\) they are given by Eq.~\eqref{eq:x2-crossed-pw}.
They grow as \(s^4\log s\) at \(\ell\le4\) and as \(s^4\) at \(\ell\ge6\), and they
do not terminate at a fixed spin: the counterterm \(P_4\), supported at \(b=0\),
adds only to \(\ell\le4\), Eq.~\eqref{eq:x2-P4}, and is included with them, while the partial waves
\(\ell<\ell_*\) number \(\ell_*\propto s^{1/2+\gamma}\).  Each is bounded by
\(|f_\ell|\le2\), and Eq.~\eqref{eq:method-low-spin-bound} gives for the \(t\)-
and \(u\)-channel low-spin part (Table~\ref{tab:x2-assignment})
\begin{equation}
\adjustbox{max width=202.8pt}{$\displaystyle
\begin{gathered}\bigl|\MlowIItu\bigr|
 \lesssim\frac{4}{3\sqrt q}s^{1/4}\ell_*^{3/2}
 =\frac{\sqrt2}{3}\frac{s\,b_*^{3/2}}{\sqrt q}, \\
\MlowII=\MlowIIs+\MlowIItu ,\end{gathered}
$}\label{eq:x2-low-master}
\end{equation}
which grows with \(s\), exceeds the constant~\eqref{eq:x2-pwu-explicit-bound}
and therefore bounds the entire low-spin part.

\begin{table*}[!t]
\centering
\caption{Assignment of the \((\partial\phi)^4\) partial waves to the two regions of
the partial-wave sum.}
\label{tab:x2-assignment}
\small
\begin{tabular}{>{\raggedright\arraybackslash}p{5.8cm}>{\raggedright\arraybackslash}p{4.9cm}>{\raggedright\arraybackslash}p{5.3cm}}
\hline
contribution & \(\MlowII\) (\(\ell<\ell_*\)) &
\(\MeikII\) (\(\ell>\ell_*\))\\
\hline
tree and \(s\)-channel bubble, \(\ell=0,2\) & all, complex, elastic, \(\le192\pi\) & none, support at \(b=0\)\\
\(t\)- and \(u\)-channel bubbles and \(P_4\), \(\ell\le4\) & all, real & none\\
\(t\)- and \(u\)-channel bubbles, even \(\ell\ge6\), \(\propto\ell^{-6}\) & \(\ell<\ell_*\), real &
high-spin part, real, \(\delta_{(\partial\phi)^4}=\kappa_2s/b^6\)\\
\hline
treatment & unitarity bound & phase matching and stationary phase\\
\hline
\end{tabular}
\end{table*}

\subsection{\texorpdfstring{$t$}{t}- and \texorpdfstring{$u$}{u}-channel bubbles: eikonal}
\label{sec:x2-eikonal-matching}
\label{sec:x2-u-eikonal}
\label{sec:x2-matching}
The high-spin partial waves are real, Eq.~\eqref{eq:x2-crossed-pw}, and define the phase; for even
\(\ell\ge6\),
\begin{equation}
 \delta_\ell(s)=\tfrac12f_{\ell,(\partial\phi)^4}^{\rm bubble}
 \simeq\frac{g_4^2s^4}{240\pi^3\Lambda^8\,\ell^6}.\label{eq:x2-delta-ell}
\end{equation}
In impact parameter the phase~\eqref{eq:x2-impact-phase},
which grows as \(s\) at fixed \(b\) and returns the fixed-order
\(s^2q^4\log q^2\) when the exponential is expanded once, is stationary where
the classical deflection equals the momentum transfer,
Eqs.~\eqref{eq:general-saddle} and~\eqref{eq:general-saddle-amplitude} with
\(n=6\),
\begin{equation}
\adjustbox{max width=202.8pt}{$\displaystyle
\begin{gathered}b_{2,\mathrm{sad}}^7=\frac{12\kappa_2s}{q}, \\
\left|\MeikII(s,-q^2)\right|
 \simeq\frac{4\pi}{\sqrt7}\,\frac{s\,b_{2,\mathrm{sad}}}{q}
 \ \propto\ \kappa_2^{1/7}s^{8/7}q^{-8/7}.\end{gathered}
$}\label{eq:x2-saddle-scaling}
\end{equation}
The corresponding spin, \(\ell_{\rm sad}=b_{2,\mathrm{sad}}\sqrt s/2\propto s^{9/14}\),
grows faster than \(\ell_*\propto s^{25/42}\), so the saddle lies in the
high-spin region.  Since only even partial waves
occur, \(P_\ell(-x)=P_\ell(x)\), Eq.~\eqref{eq:int-orth}, and the same peak appears at fixed
\(u=-q_u^2\),
\begin{equation}
 \left|\MeikII(s,u)\right|
 \sim\kappa_2^{1/7}\,s^{8/7}(-u)^{-4/7},\label{eq:u-regge-x2}
\end{equation}
the forward and backward peaks being mirror images, as Bose symmetry
requires.

The transition is fixed by \(|\MlowII(s;b_*)|\sim|\MeikII(s;b_*)|\).  With
\(b_*=B_2s^{\gamma_2}\), Eq.~\eqref{eq:x2-low-master} gives
\(s^{1+3\gamma_2/2}\), equal to \(s^{8/7}\) for
\begin{equation}
\adjustbox{max width=202.8pt}{$\displaystyle
\begin{gathered}\gamma_2=\frac{2}{21},\qquad b_*=B_2s^{2/21},\qquad \ell_*=\frac{B_2}{2}s^{25/42}, \\
|\MlowII|\lesssim\frac{\sqrt2}{3}B_2^{3/2}\frac{s^{8/7}}{\sqrt q},\end{gathered}
$}\label{eq:x2-low-high-matched}
\end{equation}
Since \(\gamma_2<1/7\), the saddle \(b_{2,\mathrm{sad}}\propto s^{1/7}\) stays in
the eikonal region.  \(B_2\) is an undetermined \(O(1)\) constant, and \(b_*\) is defined by the matching prescription.  The \((\partial\phi)^4\) sector thus obeys
\begin{equation}
\adjustbox{max width=202.8pt}{$\displaystyle
\begin{gathered}
 |\MlowII|\lesssim s^{8/7},\qquad
 |\MeikII|\sim s^{8/7},\\
 |\MlowII+\MeikII|\lesssim s^{8/7}.
\end{gathered}
$}\label{eq:x2-final-powers}
\end{equation}
The assignment of the partial waves is summarized in
Table~\ref{tab:x2-assignment}; with the transition scale fixed by the cutoff
both parts are at most \(s^{7/4}\), Eq.~\eqref{eq:eft-controlled-bound}.

\section{Eikonal estimate of the Regge growth: the \texorpdfstring{$(\partial\phi)^6$}{(∂φ)⁶} sector}
\label{sec:x3-low-eikonal}

\subsection{\texorpdfstring{$s$}{s}-channel banana: low spin}
\label{sec:x3-low-pwu}
\label{sec:x3-low}
\label{sec:x3-low-spin}

At \(\ell=0,2\) the imaginary part of the banana is the four-particle
production rate, the absorption term of Eq.~\eqref{eq:pwu-absorption}.
With a complex phase,
\begin{equation}
\begin{gathered}S_\ell=\eta_\ell\,e^{2\ii\delta_\ell}=e^{2\ii(\delta_\ell+\ii\chi_\ell)}, \\
f_\ell\simeq2\delta_\ell+2\ii\chi_\ell, \\
1-\eta_\ell^2\simeq4\chi_\ell,\end{gathered}\label{eq:x3-complex-phase}
\end{equation}
the banana gives \(\chi_\ell\neq0\) at \(\ell=0,2\), so a real phase cannot describe these partial waves~\cite{Collins}.  There is no elastic tree, and the \(s\)-channel banana, a function of \(s\) and
\(tu\) of degree two in \(x\), contributes to only \(\ell=0,2\),
\begin{equation}
\begin{gathered}f_{\ell,s,(\partial\phi)^6}^{\rm banana}
 =\frac{\ii\pi-\log(s/\mu^2)}{\pi}\operatorname{Im}f_{\ell,(\partial\phi)^6}^{\rm banana}, \\
\ell=0,2, \\
f_{\ell\ge4,s,(\partial\phi)^6}^{\rm banana}=0,\end{gathered}\label{eq:x3-s-channel-pw}
\end{equation}
with the imaginary parts of Eq.~\eqref{eq:x3-s-pw-main}.  It contains the entire
absorption, grows as \(s^8\log s\), and at sufficiently large \(s\) its \(s\) wave saturates the unitarity bound, Eq.~\eqref{eq:pwu-unitarity}; as in the quartic sector,
unitarity, applied as in Sec.~\ref{sec:x2-low} to the partial waves of
\(\M_{\rm UV}\) and not to the truncation, for which
\(|S_\ell|^2-1=-2\operatorname{Im}f^{\rm banana}_{\ell,(\partial\phi)^6}+|f^{\rm banana}_{\ell,(\partial\phi)^6}|^2\) is negative only while the banana is small, replaces this growth by a constant,
\begin{equation}
 \bigl|\MlowIIIs\bigr|\le16\pi\,(1+5)\cdot2=192\pi .\label{eq:x3-pwu-explicit-bound}
\end{equation}

\subsection{\texorpdfstring{$t$}{t}- and \texorpdfstring{$u$}{u}-channel bananas: eikonal}
\label{sec:x3-high-pw}
\label{sec:x3-u-backward}
The high-spin partial waves are real, the absorption residing at
\(\ell\le2\).  For even \(\ell\ge10\),
\begin{equation}
 \delta_\ell(s)=\tfrac12f_{\ell,(\partial\phi)^6}^{\rm banana}
 \simeq\frac{31\,g_6^2s^8}{57\,344\,\pi^7\Lambda^{16}\,\ell^{14}},\label{eq:x3-delta-ell}
\end{equation}
and in impact parameter the three terms of Eq.~\eqref{eq:x3-t-channel-exact}
give
\begin{equation}
\adjustbox{max width=202.8pt}{$\displaystyle
\begin{gathered}\begin{aligned}\delta_{(\partial\phi)^6}(s,b) &=\frac{g_6^2}{\pi^7\Lambda^{16}}
 \left(\frac{62s}{7b^{14}}+\frac{1736}{b^{16}}
 +\frac{17346560}{sb^{18}}\right) \\
 &\simeq\kappa_3\frac{s}{b^{14}},\end{aligned} \\
\kappa_3=\frac{62g_6^2}{7\pi^7\Lambda^{16}},\end{gathered}
$}\label{eq:x3-phase-boundary}
\end{equation}
the leading term being \(\delta_\ell\) at \(\ell=b\sqrt s/2\).  Exponentiated,
Eqs.~\eqref{eq:general-saddle} and~\eqref{eq:general-saddle-amplitude} with
\(n=14\) give
\begin{equation}
\adjustbox{max width=202.8pt}{$\displaystyle
\begin{gathered}b_0^{15}=\frac{28\kappa_3s}{q}, \\
|\MeikIII|
 \simeq\frac{4\pi}{\sqrt{15}}\,\frac{s\,b_0}{q}
 \ \propto\ \kappa_3^{1/15}s^{16/15}q^{-16/15},\end{gathered}
$}\label{eq:x3-high-scaling}
\end{equation}
independent of \(b_*\) as long as the saddle lies in \(b>b_*\), which holds
for \(\gamma_3=2/45<1/15\) but not for \(\gamma_2=2/21\); the same peak
appears at fixed \(u=-q_u^2\),
\begin{equation}
 \left|\MeikIII(s,u)\right|
 \sim\kappa_3^{1/15}\,s^{16/15}(-u)^{-8/15}.\label{eq:u-regge-x3}
\end{equation}

With both operators the transition is fixed by the quartic matching of
Sec.~\ref{sec:x2-matching}, \(\gamma=\gamma_2=2/21\), and
\(|\MlowIII|\lesssim s^{8/7}\); for \(g_4\to0\) the self-consistent choice is
matching to the high-spin result~\eqref{eq:x3-high-scaling},
\begin{equation}
\begin{gathered}\gamma_3=\frac{2}{45}, \\
b_*=B_3s^{2/45}, \\
|\MlowIII|
 \lesssim
 \frac{\sqrt2}{3}B_3^{3/2}\frac{s^{16/15}}{\sqrt q},\end{gathered}\label{eq:x3-self-matching}
\end{equation}
which keeps the saddle in the eikonal region, \(\gamma_3<1/15\).

\begin{figure}[!tbp]
\centering
\adjustbox{max width=\linewidth}{%
\begin{tikzpicture}[x=1cm,y=1cm,font=\footnotesize]
 \fill[gray!18] (0,0) rectangle (2.0,2.6);
 \clip (-0.6,-1.1) rectangle (8.4,2.85);
 \fill[pattern=north east lines,pattern color=gray!60] (5.6,0) rectangle (7.9,2.6);
 \draw[->,thick] (0,0) -- (8.1,0) node[below]{$b$};
 \draw[->,thick] (0,0) -- (0,2.8) node[left]{$\delta$};
 \draw[thick,dotted,domain=1.95:2.0,samples=8] plot (\x,{2.4*pow(2/\x,7)});
 \draw[thick,domain=2.0:7.8,samples=120,smooth] plot (\x,{2.4*pow(2/\x,7)});
 \node[anchor=west] at (2.75,1.25) {$\delta\simeq\kappa_3\,s/b^{14}$};
 \draw[dashed] (2.0,0) -- (2.0,2.6);
 \draw[dashed] (5.6,0) -- (5.6,2.6);
 \draw[densely dotted] (2.45,0) -- (2.45,{2.4*pow(2/2.45,7)});
 \fill (2.45,{2.4*pow(2/2.45,7)}) circle (1.5pt);
 \node[below] at (1.75,0) {$b_*\propto s^{2/45}$};
 \node[below] at (3.3,0) {$b_0\propto s^{1/15}$};
 \node[below] at (5.75,0) {$\sqrt s/\Lambda^2$};
 \node[align=center] at (1.0,2.05) {low spin\\[-1pt]$|S_\ell|\le1$};
 \node[align=center] at (3.8,2.3) {eikonal};
 \node[align=center,fill=white,inner sep=1.5pt] at (6.75,2.05) {EFT phase\\[-1pt]controlled};
 \node[align=center] at (2.0,-0.72) {$\Rightarrow s^{16/15}$};
 \node[align=center] at (5.75,-0.72) {$b_*^{\rm EFT}\Rightarrow s^{7/4}$};
\end{tikzpicture}}
\caption{Impact-parameter picture of the \((\partial\phi)^6\) estimate for \(g_4\to0\)
at fixed \(s\) and \(t=-q^2\).  Partial waves with \(b<b_*\) are bounded by
unitarity, Eq.~\eqref{eq:method-low-spin-bound}; for \(b>b_*\) the phase of
Eq.~\eqref{eq:x3-phase-boundary} is exponentiated, its stationary point \(b_0\)
follows from Eq.~\eqref{eq:x3-high-scaling}, and matching at
\(b_*\propto s^{2/45}\) gives \(s^{16/15}\), Eq.~\eqref{eq:x3-final-powers}.}
\label{fig:overview-x3}
\end{figure}
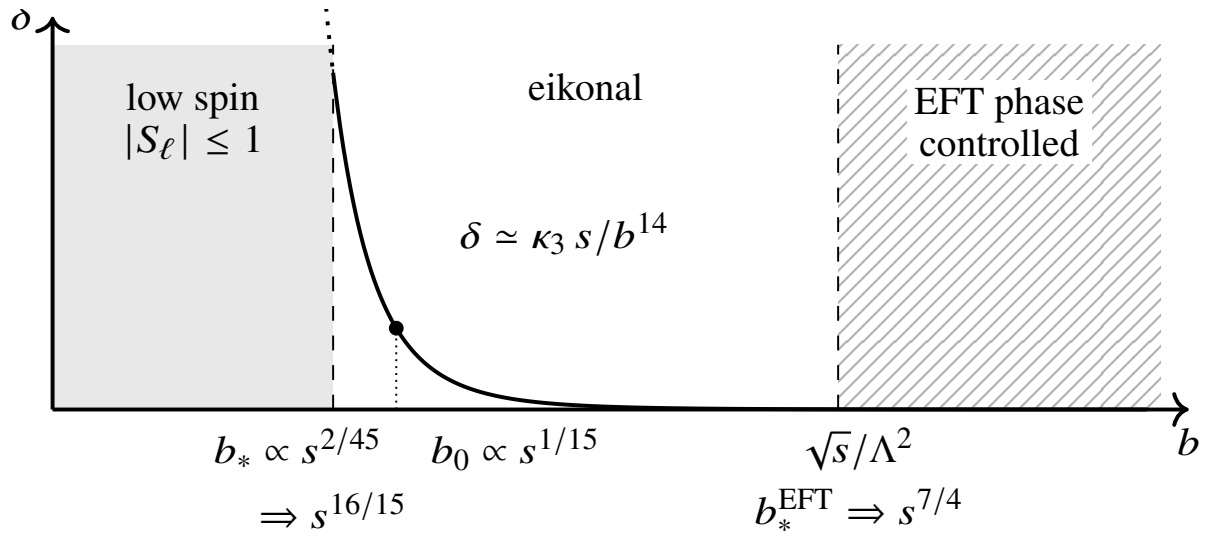

Since the low-spin result, Eq.~\eqref{eq:x3-self-matching}, is a bound and the eikonal result, Eq.~\eqref{eq:x3-high-scaling}, an estimate, the larger exponent sets the total.  With the quartic-determined scale, and
for \(g_4\to0\),
\begin{equation}
\begin{gathered}|\MlowIII|,\ |\M^{\rm banana}_{(\partial\phi)^6}|\lesssim s^{8/7}; \\
|\MlowIII|,\ |\M^{\rm banana}_{(\partial\phi)^6}|\lesssim s^{16/15}\quad (g_4\to0),\end{gathered}\label{eq:x3-final-powers}
\end{equation}
so \(s^{16/15}\) bounds the total only in the truncation \(g_4\to0\). The corresponding impact-parameter picture is Fig.~\ref{fig:overview-x3}.
  The
partial-wave contributions are collected in Table~\ref{tab:x3-growth}, the
constant in the first row bounding the complete amplitude.
\begin{table}[!t]
\centering
\caption{Growth of the \((\partial\phi)^6\) partial waves and of the
corresponding parts of the amplitude; the \(t,u\)-channel exponents hold for
\(g_4\to0\).}
\label{tab:x3-growth}
\footnotesize\setlength{\tabcolsep}{3.5pt}
\renewcommand{\arraystretch}{1.3}
\begin{tabular}{lll}
\hline
channel, spin & \(f_\ell\) & growth satisfying unitarity\\
\hline
\(s\), \(\ell=0,2\) & \(s^8\log s\) & \(\bigl|\MlowIIIs\bigr|\le192\pi\)\\
\(t,u\), \(\ell<\ell_*\) & \(s^8\log s,\ s^8/\ell^{14}\) & \(\bigl|\MlowIIItu\bigr|\propto s^{16/15}\)\\
\(t,u\), \(\ell>\ell_*\) & \(s^8/\ell^{14}\) & \(\bigl|\MeikIII\bigr|\propto s^{16/15}\)\\
\hline
\end{tabular}
\end{table}

\section{Combined phases, the EFT-controlled bound and the forward limit}
\label{sec:growth}

\subsection{Combined phases}
\label{sec:x3-joint-phase}

\begin{figure}[!tbp]
\centering
\adjustbox{max width=\linewidth}{%
\begin{tikzpicture}[x=1cm,y=1cm,font=\footnotesize]
 \clip (-0.6,-1.1) rectangle (8.4,2.85);
 \fill[blue!8] (0,0) rectangle (2.6,2.6);
 \draw[->,thick] (0,0) -- (8.1,0) node[below]{$b$};
 \draw[->,thick] (0,0) -- (0,2.8) node[left]{$\delta$};
 \draw[thick,domain=1.75:7.8,samples=120,smooth] plot (\x,{2.4*pow(2/\x,3)});
 \draw[thick,dashed,blue!70!black,domain=2.05:4.6,samples=120,smooth] plot (\x,{2.4*pow(2/\x,3)*pow(2.6/\x,8)});
 \node[anchor=west] at (4.6,0.55) {$\kappa_2\,s/b^6$};
 \node[blue!70!black,anchor=west] at (2.8,2.2) {$\kappa_3\,s/b^{14}$};
 \draw[dashed] (2.6,0) -- (2.6,2.6);
 \draw[densely dotted] (4.2,0) -- (4.2,{2.4*pow(2/4.2,3)});
 \fill (4.2,{2.4*pow(2/4.2,3)}) circle (1.5pt);
 \fill (2.6,{2.4*pow(2/2.6,3)}) circle (1.2pt);
 \node[below] at (2.6,0) {$b_\times\propto s^0$};
 \node[below] at (4.35,0) {$b_{2,\mathrm{sad}}\propto s^{1/7}$};
 \node[align=center] at (1.25,2.2) {sextic\\[-1pt]dominant};
 \node[align=center] at (6.4,2.2) {quartic dominant};
 \node[align=center] at (4.35,-0.72) {$(b_\times/b_{2,\mathrm{sad}})^8\propto s^{-8/7}\Rightarrow s^{8/7}$};
\end{tikzpicture}}
\caption{The quartic and sextic phases in impact parameter at fixed \(s\) and
\(t=-q^2\).  The two are equal at \(b_\times\), Eq.~\eqref{eq:crossover},
independent of \(s\); the stationary point of the quartic phase,
Eq.~\eqref{eq:x2-saddle-scaling}, moves to larger \(b\) as \(s^{1/7}\), so that
the sextic correction to it vanishes as \(s^{-8/7}\) and the combined phase
model scales as \(s^{8/7}\).}
\label{fig:combined-phases}
\end{figure}
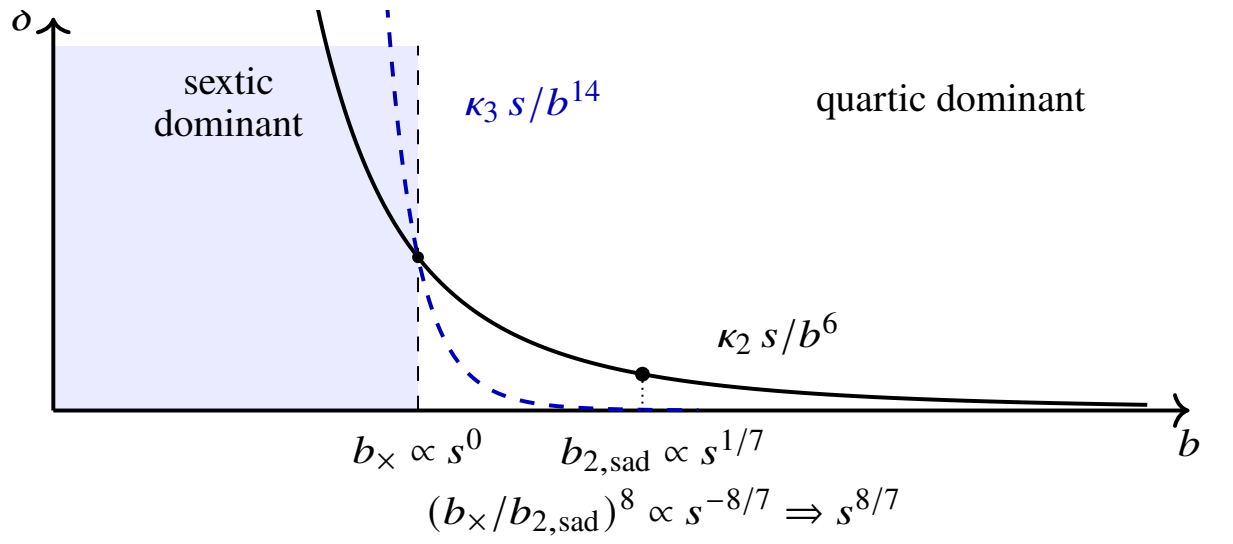

With both operators the two phases add,
\(\delta=\kappa_2s/b^6+\kappa_3s/b^{14}\), Eqs.~\eqref{eq:x2-impact-phase}
and~\eqref{eq:x3-phase-boundary}, and are equal at the impact parameter
\begin{equation}
 b_\times^8=\frac{\kappa_3}{\kappa_2}
 =\frac{465}{14\pi^4}\,\frac{g_6^2}{g_4^2\Lambda^8},\label{eq:crossover}
\end{equation}
\(b_\times\simeq0.87\,|g_6/g_4|^{1/4}/\Lambda\), independent of \(s\), below
which the sextic term dominates.  For the combined phase the
condition~\eqref{eq:saddle-condition} reads
\(q=(12\kappa_2s/b^7)[1+\tfrac73(b_\times/b)^8]\); since
\(b_{2,\mathrm{sad}}\propto s^{1/7}\) grows while \(b_\times\) is fixed, the
correction is of order \((b_\times/b_{2,\mathrm{sad}})^8\propto s^{-8/7}\), the
quartic phase fixes the saddle, and the combined phase model scales as
\(s^{8/7}\), Eq.~\eqref{eq:x2-saddle-scaling}. The two phases are shown in Fig.~\ref{fig:combined-phases}.

\subsection{The EFT-controlled bound}
\label{sec:eft-controlled}

A contact term of degree \(n\) contributes the phase
\begin{equation}
 \delta_n(s,b)\sim\frac{s^{n-1}}{\Lambda^{4n}b^{2n+2}},
 \qquad
 \frac{\delta_{n+1}}{\delta_n}\sim\frac{s}{\Lambda^4b^2},\label{eq:higher-degree-phase}
\end{equation}
so that the EFT controls the phase only for \(b\gg\sqrt s/\Lambda^2\).  The
saddles of the phase models, \(b_{2,\mathrm{sad}}\propto s^{1/7}\) and
\(b_0\propto s^{1/15}\), lie outside this region at large \(s\), so that
\(8/7\) and \(16/15\) are exponents of the models.  The transition is
therefore placed at the boundary of the controlled region,
\begin{equation}
 b_*^{\rm EFT}=\frac{\beta\sqrt s}{\Lambda^2},
 \qquad
 \frac{\delta_{n+1}}{\delta_n}\bigg|_{b_*^{\rm EFT}}\sim\frac1{\beta^2},\label{eq:bstar-eft}
\end{equation}
with a constant \(\beta>1\).  The low-spin bound~\eqref{eq:method-low-spin-bound}
then gives \(|\M_{\rm low}|\lesssim\frac{\sqrt2}{3}\beta^{3/2}s^{7/4}/(\Lambda^3\sqrt q)\).
The high-spin integral~\eqref{eq:general-eikonal} has no stationary point,
since \(b_{2,\mathrm{sad}}/b_*^{\rm EFT}\propto s^{-5/14}\to0\), and its phase is
small, \(\delta\le\kappa_2\Lambda^{12}/(\beta^6s^2)\), the sextic term being
smaller by \((b_\times/b_*^{\rm EFT})^8\propto s^{-4}\).  Expanding
\(\ii[1-e^{2\ii\delta}]=2\delta+O(\delta^2)\),
\begin{equation}
 \M_{\rm eikonal}(s,-q^2)\simeq8\pi\kappa_2s^2\int_{b_*^{\rm EFT}}^\infty\dd b\,\frac{J_0(qb)}{b^5},\label{eq:eft-high-linear}
\end{equation}
\noindent and with \(\dd[bJ_1(qb)]/\dd b=qbJ_0(qb)\), one integration by parts and
\(|J_1(z)|\le(2/\pi z)^{1/2}[1+O(1/z)]\)~\cite{AbramowitzStegun},
\(|\M_{\rm eikonal}|\lesssim8\sqrt{2\pi}\,\kappa_2\Lambda^{11}s^{-3/4}/(\beta^{11/2}q^{3/2})\),
which decreases with \(s\); the sextic phase gives \(s^{-19/4}\).  Both
sectors therefore obey
\begin{equation}
 \bigl|\M_{\rm low}+\M_{\rm eikonal}\bigr|
 \lesssim\frac{\sqrt2}{3}\,\frac{\beta^{3/2}}{\Lambda^3\sqrt q}\,s^{7/4}\label{eq:eft-controlled-bound}
\end{equation}
at fixed \(q\neq0\), which grows more slowly than \(s^2\), provided the EFT
describes \(b>b_*^{\rm EFT}\). The construction is summarized in Fig.~\ref{fig:eft-controlled}.

\subsection{The forward limit and Regge boundedness}
\label{sec:forward-regge}

At \(t=0\), \(P_\ell(1)=1\)~\cite{Szego} and \(J_0(0)=1\), so that the angle
gives no suppression.  With \(\ell_*=\beta s/(2\Lambda^2)\),
\(|f_\ell|\le2\), Eq.~\eqref{eq:pwu-unitarity}, and
\(\sum_{\ell<\ell_*,\,\ell\ {\rm even}}(2\ell+1)\simeq\ell_*^2/2\), the
low-spin part is bounded only by the number of partial waves,
\begin{equation}
 |\M_{\rm low}(s,0)|\lesssim16\pi\,\ell_*^2=\frac{4\pi\beta^2}{\Lambda^4}\,s^2 ,\label{eq:forward-eft-controlled}
\end{equation}
while Eq.~\eqref{eq:eft-high-linear} with \(J_0=1\) gives the constant
\(\M_{\rm eikonal}(s,0)\simeq2\pi\kappa_2s^2/(b_*^{\rm EFT})^4=8g_4^2/(15\pi^2\beta^4)\).
The forward bound is therefore marginal: in the arc estimate,
Eqs.~\eqref{eq:large-circle} and~\eqref{eq:app-arc-bound}, it bounds the arc
only by \(4\pi\beta^2/\Lambda^4\), a constant of the order of
\(2c=4g_4/\Lambda^4\), Eq.~\eqref{eq:eft-arc}, which does not vanish as
\(R\to\infty\), whereas Eq.~\eqref{eq:regge-bounded-condition} requires
\(\M(s,0)/s^2\to0\).  The EFT is thus compatible with a Regge-bounded
completion at fixed \(q\neq0\), while in the forward limit Regge boundedness
remains an assumption, satisfied only through the infinite set of spins of
Sec.~\ref{sec:regge}.  The four estimates are collected in
Table~\ref{tab:growth-summary}.
\begin{table}[!t]
\centering
\caption{Growth of the bound on the amplitude for the four choices of the
transition scale \(b_*\).  The first two rows are exponents of the phase
models, the last two hold where the EFT controls the phase,
Eq.~\eqref{eq:bstar-eft}.}
\label{tab:growth-summary}
\footnotesize\setlength{\tabcolsep}{3pt}
\renewcommand{\arraystretch}{1.3}
\begin{tabular}{llll}
\hline
estimate & \(b_*\) & growth & Eq.\\
\hline
\((\partial\phi)^4\) phase model, \(t<0\) & \(B_2s^{2/21}\) & \(s^{8/7}\) & \eqref{eq:x2-final-powers}\\
\((\partial\phi)^6\) phase model, \(g_4\to0\), \(t<0\) & \(B_3s^{2/45}\) & \(s^{16/15}\) & \eqref{eq:x3-final-powers}\\
EFT-controlled, \(t<0\) & \(\beta\sqrt s/\Lambda^2\) & \(s^{7/4}\) & \eqref{eq:eft-controlled-bound}\\
EFT-controlled, \(t=0\) & \(\beta\sqrt s/\Lambda^2\) & \(s^{2}\) & \eqref{eq:forward-eft-controlled}\\
\hline
\end{tabular}
\end{table}
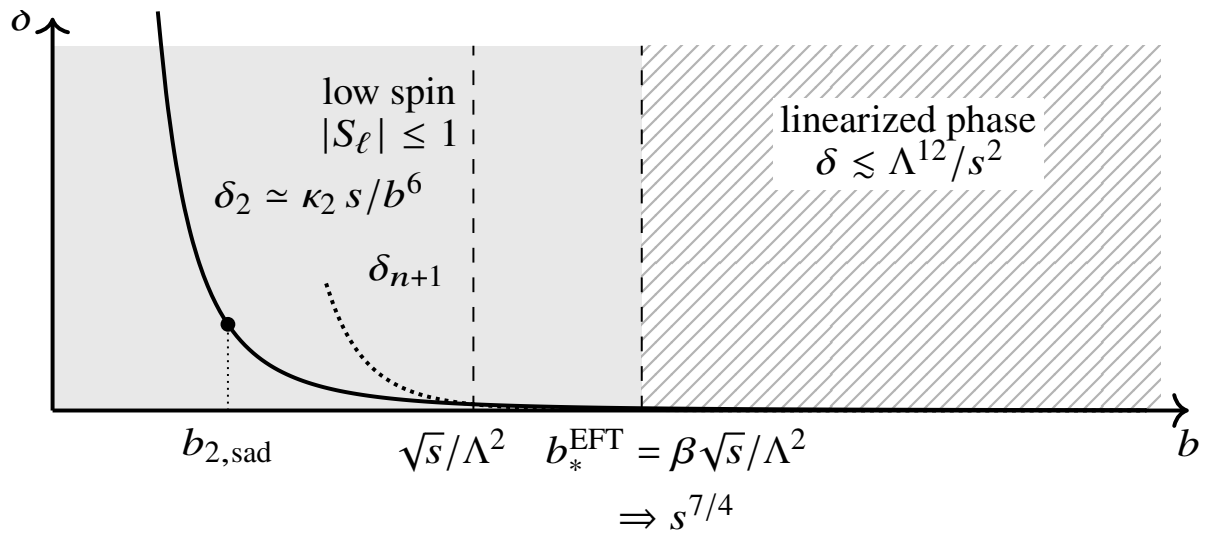
\begin{figure}[!tbp]
\centering
\adjustbox{max width=\linewidth}{%
\begin{tikzpicture}[x=1cm,y=1cm,font=\footnotesize]
 \fill[gray!18] (0,0) rectangle (4.2,2.6);
 \clip (-0.6,-1.1) rectangle (8.4,2.85);
 \fill[pattern=north east lines,pattern color=gray!60] (4.2,0) rectangle (7.9,2.6);
 \draw[->,thick] (0,0) -- (8.1,0) node[below]{$b$};
 \draw[->,thick] (0,0) -- (0,2.8) node[left]{$\delta$};
 \draw[thick,domain=0.75:7.8,samples=120,smooth] plot (\x,{1.2*pow(1/\x,3)});
 \draw[thick,densely dotted,domain=1.95:7.8,samples=120,smooth] plot (\x,{1.2*pow(1/\x,3)*pow(3.0/\x,4)});
 \node[anchor=west] at (1.0,1.55) {$\delta_2\simeq\kappa_2\,s/b^6$};
 \node[anchor=west] at (2.1,1.0) {$\delta_{n+1}$};
 \draw[dashed] (3.0,0) -- (3.0,2.6);
 \draw[dashed] (4.2,0) -- (4.2,2.6);
 \draw[densely dotted] (1.25,0) -- (1.25,{1.2*pow(1/1.25,3)});
 \fill (1.25,{1.2*pow(1/1.25,3)}) circle (1.5pt);
 \node[below] at (1.25,0) {$b_{2,\mathrm{sad}}$};
 \node[below] at (2.85,0) {$\sqrt s/\Lambda^2$};
 \node[below] at (4.45,0) {$b_*^{\rm EFT}=\beta\sqrt s/\Lambda^2$};
 \node[align=center] at (2.4,2.1) {low spin\\[-1pt]$|S_\ell|\le1$};
 \node[align=center,fill=white,inner sep=1.5pt] at (6.1,1.9) {linearized phase\\[-1pt]$\delta\lesssim\Lambda^{12}/s^2$};
 \node[align=center] at (4.45,-0.72) {$\Rightarrow s^{7/4}$};
\end{tikzpicture}}
\caption{Impact-parameter picture of the EFT-controlled estimate at fixed \(s\) and
\(t=-q^2\).  The phase of the next contact term exceeds the quartic phase for
\(b<\sqrt s/\Lambda^2\), Eq.~\eqref{eq:higher-degree-phase}; with the transition at
\(b_*^{\rm EFT}\), Eq.~\eqref{eq:bstar-eft}, the stationary point lies in the
low-spin region, the high-spin phase is small and may be linearized,
Eq.~\eqref{eq:eft-high-linear}, and the bound grows as \(s^{7/4}\),
Eq.~\eqref{eq:eft-controlled-bound}.}
\label{fig:eft-controlled}
\end{figure}

\section{Bounds on \texorpdfstring{$g_4$}{g4} and \texorpdfstring{$g_6$}{g6}}
\label{sec:bounds}

The bounds follow from the dispersion relation~\eqref{eq:intro-dispersion},
read from the ultraviolet in the form~\eqref{eq:uv-transfer}.  The input on
the ultraviolet amplitude is Regge boundedness, \(\M(s,0)/s^2\to0\), which removes the arc, Eq.~\eqref{eq:large-circle}~\cite{Froissart,Martin,JinMartin}.  At \(t=0\) the crossed-channel terms \(t^2\log(-t)\)
vanish, while their higher \(t\)-derivatives, which the twice-subtracted
forward relation does not use, diverge~\cite{TokarevaXu}.  With
\(s_{\rm th}=0\) for massless particles, Eq.~\eqref{eq:intro-dispersion} reads
\begin{equation}
\begin{gathered}\frac{2g_4}{\Lambda^4}
 =\frac{1}{\pi}\int_0^\infty\dd s\,\frac{\operatorname{Im}\M(s,0)}{s^3}, \\
\operatorname{Im}\M(s,0)=s\,\sigma_{\rm tot}(s),\end{gathered}\label{eq:positivity-a2}
\end{equation}
the second relation being the optical theorem.  The integrand is
non-negative at every \(s\), and since \(\sigma_{\rm tot}\) cannot vanish
identically in an interacting theory~\cite{Adams,Distler},
\begin{equation}
 g_4>0 .\label{eq:g4-positive}
\end{equation}

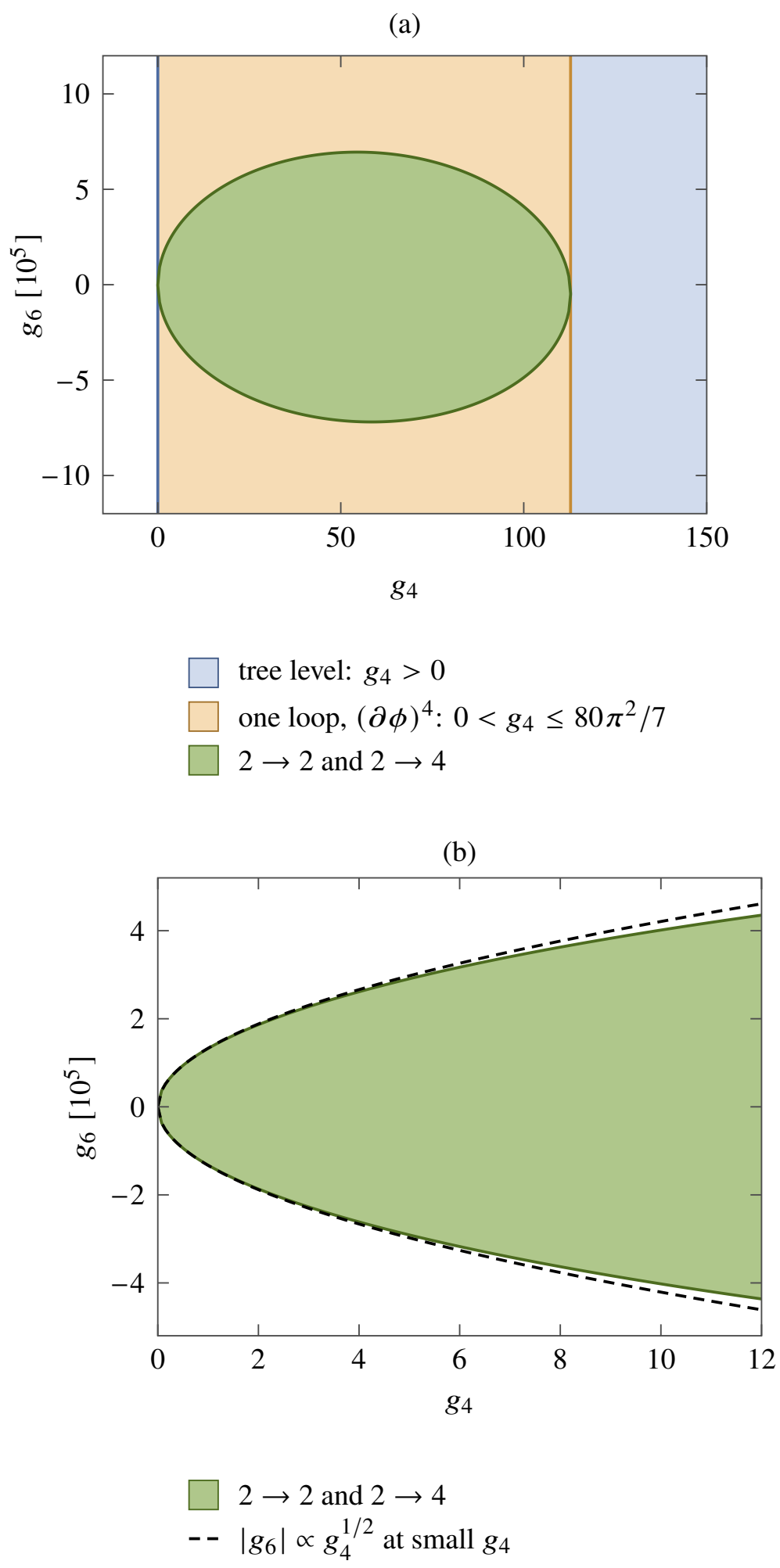
\begin{figure}[!tbp]
\centering
\adjustbox{scale=0.88}{%
\begin{tikzpicture}[font=\footnotesize]
\definecolor{cB}{HTML}{5B7FBF}\definecolor{cO}{HTML}{E8A33D}\definecolor{cG}{HTML}{6E9A2C}
\begin{scope}[x=0.035152cm,y=0.183333cm]
 \begin{scope}
  \clip (-15,-12) rectangle (150,12);
  \fill[cB!28] (112.795,-12) rectangle (150,12);
  \fill[cO!38] (0,-12) rectangle (112.795,12);
  \draw[cO!85!black,thick] (112.795,-12)--(112.795,12);
  \draw[cB!85!black,thick] (0,-12)--(0,12);
  \fill[cG!55] plot[domain=0:112.795,samples=240,variable=\g] (\g,{-0.3847284*(\g/100)^2+sqrt(max(0,49.9524*(1-(1-\g/56.3977)^2)-0.00039968*(\g/100)^4))}) -- plot[domain=112.795:0,samples=240,variable=\g] (\g,{-0.3847284*(\g/100)^2-sqrt(max(0,49.9524*(1-(1-\g/56.3977)^2)-0.00039968*(\g/100)^4))}) -- cycle;
  \draw[cG!70!black,thick] plot[domain=0:112.795,samples=240,variable=\g] (\g,{-0.3847284*(\g/100)^2+sqrt(max(0,49.9524*(1-(1-\g/56.3977)^2)-0.00039968*(\g/100)^4))}) -- plot[domain=112.795:0,samples=240,variable=\g] (\g,{-0.3847284*(\g/100)^2-sqrt(max(0,49.9524*(1-(1-\g/56.3977)^2)-0.00039968*(\g/100)^4))}) -- cycle;
 \end{scope}
 \draw[black!70] (-15,-12) rectangle (150,12);
 \foreach \xx in {0,50,100,150} {\draw[black!70] (\xx,-12)--++(0,0.6) (\xx,12)--++(0,-0.6); \node[below] at (\xx,-12) {$\xx$};}
 \foreach \yy in {-10,-5,0,5,10} {\draw[black!70] (-15,\yy)--++(3,0) (150,\yy)--++(-3,0); \node[left] at (-15,\yy) {$\yy$};}
 \node at (67.5,-16) {$g_4$};
 \node[rotate=90] at (-36,0) {$g_6\ [10^5]$};
 \node at (67.5,13.6) {(a)};
\end{scope}
\begin{scope}[shift={(0,-7.9cm)},x=0.483333cm,y=0.423077cm]
 \begin{scope}
  \clip (0,-5.2) rectangle (12,5.2);
  \fill[cG!55] plot[domain=0:12,samples=160,variable=\g] (\g,{-0.3847284*(\g/100)^2+sqrt(max(0,49.9524*(1-(1-\g/56.3977)^2)-0.00039968*(\g/100)^4))}) -- plot[domain=12:0,samples=160,variable=\g] (\g,{-0.3847284*(\g/100)^2-sqrt(max(0,49.9524*(1-(1-\g/56.3977)^2)-0.00039968*(\g/100)^4))}) -- cycle;
  \draw[cG!70!black,thick] plot[domain=0:12,samples=160,variable=\g] (\g,{-0.3847284*(\g/100)^2+sqrt(max(0,49.9524*(1-(1-\g/56.3977)^2)-0.00039968*(\g/100)^4))});
  \draw[cG!70!black,thick] plot[domain=0:12,samples=160,variable=\g] (\g,{-0.3847284*(\g/100)^2-sqrt(max(0,49.9524*(1-(1-\g/56.3977)^2)-0.00039968*(\g/100)^4))});
  \draw[densely dashed,thick] plot[domain=0:12,samples=160,variable=\g] (\g,{7.0677*sqrt(2*\g/56.3977)});
  \draw[densely dashed,thick] plot[domain=0:12,samples=160,variable=\g] (\g,{-7.0677*sqrt(2*\g/56.3977)});
 \end{scope}
 \draw[black!70] (0,-5.2) rectangle (12,5.2);
 \foreach \xx in {0,2,4,6,8,10,12} {\draw[black!70] (\xx,-5.2)--++(0,0.26) (\xx,5.2)--++(0,-0.26); \node[below] at (\xx,-5.2) {$\xx$};}
 \foreach \yy in {-4,-2,0,2,4} {\draw[black!70] (0,\yy)--++(0.22,0) (12,\yy)--++(-0.22,0); \node[left] at (0,\yy) {$\yy$};}
 \node at (6,-6.77) {$g_4$};
 \node[rotate=90] at (-1.55,0) {$g_6\ [10^5]$};
 \node at (6,5.75) {(b)};
\end{scope}
\begin{scope}[shift={(0.3cm,-3.45cm)}]
 \foreach \c/\o/\l [count=\k] in {cB/28/{tree level: $g_4>0$},cO/38/{one loop, $(\partial\phi)^4$: $0<g_4\le80\pi^2/7$},cG/55/{$2\to2$ and $2\to4$}} {
  \fill[\c!\o] (0,{-0.42*\k}) rectangle ++(0.28,0.28);
  \draw[\c!70!black] (0,{-0.42*\k}) rectangle ++(0.28,0.28);
  \node[anchor=west] at (0.35,{0.14-0.42*\k}) {\l};}
\end{scope}
\begin{scope}[shift={(0.3cm,-11.35cm)}]
 \fill[cG!55] (0,-0.42) rectangle ++(0.28,0.28); \draw[cG!70!black] (0,-0.42) rectangle ++(0.28,0.28);
 \node[anchor=west] at (0.35,-0.28) {$2\to2$ and $2\to4$};
 \draw[densely dashed,thick] (0,-0.70)--(0.28,-0.70); \node[anchor=west] at (0.35,-0.70) {$|g_6|\propto g_4^{1/2}$ at small $g_4$};
\end{scope}
\end{tikzpicture}}
\caption{(color online) Bounds on \(g_4\) and \(g_6\).  (a) The tree-level condition
\(g_4>0\), Eq.~\eqref{eq:g4-positive}; the one-loop \((\partial\phi)^4\) condition
\(0<g_4\le80\pi^2/7\), obtained from Eq.~\eqref{eq:g4-g6-bound} without the
non-negative \(2\to4\) term; and the allowed region with the \(2\to4\) cross section, centred on \(g_6=-c\,g_4^2\),
Eq.~\eqref{eq:allowed-region}.  (b) The allowed region at small couplings, with the
small-\(g_4\) form \(|g_6|\simeq\frac{46\,080\sqrt3\,\pi^4}{11}(7g_4/20\pi^2)^{1/2}\)
(dashed) that follows from Eq.~\eqref{eq:allowed-region}.}
\label{fig:g4g6}
\end{figure}

Below the cutoff the forward absorptive part is supplied by the EFT
through the optical theorem: it is the sum of the \(2\to2\) and \(2\to4\) cross
sections, each evaluated with the tree amplitudes, which are the leading terms
of the respective amplitudes.  The \(2\to2\) tree is the quartic contact term,
Eq.~\eqref{eq:general-tree-ansatz}, whose cross section is the absorptive part
of the \(s\)-channel bubble, Eq.~\eqref{eq:x2-optical}.  The \(2\to4\) tree
contains the sextic contact term, Eq.~\eqref{eq:x3-tree-permutation}, and the
exchange of one line between two quartic vertices, of the same degree, six, in
the momenta,
\begin{equation}
\adjustbox{max width=0.86\linewidth}{$\displaystyle
\begin{aligned}
 \M^{\rm tree}_{2\to4}&=\frac{g_6}{\Lambda^8}\,A_6+\frac{g_4^2}{\Lambda^8}\,A_{44},\\
 A_{44}&=\sum_{\{a,b,c\}\ni1}\frac{V_4(p_a,p_b,p_c,-P)\,V_4(p_d,p_e,p_f,P)}{P^2},
\end{aligned}
$}\label{eq:x6-exchange}
\end{equation}
where \(A_6\) is the sum of Eq.~\eqref{eq:x3-tree-permutation} without
\(g_6/\Lambda^8\), the sum runs over the ten partitions of the six legs into
\(\{a,b,c\}\ni1\) and \(\{d,e,f\}\), \(P=p_a+p_b+p_c\), and
\(V_4(p_1,\dots,p_4)=8[(p_1\!\cdot\!p_2)(p_3\!\cdot\!p_4)+(p_1\!\cdot\!p_3)(p_2\!\cdot\!p_4)+(p_1\!\cdot\!p_4)(p_2\!\cdot\!p_3)]\)
is the quartic vertex, equal to Eq.~\eqref{eq:general-tree-ansatz} on shell.
The propagator is singular only where three external momenta are collinear,
where both vertices vanish, so that the phase-space integral converges.  The
\(2\to4\) cross section is a quadratic form in \(g_6\) and \(g_4^2\), evaluated
in Appendix~\ref{app:rambo}, Eq.~\eqref{eq:rambo-2to4}, and
\begin{equation}
\adjustbox{max width=0.86\linewidth}{$\displaystyle
\begin{aligned}\operatorname{Im}\M(s,0)
 &=\frac{7g_4^2s^4}{20\pi\Lambda^8}
 +\frac{121\,s^8}{185\,794\,560\,\pi^5\Lambda^{16}}\\
 &\quad\times\Bigl[\bigl(g_6+c\,g_4^2\bigr)^2+d\,g_4^4\Bigr],\qquad s\le\Lambda^2,\\
 c&=3.84728(1),\qquad d=0.039968(4),
\end{aligned}
$}\label{eq:improved-positivity-im}
\end{equation}
each term non-negative; the term in \(g_6^2\) is the banana,
Eq.~\eqref{eq:banana-imaginary}, and the terms in \(g_4^2g_6\) and \(g_4^4\)
are the interference with the exchange and its square.  The order of each
term follows from the loop order of the elastic amplitude whose cut it is:
odd-particle cuts vanish by \(\phi\to-\phi\), Sec.~\ref{sec:symmetry}, and
with \(\hat s=s/\Lambda^2\) and the coefficients omitted,
\begin{equation}
\adjustbox{max width=0.86\linewidth}{$\displaystyle
\begin{aligned}\operatorname{Im}\M(s,0)\sim{}&\underbrace{g_4^2\hat s^4}_{\text{1 loop, 2-body}}
 +\underbrace{g_4^3\hat s^6\log\hat s}_{\text{2 loops, 2-body}}\\
 &+\underbrace{\bigl(g_6^2,\,g_4^2g_6,\,g_4^4\bigr)\hat s^8}_{\text{3 loops, 4-body}}
 +\underbrace{g_4^4\hat s^8\log^2\hat s}_{\text{3 loops, 2-body}}+\cdots .\end{aligned}
$}\label{eq:x6-power-counting}
\end{equation}
Equation~\eqref{eq:improved-positivity-im} contains the complete one-loop term
and the complete four-particle cut at three loops, which carries the entire
dependence on \(g_6\) at this order: the two-particle cut of the
\(g_4^2g_6\) amplitude is the product of the quartic tree and the \(g_4g_6\)
graphs, which vanish, and the elastic \(g_6\) tadpole vanishes,
Eq.~\eqref{eq:x3-tree-permutation}.  The two-particle cuts at two and three
loops depend on \(g_4\) alone and are not included; they are smaller than the
one-loop term by the ratio of the one-loop to the tree amplitude,
\(7g_4s^2/(80\pi\Lambda^4)\) at \(t=0\), Eqs.~\eqref{eq:general-tree-ansatz}
and~\eqref{eq:x2-optical}, times a logarithm, and the term \(d\,g_4^4\) is not the
complete coefficient of \(g_4^4\hat s^8\).  The dependence of the bound on
\(g_6\), the centre line \(g_6=-c\,g_4^2\) and the width set by \(d\), is
therefore complete at this order, whereas the omitted terms shift only the
dependence on \(g_4\) alone, by a relative amount that is small at small
\(g_4\) and of order one near the edge \(g_4\simeq80\pi^2/7\) discussed
below.  Inserting
Eq.~\eqref{eq:improved-positivity-im} for \(s\le\Lambda^2\) in
Eq.~\eqref{eq:positivity-a2} and dropping the non-negative contribution from
\(s>\Lambda^2\),
\begin{equation}
\adjustbox{max width=0.86\linewidth}{$\displaystyle
\begin{aligned}\frac{2g_4}{\Lambda^4}\ &\ge\ \frac1\pi\int_0^{\Lambda^2}\dd s\,
 \frac{\operatorname{Im}\M(s,0)}{s^3} \\
 &=\frac{7g_4^2}{40\pi^2\Lambda^4}
 +\frac{121\,\bigl[(g_6+c\,g_4^2)^2+d\,g_4^4\bigr]}{1\,114\,767\,360\,\pi^6\Lambda^4}.\end{aligned}
$}\label{eq:improved-positivity}
\end{equation}
After multiplication by \(\Lambda^4/2\), the tree coefficient on the left must exceed the absorptive contributions on the right, which are of second order in the couplings,
\begin{equation}
\adjustbox{max width=0.86\linewidth}{$\displaystyle
\begin{aligned}g_4-\frac{7}{80\pi^2}\,g_4^2
 &\ge\frac{121\,\bigl[(g_6+c\,g_4^2)^2+d\,g_4^4\bigr]}{2\,229\,534\,720\,\pi^6}\\
 &\simeq5.6\times10^{-11}\,\bigl[(g_6+c\,g_4^2)^2+d\,g_4^4\bigr] .\end{aligned}
$}\label{eq:g4-g6-bound}
\end{equation}
Completing the square in \(g_4\), the left-hand side of Eq.~\eqref{eq:g4-g6-bound} is a parabola that vanishes at \(g_4=0\) and \(g_4=80\pi^2/7\) and reaches its maximum \(20\pi^2/7\) at \(g_4=40\pi^2/7\),
\begin{equation}
 g_4-\frac{7}{80\pi^2}\,g_4^2
 =\frac{20\pi^2}{7}\left[1-\Bigl(1-\frac{7g_4}{40\pi^2}\Bigr)^2\right],\label{eq:g4-bar}
\end{equation}
so that, after division of Eq.~\eqref{eq:g4-g6-bound} by \(20\pi^2/7\), with \(2\,229\,534\,720\cdot20/7=6\,370\,099\,200\), it becomes
\begin{equation}
\adjustbox{max width=0.86\linewidth}{$\displaystyle
 \Bigl(1-\frac{7g_4}{40\pi^2}\Bigr)^2
 +\frac{121\,\bigl[(g_6+c\,g_4^2)^2+d\,g_4^4\bigr]}{6\,370\,099\,200\,\pi^8}\ \le\ 1 .
$}\label{eq:g6-requires-g4}
\end{equation}
Together with Eq.~\eqref{eq:g4-positive}, the allowed region is
\begin{equation}
\adjustbox{max width=0.86\linewidth}{$\displaystyle
\begin{gathered}0<g_4\le\frac{80\pi^2}{7},\qquad
G=\frac{46\,080\sqrt3\,\pi^4}{11}\simeq7.07\times10^5,\\
\bigl|g_6+c\,g_4^2\bigr|\le\Bigl\{G^2\Bigl[1-\Bigl(1-\frac{7g_4}{40\pi^2}\Bigr)^2\Bigr]-d\,g_4^4\Bigr\}^{1/2},
\end{gathered}
$}\label{eq:allowed-region}
\end{equation}
in which \(-7.2\times10^5\lesssim g_6\lesssim6.9\times10^5\); the term
\(d\,g_4^4\) moves the edge \(g_4=80\pi^2/7\) inwards by \(4\times10^{-4}\).  For
\(g_4\to0\) the bound vanishes as
\(|g_6|\simeq G\bigl(7g_4/20\pi^2\bigr)^{1/2}\), the terms \(c\,g_4^2\) and
\(d\,g_4^4\) being of higher order: a sextic coupling requires a quartic one,
Fig.~\ref{fig:g4g6}(b).  The interference with the exchange centres the region
on \(g_6=-c\,g_4^2\), so that negative \(g_6\) extend further, by at most
\(c\,(80\pi^2/7)^2\simeq4.9\times10^4\) at the edge; without the exchange the
sign of \(g_6\) would not be constrained, since \((\partial\phi)^6\) has no
four-point tree, Eq.~\eqref{eq:x3-tree-permutation}.  The region is shown in
Fig.~\ref{fig:g4g6}(a), together with the weaker tree-level and quartic-only
conditions.

The derivation uses the loop absorptive parts~\eqref{eq:improved-positivity-im}
as \(\operatorname{Im}\M(s,0)\) up to \(s=\Lambda^2\), so the region applies
where this expansion is reliable up to the cutoff.  Near the boundary
\(g_4\simeq80\pi^2/7\) the two terms on the left of
Eq.~\eqref{eq:g4-g6-bound} are equal, the one-loop absorptive part is as
large as the tree contribution it is compared with, and higher orders are
no longer negligible; the correlation is most robust at small couplings,
where it matters.  Moving the splitting scale to \(s=\xi\Lambda^2\) replaces
\(g_4^2\to\xi^2g_4^2\) on the left and the bracket by \(\xi^6\) times itself on
the right of Eq.~\eqref{eq:g4-g6-bound}, which maps the region onto itself under
\(g_4\to\xi^2g_4\), \(g_6\to\xi^4g_6\) and rescales it to
\(g_4\le80\pi^2/(7\xi^2)\) and \(G\to G\,\xi^{-4}\).  Further constraints from full crossing on the
coefficients of \(stu\) and \((s^2+t^2+u^2)^2\) relative to that of
\(s^2+t^2+u^2\)~\cite{Tolley,CaronHuotVanDuong,EFThedron} are satisfied at
tree level, where both vanish for the Lagrangian~\eqref{eq:lagrangian}, the
second on the boundary; at loop level the \(s^4\) sum rule contains the
\(\mu\) dependence of \(P_4\), and the crossed logarithm \(C_2\) enters the
non-forward sum rules, which must be formulated at a scale or at finite
\(t\)~\cite{BellazziniRiembauRiva,PengShaoTokarevaXu,TokarevaXu}.

\section{Conclusions}
\label{sec:conclusions}

In a massless theory the dispersion relation for the coefficient \(c\)
follows from Cauchy's theorem on the cut plane, the origin being a branch
point at which the limit of a small loop plays the part of the residue,
Eq.~\eqref{eq:cauchy-cut}.  The truncated EFT does not satisfy it on its
own, its arc growing as \(R^2\) at one loop, Eq.~\eqref{eq:eft-arc}, and the twice-subtracted relation rests on Regge boundedness of the completion.  In the \(t\) channel the contact interaction has spin two and
grows as \(s^2\), Eq.~\eqref{eq:t-channel-spin}, so that a Regge-bounded completion must supply an infinite set of spins~\cite{Regge,Collins}.

We have studied the large-\(s\) growth of the one-loop \(g_4^2\) and the
three-loop \(g_6^2\) sectors of the EFT~\eqref{eq:lagrangian} at fixed
\(t<0\), following Eq.~\eqref{eq:logic}: at low spin only \(|S_\ell|\le1\) is
used, and at high spin the EFT enters through the eikonal phase of the
crossed logarithms, \(\delta=\kappa_2s/b^6\) and \(\kappa_3s/b^{14}\), evaluated by
stationary phase.  The \(s\)-channel tree and loops are confined to
\(\ell\le2\) and their growth is replaced by the constant \(192\pi\), Eq.~\eqref{eq:x2-pwu-explicit-bound}.
Matching the two regions gives a transition radius growing with energy, \(b_*\sim s^{2/21}\), as the classicalization radius of derivatively self-coupled scalars does~\cite{Dvali}, and the scaling \(s^{8/7}\) for the \((\partial\phi)^4\) phase model, Eq.~\eqref{eq:x2-low-high-matched}; in the \((\partial\phi)^6\) sector the high-spin part
scales as \(s^{16/15}\), but with the quartic transition scale the low-spin
bound \(s^{8/7}\) is the larger, and only for \(g_4\to0\), with
\(b_*\sim s^{2/45}\), do both parts scale as \(s^{16/15}\).  These exponents
are properties of the phase models, since at the saddles the higher-degree contact terms are not suppressed, Eq.~\eqref{eq:higher-degree-phase}.  Placing
the transition at the boundary of the region where the EFT controls the phase,
\(b_*\sim\sqrt s/\Lambda^2\), gives \(s^{7/4}<s^2\) for both sectors at fixed
\(q\neq0\), Eq.~\eqref{eq:eft-controlled-bound}, conditional on the EFT
describing \(b>b_*\) with no additional light states.  In the forward limit,
which the dispersion relation requires, the same construction gives the
marginal \(s^2\), Eq.~\eqref{eq:forward-eft-controlled}.  The powers \(s^{8/7}\), \(s^{16/15}\) and \(s^{7/4}\) bound the growth of the EFT contributions and are compatible with a Regge-bounded completion at fixed \(q\neq0\), while Regge boundedness itself, Eq.~\eqref{eq:regge-bounded-condition}, remains an assumption on the completion.

Used as an input, Regge boundedness constrains the couplings: the \(2\to2\) and
\(2\to4\) absorptive parts, integrated up to the cutoff, give \(g_4>0\) and the
region of Eq.~\eqref{eq:allowed-region}, centred on \(g_6=-c\,g_4^2\) by the
interference of the sextic contact term with the quartic exchange, in which
\(|g_6|\) vanishes as \(g_4^{1/2}\) for \(g_4\to0\).

\section{Outlook}
\label{sec:outlook}

This section collects conjectures that go beyond the computed bubble and
banana; they are not used elsewhere in the paper.  Power counting suggests
that the exponents of the phase models follow a simple rule.  A term
\(s^Jt^m\log(-t)\) of an \(L\)-loop amplitude, with \(J+m=2L+2\), gives by
Eq.~\eqref{eq:log-transform} the phase \(\delta\propto s^{J-1}/b^{4L+6-2J}\),
and stationary phase, Eq.~\eqref{eq:general-saddle}, with the low-spin matching would give
\begin{equation}
 \alpha=1+\frac{J-1}{4L+7-2J},
 \qquad
 \gamma=\frac23\,(\alpha-1).\label{eq:alpha-topology}
\end{equation}
For the crossed logarithms \(J=2\) and \(\alpha=1+1/(4L+3)\): \(L=1\) gives
\(8/7\) and \(L=3\) gives \(16/15\).  We conjecture that the rule extends to an operator \(g_{2k}(\partial\phi)^{2k}/\Lambda^{4k-4}\) with \(k\ge3\), whose leading elastic contribution would be the banana of two vertices joined by \(2k-2\) lines, with
\(L=2k-3\) loops and degree \(D=8k-8\) in the momenta.  If each vertex, linear in every external momentum, contains only the spins \(\ell,J\le2\), the crossed logarithm grows as \(s^2\), \(J=2\), and Eq.~\eqref{eq:alpha-topology} gives the phase \(\delta\propto s/b^{8k-10}\) and
the exponents \(\alpha_k=D/(D-1)\) and \(\gamma_k=2/[3(8k-9)]\) of
Table~\ref{tab:xk}.  Only the rows \(k=2,3\) rest on computed amplitudes, Secs.~\ref{sec:x2-low} and~\ref{sec:x3-low-eikonal}; the rows \(k\ge4\) assume non-vanishing crossed-logarithm coefficients, which have not been computed.
\begin{table}[!t]
\centering
\caption{Conjectured phase-model exponents of the leading elastic contribution of \((\partial\phi)^{2k}\); only the rows \(k=2,3\) are computed.}
\label{tab:xk}
\footnotesize\setlength{\tabcolsep}{4pt}
\begin{tabular}{lccccc}
\hline
operator & \(L\) & \(D\) & \(\delta\propto\) & \(\alpha_k\) & \(\gamma_k\)\\
\hline
\((\partial\phi)^4\) & 1 & 8 & \(s/b^6\) & \(8/7\) & \(2/21\)\\
\((\partial\phi)^6\) & 3 & 16 & \(s/b^{14}\) & \(16/15\) & \(2/45\)\\
\((\partial\phi)^8\) & 5 & 24 & \(s/b^{22}\) & \(24/23\) & \(2/69\)\\
\((\partial\phi)^{10}\) & 7 & 32 & \(s/b^{30}\) & \(32/31\) & \(2/93\)\\
\((\partial\phi)^{16}\) & 13 & 56 & \(s/b^{54}\) & \(56/55\) & \(2/165\)\\
\((\partial\phi)^{20}\) & 17 & 72 & \(s/b^{70}\) & \(72/71\) & \(2/213\)\\
\hline
\end{tabular}
\end{table}
Two extensions would sharpen the
comparison with Regge boundedness.  The high-spin phase used here is real;
a complete eikonal requires the absorptive part \(\chi\) of the phase at high spin~\cite{ACV,DiVecchiaEikonal}.  In the forward limit, where the fixed-\(q\) bounds contain
\(q^{-1/2}\) and the EFT-controlled estimate is marginal, an infrared
treatment compatible with the loop-level positivity and null
constraints~\cite{BellazziniRiembauRiva,PengShaoTokarevaXu,TokarevaXu,AlberteQED,HerreroValea} is
needed.

\appendix
\titleformat{\section}[block]{\normalfont\large\bfseries\filcenter}{\MakeUppercase{\appendixname}~\thesection:}{0.5em}{\MakeUppercase}

\section{Partial-wave projections}
\label{app:pw}
\setcounter{equation}{0}
\gdef\theequation{A\arabic{equation}}

The projections use the angle variable and the projection of
Eq.~\eqref{eq:pwu-convention},
\begin{equation}
\begin{gathered}
 x=1+\frac{2t}{s},\\
 t=-\frac s2\,(1-x),\qquad u=-\frac s2\,(1+x),\\
 f_\ell=\frac{1}{32\pi}\int_{-1}^1\dd x\,P_\ell(x)\,\M ,
\end{gathered}\label{eq:app-pw-def}
\end{equation}
and the integrals of
Appendix~\ref{app:integrals}.  Every loop diagram is treated channel by
channel: in the \(s\) channel the logarithm is constant in \(x\), in the \(t\)
and \(u\) channels it is non-analytic at \(x=1\) and \(x=-1\), Eq.~\eqref{eq:x2-logs-angle}.

\paragraph*{Tree.}  By Eqs.~\eqref{eq:app-pw-def} and~\eqref{eq:int-xn},
\begin{equation}
\adjustbox{max width=0.86\linewidth}{$\displaystyle
\begin{gathered}
 t^2+u^2=\frac{s^2}{2}\,(1+x^2),\\
 s^2+t^2+u^2=\frac{s^2}{2}\,(3+x^2)=s^2\Bigl[\frac53P_0(x)+\frac13P_2(x)\Bigr],
\end{gathered}
$}\label{eq:app-tree-angle}
\end{equation}
which gives Eq.~\eqref{eq:x2-tree-pw-main}.  In the crossed process, with
\(1,3\) incoming,
\begin{equation}
 z=1+\frac{2s}{t},\qquad s=-\frac t2\,(1-z),\qquad u=-\frac t2\,(1+z),\label{eq:app-crossed-angle}
\end{equation}
and the same steps give Eq.~\eqref{eq:x2-tree-cyclic}, with \(t\)-channel
spins \(J=0,2\) only.

\paragraph*{Bubbles.}  The three bubbles are the terms \(X=s,t,u\) of
Eq.~\eqref{eq:x2-channel-decomposition}, denoted
\(\M^{\rm bubble}_{X,(\partial\phi)^4}\).  The normalization follows from the two-particle cut.  With the tree on both
sides and the two-body phase space of Eq.~\eqref{eq:rambo-w2},
\begin{equation}
 \M^{\rm tree}=c\,(s^2+t^2+u^2),\qquad \PhiPS_2=\frac{\dd\Omega_k}{32\pi^2},\label{eq:app-bubble-input}
\end{equation}
the integral over the direction of \(k_1\) gives
\begin{equation}
\begin{aligned}
 2\operatorname{Im}\M_s
 &=\frac1{2!}\int\frac{\dd\Omega_k}{32\pi^2}\,
 \M^{\rm tree}(12\to k_1k_2)\\
 &\qquad\times\M^{\rm tree}(k_1k_2\to34)\\
 &=\frac{g_4^2}{30\pi\Lambda^8}\,s^2\bigl(21s^2-tu\bigr),
\end{aligned}\label{eq:app-bubble-cut}
\end{equation}
and since for \(s>0\) the argument \(-s-\ii0\) lies just below the negative real axis, where the principal logarithm has the phase \(-\pi\),
\begin{equation}
 \log\frac{-s-\ii0}{\mu^2}=\log\frac{s}{\mu^2}-\ii\pi ,\label{eq:app-log-s}
\end{equation}
the coefficient of the logarithm is \(-\operatorname{Im}\M_s/\pi\), the term
\(X=s\) of Eq.~\eqref{eq:x2-channel-decomposition}; crossing gives the other
two.  The \(\log(s/\mu^2)\) part of the sum is, by Eq.~\eqref{eq:x2-sigma-sum},
\begin{equation}
\adjustbox{max width=0.86\linewidth}{$\displaystyle
 -\frac{g_4^2}{60\pi^2\Lambda^8}\sum_{X=s,t,u}X^2\bigl(21X^2-YZ\bigr)=-\frac{7g_4^2}{10\pi^2\Lambda^8}\,\sigma_2^{\,2},
$}\label{eq:app-L-coeff}
\end{equation}
the first term of Eq.~\eqref{eq:x2-L-part}.  With \(\sigma_2\) of
Eq.~\eqref{eq:sigma-angle} and
\begin{equation}
\begin{gathered}
 \int_{-1}^1\dd x\,P_\ell(x)\,(3+x^2)^2=\frac{112}{5},\ \frac{64}{35},\ \frac{16}{315},\\
 \ell=0,2,4,
\end{gathered}\label{eq:app-proj-sigma2}
\end{equation}
its projections, obtained with \(\sigma_2^2=s^4(3+x^2)^2/16\), the coefficient \(-7/10\pi^2\) of Eq.~\eqref{eq:x2-L-part} and the factor \(1/32\pi\) of the projection~\eqref{eq:pwu-convention}, in units of \(g_4^2s^4/(\pi^3\Lambda^8)\),
\begin{equation}
\begin{gathered}
 -\frac{7}{10\cdot16\cdot32}\Bigl(\frac{112}{5},\ \frac{64}{35},\ \frac{16}{315}\Bigr)\\
 =\Bigl(-\frac{49}{1600},\ -\frac{1}{400},\ -\frac{1}{14\,400}\Bigr),
\end{gathered}\label{eq:app-L-proj}
\end{equation}
are the coefficients of \(\log(s/\mu^2)\) in Eq.~\eqref{eq:x2-full-pw-main}.

\emph{\(s\) channel.}  The logarithm is Eq.~\eqref{eq:app-log-s} and the
angle enters only through \(tu\),
\begin{equation}
\adjustbox{max width=0.86\linewidth}{$\displaystyle
\begin{gathered}
 tu=\frac{s^2}{4}\,(1-x^2),\\
 s^2\bigl(21s^2-tu\bigr)=\frac{s^4}{4}\,(83+x^2)\\
 =s^4\Bigl[\frac{125}{6}P_0(x)+\frac16P_2(x)\Bigr],
\end{gathered}
$}\label{eq:app-s-angle}
\end{equation}
so that the orthogonality relation~\eqref{eq:int-orth} picks out the coefficients of \(P_0\) and \(P_2\), and with the logarithm~\eqref{eq:app-log-s} and the coefficient of Eq.~\eqref{eq:x2-channel-decomposition}
\begin{equation}
\adjustbox{max width=0.86\linewidth}{$\displaystyle
\begin{gathered}f_{\ell,s,(\partial\phi)^4}^{\rm bubble}
 =-\frac{g_4^2s^4}{60\pi^2\Lambda^8}\,\frac{d_\ell}{16\pi(2\ell+1)}
 \Bigl(\log\frac{s}{\mu^2}-\ii\pi\Bigr), \\
d_0=\frac{125}{6},\quad d_2=\frac16,\quad d_{\ell\ge4}=0 ,\end{gathered}
$}\label{eq:app-s-bubble-pw}
\end{equation}
which, with the tree partial waves of Eq.~\eqref{eq:x2-tree-pw-main}, equals
\begin{equation}
 f_{\ell,s,(\partial\phi)^4}^{\rm bubble}=-\frac{1}{2\pi}\bigl(f_{\ell,(\partial\phi)^4}^{\rm tree}\bigr)^2\log\frac{-s-\ii0}{\mu^2},\label{eq:app-s-tree-sq}
\end{equation}
Eq.~\eqref{eq:x2-s-channel-pw}.  Its imaginary part is fixed by the
two-particle cut: inserting
\begin{equation}
\begin{gathered}
 \M=16\pi\sum_\ell(2\ell+1)f_\ell P_\ell(x)\quad\text{in}\\
 2\operatorname{Im}\M=\frac{1}{2!}\int\PhiPS_2\,\M\M^*
\end{gathered}\label{eq:app-cut-insert}
\end{equation}
and using
Eqs.~\eqref{eq:int-addition} and~\eqref{eq:rambo-w2},
\begin{equation}
\adjustbox{max width=0.86\linewidth}{$\displaystyle
\begin{gathered}
 2\operatorname{Im}\M=\frac12\cdot\frac{(16\pi)^2}{32\pi^2}\cdot4\pi
 \sum_\ell(2\ell+1)|f_\ell|^2P_\ell(x)\\
 \Longrightarrow\quad
 \operatorname{Im}f_\ell=\tfrac12|f_\ell|^2 ,
\end{gathered}
$}\label{eq:app-elastic}
\end{equation}
Eq.~\eqref{eq:pwu-one-loop-pw} at lowest order.

\emph{\(t\) channel.}  With \(y=1-x\), \(t=-\tfrac s2y\), \(u=-\tfrac s2(2-y)\),
\begin{equation}
\begin{gathered}t^2\bigl(21t^2-us\bigr)=\frac{s^4}{4}\Bigl[y^2-\frac12y^3+\frac{21}{4}y^4\Bigr], \\
\log\frac{-t}{\mu^2}=\log\frac{s}{\mu^2}-\log2+\log y .\end{gathered}\label{eq:app-t-poly}
\end{equation}
The constant \(\log(s/\mu^2)-\log2\) multiplies a polynomial of degree four and
contributes only to \(\ell\le4\), where Eq.~\eqref{eq:int-ylog} completes the
projection, the \(\log2\) terms cancelling.  For \(\ell\ge5\) only \(\log y\)
contributes, and Eq.~\eqref{eq:app-endpoint} with
\begin{equation}
 I_2=-32F_2,\qquad I_3=576F_3,\qquad I_4=-18\,432F_4\label{eq:app-I234}
\end{equation}
inserted in the projection~\eqref{eq:pwu-convention} of the \(t\)-channel term of Eq.~\eqref{eq:x2-channel-decomposition} gives, for \(\ell\) above the degree of its polynomial, Eq.~\eqref{eq:app-t-high}.
\begin{figure*}[!t]
\hrule\vspace{0.1\baselineskip}
\begin{equation}
\begin{aligned}
 f_{\ell,t,(\partial\phi)^4}^{\rm bubble}
 &=-\frac{1}{32\pi}\,\frac{g_4^2}{60\pi^2\Lambda^8}\,\frac{s^4}{4}
 \Bigl[I_2-\frac12I_3+\frac{21}{4}I_4\Bigr]\\
 &=\frac{g_4^2s^4}{240\pi^3\Lambda^8}\bigl[F_2(\ell)+9F_3(\ell)+3024F_4(\ell)\bigr],
 \qquad \ell\ge5 .
\end{aligned}
 \label{eq:app-t-high}
\end{equation}
\vspace{0.1\baselineskip}\hrule
\end{figure*}
\noindent \emph{\(u\) channel.}  With \(w=1+x\), the variable that vanishes in the backward direction, where the \(u\)-channel logarithm is singular,
\begin{equation}
\begin{gathered}
 u=-\frac s2\,w,\qquad t=-\frac s2\,(2-w),\\
 \log\frac{-u}{\mu^2}=\log\frac{s}{\mu^2}-\log2+\log w,
\end{gathered}\label{eq:app-u-angle}
\end{equation}
\(u^2(21u^2-st)\) is the polynomial of Eq.~\eqref{eq:app-t-poly} with \(y\to w\),
so that
\begin{equation}
 \M^{\rm bubble}_{u,(\partial\phi)^4}(s,x)=\M^{\rm bubble}_{t,(\partial\phi)^4}(s,-x),\label{eq:app-u-t-sym}
\end{equation}
and Eq.~\eqref{eq:int-parity}, which moves the reflection \(x\to-x\) onto the Legendre polynomial, gives
\begin{equation}
 f_{\ell,u,(\partial\phi)^4}^{\rm bubble}=(-1)^\ell f_{\ell,t,(\partial\phi)^4}^{\rm bubble}\label{eq:app-u-bubble-pw}
\end{equation}
at every \(\ell\): the odd partial waves of the \(t\) and \(u\) channels cancel,
the even ones double, and for \(\ell\ge6\) the sum is
Eq.~\eqref{eq:x2-crossed-pw}; at \(\ell\le4\) it is Eq.~\eqref{eq:x2-tu-low-pw}.

\paragraph*{Crossed logarithm.}  A contact tree of degree \(n\) growing as
\(\kappa_ns^n\) at fixed \(t\) has, by Eq.~\eqref{eq:int-orth}, the \(t\)-channel expansion
\begin{equation}
 \M^{\rm tree}_{[n]}=t^n\sum_{J\le n}b_J\,P_J(z).\label{eq:app-tree-n}
\end{equation}
Matching \(z^n\) with
Eq.~\eqref{eq:app-PJ-asymptotic}, \(z\simeq2s/t\), and projecting, the \(J=n\) term of the \(t\)-channel
unitarity sum gives the leading power of \(s\) at fixed \(t\),
\begin{equation}
\begin{gathered}
 b_n=\frac{\kappa_n(n!)^2}{(2n)!},
 \qquad
 f_J^{(t)}=\frac{t^nb_J}{16\pi(2J+1)},\\
 \operatorname{Im}_t\M=\sum_{J\le n}16\pi(2J+1)\tfrac12\bigl(f_J^{(t)}\bigr)^2P_J(z)\\
 =\frac{\kappa_n^2(n!)^2}{32\pi(2n+1)(2n)!}\,s^nt^n+O(s^{n-1}) ,
\end{gathered}\label{eq:app-Imt}
\end{equation}
and since \(\operatorname{Im}\log(-t-\ii0)=-\pi\) for \(t>0\), the coefficient of the logarithm is \(-\operatorname{Im}_t\M/\pi\),
\begin{equation}
\begin{gathered}\M^{\rm bubble}_{t,[n]}=C_n\,s^nt^n\log\frac{-t}{\mu^2}+O\Bigl(s^{n-1}\log\frac{-t}{\mu^2}\Bigr), \\
C_n=-\frac{\kappa_n^2\,(n!)^2}{32\pi^2(2n+1)\,(2n)!}.\end{gathered}\label{eq:app-Cn}
\end{equation}
For \(n=2\) with \(\kappa_2=2c\), and for the structure \((s^2+t^2+u^2)^2\)
with \(\kappa_4=4\),
\begin{equation}
\begin{gathered}
 b_2=\frac c3,\qquad C_2=-\frac{c^2}{240\pi^2}=-\frac{g_4^2}{60\pi^2\Lambda^8},\\
 C_4=-\frac{1}{1260\pi^2},
\end{gathered}\label{eq:app-C2}
\end{equation}
\(b_2\) as in Eq.~\eqref{eq:x2-tree-cyclic} and \(C_2\) the \(C(s)/s^2\) of
Eq.~\eqref{eq:x2-regge}.

\paragraph*{Bananas.}  The three bananas are the terms \(X=s,t,u\) of
Eq.~\eqref{eq:x3-elastic-full}, with the common prefactor
\begin{equation}
 K\equiv-\frac{g_6^2}{512\,10!\,\pi^6\Lambda^{16}} .\label{eq:app-K}
\end{equation}

\emph{\(s\) channel.}  With the angular dependence of
Eq.~\eqref{eq:x3-coeff-angle},
\begin{equation}
\begin{gathered}\begin{aligned}f_{\ell,s,(\partial\phi)^6}^{\rm banana}
 &=-\frac{g_6^2s^8}{512\,10!\,\pi^6\Lambda^{16}}\,\frac{e_\ell}{16\pi(2\ell+1)}\\
 &\qquad\times\Bigl(\log\frac{s}{\mu^2}-\ii\pi\Bigr),\end{aligned}\\
e_0=\frac{7229}{6},\quad e_2=\frac{31}{6},\quad e_{\ell\ge4}=0 ,\end{gathered}\label{eq:app-x3-s}
\end{equation}
Eq.~\eqref{eq:x3-s-channel-pw}; its imaginary part is the four-particle
production rate, Eq.~\eqref{eq:x3-s-pw-main}.

\emph{\(t\) channel.}  With \(y=1-x\),
\begin{equation}
\adjustbox{max width=0.86\linewidth}{$\displaystyle
\begin{gathered}t^6\bigl(1210t^2-31us\bigr)
 =\frac{s^8}{64}\left[31y^6-\frac{31}{2}y^7+\frac{605}{2}y^8\right], \\
\log\frac{-t}{\mu^2}=\log\frac{s}{\mu^2}-\log2+\log y .\end{gathered}
$}\label{eq:app-x3-t-poly}
\end{equation}
The constant \(\log(s/\mu^2)-\log2\) multiplies a polynomial of degree eight
and contributes to \(\ell\le8\), completed by Eq.~\eqref{eq:int-ylog}
(Table~\ref{tab:x3-pw}).  For \(\ell\ge9\) only \(\log y\) contributes, and
Eq.~\eqref{eq:app-endpoint} with
\begin{equation}
\begin{gathered}
 I_6=-128\,(6!)^2F_6,\qquad I_7=256\,(7!)^2F_7,\\
 I_8=-512\,(8!)^2F_8
\end{gathered}\label{eq:app-I678}
\end{equation}
inserted in the projection of the \(t\)-channel banana, Eq.~\eqref{eq:x3-elastic-full}, as for the quartic sector, Eq.~\eqref{eq:app-t-high}, gives
\begin{equation}
\begin{aligned}
 f_{\ell,t,(\partial\phi)^6}^{\rm banana}
 &=-\frac{1}{32\pi}\,\frac{g_6^2}{512\,10!\,\pi^6\Lambda^{16}}\,\frac{s^8}{64}\\
 &\quad\times\Bigl[31I_6-\frac{31}{2}I_7+\frac{605}{2}I_8\Bigr]\\
 &=\frac{g_6^2s^8}{\pi^7\Lambda^{16}}
 \Bigl[\frac{31}{57\,344}F_6(\ell)+\frac{217}{8192}F_7(\ell)\\
 &\quad+\frac{4235}{64}F_8(\ell)\Bigr],
 \qquad \ell\ge9 .
\end{aligned}\label{eq:app-x3-t-high}
\end{equation}
\emph{\(u\) channel.}  With Eq.~\eqref{eq:app-u-angle},
\(u^6(1210u^2-31st)\) is the polynomial of Eq.~\eqref{eq:app-x3-t-poly} with
\(y\to w\), so that
\begin{equation}
 f_{\ell,u,(\partial\phi)^6}^{\rm banana}=(-1)^\ell f_{\ell,t,(\partial\phi)^6}^{\rm banana} ;\label{eq:app-x3-u}
\end{equation}
the odd partial waves cancel and the even ones double, which for
\(\ell\ge10\) is Eq.~\eqref{eq:x3-t-pw-main}.  The coefficients of the three
logarithms add up to Eq.~\eqref{eq:x3-sigma-sum}.
\begin{table}[!t]
\centering
\caption{Complete partial waves \(f_{\ell,(\partial\phi)^6}^{\rm banana}\) for even
\(\ell\le8\), in units of \(g_6^2s^8/(\pi^7\Lambda^{16})\) and up to the
counterterm \(P\); the odd partial waves vanish.  The
entries are rational numbers, the imaginary parts being
\(7229\pi/178\,362\,777\,600\) and \(31\pi/891\,813\,888\,000\).}
\label{tab:x3-pw}
\footnotesize\setlength{\tabcolsep}{4pt}
\renewcommand{\arraystretch}{1.15}
\begin{tabular}{cccc}
\hline
\(\ell\) & \(\log(s/\mu^2)\) coeff. & constant & imag.\ part\\
\hline
0 & \(-4.961\times10^{-8}\) & \(1.015\times10^{-9}\) & \(1.273\times10^{-7}\)\\
2 & \(-4.642\times10^{-9}\) & \(1.556\times10^{-10}\) & \(1.092\times10^{-10}\)\\
4 & \(-8.807\times10^{-10}\) & \(-1.531\times10^{-10}\) & 0\\
6 & \(-5.017\times10^{-11}\) & \(-3.098\times10^{-11}\) & 0\\
8 & \(-3.721\times10^{-13}\) & \(-7.394\times10^{-13}\) & 0\\
\hline
\end{tabular}
\end{table}

\paragraph*{Impact-parameter phases.}  At \(t=-q^2\) and \(u=-s+q^2\) the
\(t\)-channel bubble of Eq.~\eqref{eq:x2-regge} multiplies
\(\log q^2-\log\mu^2\) by
\begin{equation}
 -\frac{g_4^2}{60\pi^2\Lambda^8}\bigl(s^2q^4-sq^6+21q^8\bigr).\label{eq:app-phase-poly}
\end{equation}
Only \(\log q^2\) has a transform at \(b>0\).  Equation~\eqref{eq:log-transform} with \(n=2,3,4\) gives
\begin{equation}
 -\frac{64}{\pi b^6},\qquad \frac{2304}{\pi b^8},\qquad -\frac{147\,456}{\pi b^{10}},\label{eq:app-ft-234}
\end{equation}
and Eq.~\eqref{eq:eikonal-definition}, with its factor \(1/4s\), then yields
Eq.~\eqref{eq:x2-impact-phase}.  For the banana the coefficient of
\(\log q^2\) is, by Eq.~\eqref{eq:x3-t-channel-exact},
\begin{equation}
 K\,q^{12}\bigl(31s^2-31sq^2+1210q^4\bigr),\label{eq:app-banana-phase-poly}
\end{equation}
and the transform~\eqref{eq:log-transform} of \(q^{2n}\log q^2\) with \(n=6,7,8\) gives
\begin{equation}
\adjustbox{max width=0.86\linewidth}{$\displaystyle
 -\frac{4^6(6!)^2}{\pi b^{14}},\qquad \frac{4^7(7!)^2}{\pi b^{16}},\qquad -\frac{4^8(8!)^2}{\pi b^{18}},
$}\label{eq:app-ft-678}
\end{equation}
hence Eq.~\eqref{eq:x3-phase-boundary}; the leading term has
\begin{equation}
 \kappa_3=-\frac{31K\,4^6(6!)^2}{4\pi}=\frac{62\,g_6^2}{7\pi^7\Lambda^{16}} .\label{eq:app-kappa3}
\end{equation}

\section{Table of integrals}
\label{app:integrals}
\setcounter{equation}{0}
\gdef\theequation{B\arabic{equation}}

This appendix collects the integrals used in the text and in the other appendices.

\paragraph*{Legendre polynomials.}  For \(-1\le x\le1\),
\begin{equation}
\begin{gathered}\int_{-1}^1\dd x\,P_\ell(x)P_{\ell'}(x)=\frac{2\,\delta_{\ell\ell'}}{2\ell+1}, \\
P_\ell(-x)=(-1)^\ell P_\ell(x),\end{gathered}\label{eq:int-orth}
\end{equation}
the orthogonality and the parity of the Legendre polynomials~\cite{AbramowitzStegun}, and the powers of \(x\) expand as
\begin{equation}
\adjustbox{max width=0.86\linewidth}{$\displaystyle
\begin{gathered}x^n=\sum_{\substack{J=n,n-2,\dots\\ J\ge0}}
 \frac{(2J+1)\,n!}{2^{(n-J)/2}\bigl(\tfrac{n-J}{2}\bigr)!\,(n+J+1)!!}\,P_J(x), \\
x^2=\tfrac13\bigl[P_0(x)+2P_2(x)\bigr],\end{gathered}
$}\label{eq:int-xn}
\end{equation}
so a polynomial of degree \(n\) contributes to only \(\ell\le n\), and the substitution \(x\to-x\) with the parity of Eq.~\eqref{eq:int-orth} gives
\begin{equation}
\adjustbox{max width=0.86\linewidth}{$\displaystyle
\int_{-1}^1\dd x\,P_\ell(x)\,g(-x)=(-1)^\ell\int_{-1}^1\dd x\,P_\ell(x)\,g(x).
$}\label{eq:int-parity}
\end{equation}
The addition theorem~\cite{AbramowitzStegun}, with \(\hat n_1\cdot\hat n_2=x\), which follows from the expansion of \(P_\ell\) in spherical harmonics and their orthonormality, is
\begin{equation}
 \int\dd\Omega_k\,P_\ell(\hat n_1\!\cdot\hat k)P_{\ell'}(\hat k\cdot\hat n_2)
 =\frac{4\pi\,\delta_{\ell\ell'}}{2\ell+1}\,P_\ell(x).\label{eq:int-addition}
\end{equation}
The large-argument form of \(P_J(z)\) is Eq.~\eqref{eq:app-PJ-asymptotic}, the
Bessel limit Eq.~\eqref{eq:app-eik-bessel}, Bernstein's inequality is used in
Eq.~\eqref{eq:method-gegenbauer-bound}, and the large-argument form of \(J_0\)
is Eq.~\eqref{eq:j0-large-argument}.

\paragraph*{Projections of powers and logarithms.}  For \(a>-1\),
\begin{equation}
\begin{aligned}&\int_{-1}^1\dd x\,(1-x)^a P_\ell(x)\\
 &\quad=(-1)^\ell\,\frac{2^{a+1}\Gamma(a+1)^2}{\Gamma(a+1-\ell)\,\Gamma(a+\ell+2)} ;\end{aligned}\label{eq:int-power}
\end{equation}
setting \(a=k+\epsilon\), \(k<\ell\), and differentiating at \(\epsilon=0\),
where only
\(1/\Gamma(k+1-\ell+\epsilon)=(-1)^{\ell-k-1}(\ell-k-1)!\,\epsilon+O(\epsilon^2)\)
contributes,
\begin{equation}
\begin{aligned}I_k(\ell) &\equiv\int_{-1}^1\dd x\,P_\ell(x)(1-x)^k\log(1-x) \\
 &=(-1)^{k+1}2^{k+1}(k!)^2F_k(\ell),\qquad \ell>k,\end{aligned}\label{eq:app-endpoint}
\end{equation}
with \(F_k(\ell)=(\ell-k-1)!/(\ell+k+1)!\simeq\ell^{-2k-2}\), and for the low
partial waves
\begin{equation}
\begin{gathered}\int_0^2\dd y\,y^n\log y=2^{n+1}\Bigl[\frac{\log2}{n+1}-\frac{1}{(n+1)^2}\Bigr], \\
\int_0^1\dd x\,\log\bigl[x(1-x)\bigr]=-2 .\end{gathered}\label{eq:int-ylog}
\end{equation}

\paragraph*{Loop and phase space.}  The scalar bubble in \(d=4-2\epsilon\) is
Eq.~\eqref{eq:app-scalar-bubble}, and the massless \(n\)-body phase-space
volume is Eq.~\eqref{eq:rambo-weight}, with
\(\int\PhiPS_2=\int\dd\Omega/32\pi^2=1/8\pi\), Eq.~\eqref{eq:rambo-w2}.

\paragraph*{Impact parameter.}  The transverse Fourier transform reduces to a
Bessel integral,
\begin{equation}
\adjustbox{max width=0.86\linewidth}{$\displaystyle
 \int\frac{\dd^2\bm q}{(2\pi)^2}e^{-\ii\bm q\cdot\bm b}g(q)
 =\frac{1}{2\pi}\int_0^\infty\dd q\,q\,J_0(qb)\,g(q),
$}\label{eq:int-fourier}
\end{equation}
which on powers times a logarithm gives Eq.~\eqref{eq:log-transform}.

\section{Dispersion relation and Regge boundedness}
\label{app:dispersion-regge}
\setcounter{equation}{0}
\gdef\theequation{C\arabic{equation}}

This appendix derives the dispersion relation of Sec.~\ref{sec:contour} with
\(n\) subtractions and the condition it imposes on the Regge exponent of
Eq.~\eqref{eq:regge-exponent}.

By real analyticity, Eq.~\eqref{eq:analytic},
the discontinuity across the right-hand cut is
\begin{equation}
\begin{aligned}\operatorname{Disc}\M(s',t) &\equiv\M(s'+\ii0,t)-\M(s'-\ii0,t) \\
 &=2\ii\operatorname{Im}\M(s',t),\qquad s'>0,\end{aligned}\label{eq:disc-def}
\end{equation}
with \(\operatorname{Im}\M(s',t)\equiv\operatorname{Im}\M(s'+\ii0,t)\).  On the
left-hand cut \(s'=-u'-t\) with \(u'>0\); since \(u=-s-t\) reverses the sign of
the imaginary part, crossing maps \(s'\pm\ii0\) to \(u'\mp\ii0\), and
\begin{equation}
\begin{aligned}\operatorname{Disc}\M(s',t) &=\M(u'-\ii0,t)-\M(u'+\ii0,t) \\
 &=-2\ii\operatorname{Im}\M(u',t).\end{aligned}\label{eq:disc-cuts}
\end{equation}

These discontinuities enter the Cauchy identity with \(n\) subtractions,
which we write between two loops of finite radius, \(|s|=r\) and \(|s|=R\),
\(0<r<R\), so that every term is finite for every \(n\).  With the arcs
\begin{equation}
 a_n(r)\equiv\frac{1}{2\pi\ii}\oint_{|s|=r}\frac{\M(s,0)}{s^{n+1}}\,\dd s ,
 \qquad A_n(R)\equiv a_n(R),\label{eq:app-arc-r}
\end{equation}
both loops being traversed counter-clockwise, \(\M(s,0)/s^{n+1}\) is analytic
between the loops of Fig.~\ref{fig:contour}(a), cut along the real axis.  The
boundary of this region consists of the large loop, counter-clockwise, the small
loop, clockwise, the upper edges of the cuts, traversed towards
\(+\operatorname{Re}s\), and the lower edges, towards \(-\operatorname{Re}s\),
so that Cauchy's theorem gives
\begin{equation}
\begin{aligned}a_n(r) &=A_n(R)\\
 &\quad+\frac{1}{2\pi\ii}\int_{r<|s'|<R}\dd s'\,
 \frac{\operatorname{Disc}\M(s',0)}{s'^{n+1}} .\end{aligned}\label{eq:app-cauchy-n}
\end{equation}
On the right-hand cut Eq.~\eqref{eq:disc-def} gives
\[\frac1\pi\int_r^R\dd s\,\frac{\operatorname{Im}\M(s,0)}{s^{n+1}}.\]
On the
left-hand cut, \(s'=-u'\),
\begin{equation}
 \dd s'=-\dd u',
 \qquad
 s'^{n+1}=(-1)^{n+1}u'^{n+1},\label{eq:app-left-sub}
\end{equation}
and \(s'\) running from \(-R\) to \(-r\) corresponds to \(u'\) running from
\(R\) to \(r\), so that Eq.~\eqref{eq:disc-cuts} gives
\begin{equation}
\begin{aligned}
 &\frac{1}{2\pi\ii}\int_{-R}^{-r}\dd s'\,
 \frac{\operatorname{Disc}\M(s',0)}{s'^{n+1}}\\
 &\quad=\frac{(-1)^n}{\pi}\int_r^R\dd u'\,
 \frac{\operatorname{Im}\M(u',0)}{u'^{n+1}} .
\end{aligned}\label{eq:app-left-n}
\end{equation}
For \(n=2\) these two steps are Eq.~\eqref{eq:cuts}.  Adding the cuts,
\begin{equation}
 a_n(r)=\frac{1+(-1)^n}{\pi}\int_r^R\dd s\,\frac{\operatorname{Im}\M(s,0)}{s^{n+1}}
 +A_n(R) .\label{eq:app-identity}
\end{equation}
For odd \(n\) both sides vanish by crossing, Eq.~\eqref{eq:analytic}; for even \(n\) the cuts add.
Equation~\eqref{eq:app-identity} holds for every \(0<r<R\) and involves no
assumption on the behaviour at small or at large \(s\); for the one-loop EFT,
\(A_2(R)=2c-7g_4^2R^2/(20\pi^2\Lambda^8)\), Eq.~\eqref{eq:eft-arc}.

The small loop can be shrunk to the origin only where the arc has a limit.
At one loop the \(s\)- and \(u\)-channel bubbles give the forward amplitude,
the \(t\)-channel term vanishing at \(t=0\) with its factor \(X^2=t^2\),
Eq.~\eqref{eq:x2-channel-decomposition},
\begin{equation}
\begin{aligned}\M(s,0) &=2c\,s^2+\rho_4\,s^4\\
 &\quad-\frac{7c^2s^4}{80\pi^2}\Bigl[\log\Bigl(\frac{-s}{\mu^2}\Bigr)+\log\Bigl(\frac{s}{\mu^2}\Bigr)\Bigr],\end{aligned}\label{eq:app-forward-one-loop}
\end{equation}
where \(7c^2/80\pi^2=21g_4^2/60\pi^2\Lambda^8\), in agreement with
\(\operatorname{Im}\log(-s-\ii0)=-\pi\) and Eq.~\eqref{eq:x2-optical}, and
\(\rho_4\) is real and depends on the counterterm, Appendix~\ref{app:P4}.  With
\(s=re^{\ii\theta}\), \(\frac{1}{2\pi\ii}\oint_{|s|=r}\log(\pm s/\mu^2)\,\dd s/s=\log(r/\mu^2)\), and
the terms \(s^{k}\log s\) with \(k\neq n\) give \(O(r^{k-n})\), so that
\begin{equation}
\begin{gathered}a_2(r)=2c+O(r^2), \\
a_4(r)=\rho_4-\frac{7c^2}{40\pi^2}\log\frac{r}{\mu^2} .\end{gathered}\label{eq:app-soft}
\end{equation}
For \(n=2\) the limit \(r\to0\) exists and defines \(c_2\equiv a_2(0)=2c\), the
coefficient of \(s^2\) at the origin; the terms beyond the tree begin at \(s^4\)
at every order, so that the limit is not affected by higher loops.  For
\(n=4\) the arc diverges logarithmically as \(r\to0\): the term
\(s^4\log(-s)\) has no fourth derivative at the origin, and on the right-hand
side of Eq.~\eqref{eq:app-identity}
\(\operatorname{Im}\M(s,0)/s^5=7c^2/(80\pi s)\) gives
\(\frac{7c^2}{40\pi^2}\log(R/r)\), the same divergence.  For \(n\ge6\) the arc
diverges as \(r^{4-n}\), and at higher orders the terms \(s^{2k}\log^js\)
enter in the same way.  Equations~\eqref{eq:app-arc-r}--\eqref{eq:app-identity}
therefore hold for every \(n\) at \(r>0\), whereas the limit \(r\to0\), and with it a
coefficient \(c_n\) of \(s^n\) at the origin, exists only for \(n=2\).  The dispersion
relation of Sec.~\ref{sec:contour}, Eqs.~\eqref{eq:cauchy-cut},~\eqref{eq:intro-dispersion}
and~\eqref{eq:uv-transfer}, uses only \(n=2\) at \(t=0\), and at fixed \(t<0\) the small loop is at
\(r>-t\), Eq.~\eqref{eq:app-fixed-t-n}.

The identity becomes a constraint only when the arc vanishes, and the arc is controlled by the Regge exponent~\cite{Froissart,JinMartin}.  On \(|s|=R\), \(|\dd s|=R\,\dd\theta\) and
\begin{equation}
|A_n(R)|\le\frac{\max_{|s|=R}|\M(s,0)|}{R^n}.\label{eq:app-arc-bound}
\end{equation}
The maximum is over all directions of \(s\), whereas the Regge
behaviour~\eqref{eq:regge-exponent} is stated along the real axis; for an
amplitude polynomially bounded in the plane, the Phragm\'en--Lindel\"of
theorem~\cite{Titchmarsh} extends a bound \(|\M|\lesssim|s|^{a}\) from the two edges of the cuts
to the entire loop.  Then \(|A_n(R)|\lesssim R^{a(0)-n}\to0\) for \(a(0)<n\),
and the arc at the finite radius \(r\) is a sum rule with an infrared-finite
integrand~\cite{BellazziniRiembauRiva,BellazziniMoments,AlberteQED,HerreroValea},
\begin{equation}
\begin{gathered}a_n(r)=\frac{2}{\pi}\int_r^\infty\dd s\,\frac{\operatorname{Im}\M(s,0)}{s^{n+1}}
 \ \ge0, \\
n\ \text{even},\quad n>a(0),\quad r>0.\end{gathered}\label{eq:app-sum-rule}
\end{equation}
For \(n\ge4\) the left-hand side is computed in the EFT at a radius
\(r\ll\Lambda^2\), with the logarithm of Eq.~\eqref{eq:app-soft}, and the right-hand
side contains no region \(s<r\); the infrared-sensitive integral over
\(0<s<r\), which diverges for \(n\ge4\), is thereby moved to the left-hand side,
where it combines with the divergence of the coefficient of \(s^n\) into the
finite \(a_n(r)\).
The sign follows from the partial-wave expansion~\eqref{eq:pwu-convention} at
\(x=1\), where \(P_\ell(1)=1\),
\begin{equation}
\begin{aligned}\operatorname{Im}\M(s,0) &=16\pi\sum_{\ell\ {\rm even}}(2\ell+1)\operatorname{Im}f_\ell(s) \\
 &=s\,\sigma_{\rm tot}(s)\ge0,\end{aligned}\label{eq:app-optical}
\end{equation}
each term being non-negative by Eq.~\eqref{eq:pwu-unitarity}.  For \(n=2\) and
\(r\to0\), \(c_2=2c\), Eq.~\eqref{eq:app-sum-rule} divided by two is
Eq.~\eqref{eq:intro-dispersion} without the arc, and splitting the integral at
\(\Lambda^2\) gives Eq.~\eqref{eq:uv-transfer}.

If instead \(a(0)\ge n\), \(A_n(\infty)\) does not
vanish, and its value is set by the asymptotics, Eq.~\eqref{eq:app-arc-power}.  Consider the
crossing-symmetric form \(\M(s,0)\simeq\beta\bigl[(-s-\ii0)^a+s^a\bigr]\), the
sum of an \(s\)- and a \(u\)-channel term with the cuts of
Fig.~\ref{fig:contour}(a).  For \(s^a\), with \(s=Re^{\ii\theta}\),
\(-\pi<\theta<\pi\), and \(\dd s=\ii s\,\dd\theta\),
\begin{equation}
\begin{aligned}\frac{1}{2\pi\ii}\oint_{|s|=R}\frac{s^{a}}{s^{n+1}}\,\dd s &=\frac{R^{a-n}}{2\pi}\int_{-\pi}^{\pi}\dd\theta\,e^{\ii(a-n)\theta} \\
 &=\frac{\sin\pi(a-n)}{\pi(a-n)}\,R^{a-n}.\end{aligned}\label{eq:app-arc-sa}
\end{equation}
For \((-s-\ii0)^a\), with \(s=Re^{\ii\varphi}\), \(0<\varphi<2\pi\),
\((-s)^a=R^ae^{\ii a(\varphi-\pi)}\) and \(e^{-\ii n\varphi}=(-1)^ne^{-\ii n(\varphi-\pi)}\),
so that the substitution \(\theta=\varphi-\pi\) gives the same result for even
\(n\), and
\begin{equation}
 A_n(R)=2\beta\,\frac{\sin\pi(a-n)}{\pi(a-n)}\,R^{a-n}.\label{eq:app-arc-power}
\end{equation}
At \(a=n\) the arc is the constant \(2\beta\), and
Eq.~\eqref{eq:app-identity} becomes
\(a_n(r)-2\beta=\frac2\pi\int_r^\infty\dd s\,\operatorname{Im}\M/s^{n+1}\) with \(\beta\)
unknown; for \(a>n\) the arc grows as \(R^{a-n}\) with a sign that depends on
\(a\).  In both cases \(\operatorname{Im}\M\ge0\) no longer fixes the sign of \(a_n(r)\), Eq.~\eqref{eq:app-sum-rule}.

The same steps apply at fixed \(t<0\), where the cuts overlap on \(0\le s\le-t\), and the small loop, at finite radius as at \(t=0\), must also enclose the overlap, \(|s|=r>-t\), Eq.~\eqref{eq:app-fixed-t-n}.  On the left-hand cut
\(s=-u-t\), Eq.~\eqref{eq:disc-cuts} holds, \(|s|>r\) corresponds to \(u>r-t\), and
\begin{equation}
 s^{n+1}=(-1)^{n+1}(u+t)^{n+1},\label{eq:app-left-t}
\end{equation}
so that the steps of Eqs.~\eqref{eq:app-cauchy-n}--\eqref{eq:app-identity} give,
for \(a(t)<n\),
\begin{equation}
\begin{aligned}&\frac{1}{2\pi\ii}\oint_{|s|=r}\frac{\M(s,t)}{s^{n+1}}\,\dd s\\
 &\quad=\frac1\pi\int_r^\infty\dd s\,\frac{\operatorname{Im}\M(s,t)}{s^{n+1}}\\
 &\qquad+\frac{(-1)^n}{\pi}\int_{r-t}^\infty\dd u\,
 \frac{\operatorname{Im}\M(u,t)}{(u+t)^{n+1}} ,\end{aligned}\label{eq:app-fixed-t-n}
\end{equation}
which reduces to Eq.~\eqref{eq:dispersion-fixed-t} for \(n=2\).

Whether the completion is Regge bounded is not decided by unitarity alone~\cite{Froissart,Martin}.  Unitarity bounds every partial wave, \(|f_\ell|\le2\), Eq.~\eqref{eq:pwu-unitarity}, but not the number of partial waves that contribute: in the massless
theory they fall only as a power of \(\ell\), Eq.~\eqref{eq:x2-crossed-pw}, and
the count of Eq.~\eqref{eq:forward-eft-controlled},
\(|\M(s,0)|\lesssim16\pi\ell_*^2\), stays below \(s^2\) only if \(\ell_*\) grows
more slowly than \(s\).

The growth is tied instead to the spins of the crossed channel~\cite{Regge,Collins}.  In the
crossed process \(z=1+2s/t\), and from the explicit form of the Legendre polynomial~\cite{Szego},
\begin{equation}
\begin{aligned}P_J(z) &=\frac{1}{2^J}\sum_{k}(-1)^k\binom Jk\binom{2J-2k}{J}z^{J-2k} \\
 &\simeq\frac{(2J)!}{2^J(J!)^2}\,z^J,\qquad z\to\infty,\end{aligned}\label{eq:app-PJ-asymptotic}
\end{equation}
which is Eq.~\eqref{eq:t-channel-spin}; a growth slower than \(s^2\)
therefore requires infinitely many \(J\) (Sec.~\ref{sec:regge}).  Since \(1/\sin\pi J\) has residue
\((-1)^k/\pi\) at \(J=k\) and \(P_k(-z)=(-1)^kP_k(z)\), the sum over even \(J\) of
Eq.~\eqref{eq:t-channel-pw} can be written as
\begin{equation}
\adjustbox{max width=0.86\linewidth}{$\displaystyle
\begin{aligned}
 &\sum_{J\ {\rm even}}(2J+1)f^{(t)}_J(t)P_J(z)\\
 &\quad=\frac{1}{4\ii}\oint_{\mathcal C}\dd J\,(2J+1)f^{(t)}(J,t)\,
 \frac{P_J(-z)+P_J(z)}{\sin\pi J},
\end{aligned}
$}\label{eq:app-sommerfeld-watson}
\end{equation}
where \(\mathcal C\) encircles the non-negative integers counter-clockwise and
\(f^{(t)}(J,t)\) is the continuation of the even partial waves to complex
\(J\) (the Sommerfeld--Watson transform~\cite{Collins}).  Deforming \(\mathcal C\) to the line
\(\operatorname{Re}J=-\frac12\), whose contribution falls as \(s^{-1/2}\)~\cite{Collins}, the
rightmost singularity of \(f^{(t)}(J,t)\), at \(J=\alpha(t)\), gives
\begin{equation}
 \M(s,t)\propto\beta(t)\,s^{\alpha(t)},
 \qquad a(t)=\alpha(t),\label{eq:app-regge-pole}
\end{equation}
and Regge boundedness requires \(\alpha(t)<2\).

\section{Impact-parameter form of the high-spin sum}
\label{app:impact}
\setcounter{equation}{0}
\gdef\theequation{D\arabic{equation}}

This appendix derives the replacements~\eqref{eq:gegenbauer-bessel} and the
integral in the last line of Eq.~\eqref{eq:general-eikonal} from the
partial-wave sum in its first line, where \(\cos\theta=x=1+2t/s\),
Eq.~\eqref{eq:pwu-convention}.

\paragraph*{Scattering angle.}  For \(t=-q^2\), \(1-\cos\theta=2q^2/s\) and
\(1-\cos\theta=2\sin^2(\theta/2)\) give
\begin{equation}
 \sin\frac\theta2=\frac{q}{\sqrt s},\qquad
 \theta=\frac{2q}{\sqrt s}\Bigl[1+O\Bigl(\frac{q^2}{s}\Bigr)\Bigr],\label{eq:app-eik-angle}
\end{equation}
so the Regge limit \(s\to\infty\) at fixed \(q\) is the limit of small angle.

\paragraph*{Impact parameter.}  The orbital angular momentum
\(\ell+\tfrac12\) corresponds to \(pb\), with the centre-of-mass momentum
\(p=\sqrt s/2\)~\cite{KabatOrtiz,DiVecchiaEikonal}, which defines
\begin{equation}
 b\equiv\frac{2\ell+1}{\sqrt s},\qquad
 2\ell+1=\sqrt s\,b,\qquad
 \dd\ell=\frac{\sqrt s}{2}\,\dd b .\label{eq:app-eik-b}
\end{equation}
These are the second and third relations of Eq.~\eqref{eq:gegenbauer-bessel},
exact with this definition; the main text uses \(b=2\ell/\sqrt s\), which
differs by a relative amount \(1/(2\ell)\).

\paragraph*{Bessel asymptotics.}  Hilb's formula~\cite{Szego} gives, for
\(0<\theta\le\pi-\epsilon\) with fixed \(\epsilon>0\),
\begin{equation}
\begin{gathered}
 P_\ell(\cos\theta)=\Bigl(\frac{\theta}{\sin\theta}\Bigr)^{1/2}
 J_0\bigl((\ell+\tfrac12)\theta\bigr)+R_\ell(\theta),\\
 R_\ell(\theta)=\begin{cases}\theta^{1/2}\,O(\ell^{-3/2}), & \theta\ge c/\ell,\\
 \theta^2\,O(1), & \theta\le c/\ell,\end{cases}
\end{gathered}\label{eq:app-eik-hilb}
\end{equation}
with \(c\) a fixed positive constant.  By Eqs.~\eqref{eq:app-eik-angle}
and~\eqref{eq:app-eik-b}, \((\ell+\tfrac12)\theta=qb\,[1+O(q^2/s)]\) and
\((\theta/\sin\theta)^{1/2}=1+O(q^2/s)\), while
\(\theta^{1/2}\ell^{-3/2}\simeq4q^{1/2}/(b^{3/2}s)\), smaller than
\(|J_0(qb)|\sim(qb)^{-1/2}\) by \(4q/(bs)\).  Hence
\begin{equation}
 P_\ell\Bigl(1-\frac{2q^2}{s}\Bigr)=J_0(qb)
 +O\Bigl(\frac{q^3b}{s}\Bigr)+O\Bigl(\frac{q^{1/2}}{b^{3/2}s}\Bigr),\label{eq:app-eik-bessel}
\end{equation}
the first relation of Eq.~\eqref{eq:gegenbauer-bessel}, whose argument
\(1-q^2b^2/(2\ell^2)\) equals \(1-2q^2/s\) for \(b=2\ell/\sqrt s\).

\paragraph*{Sum over even spins.}  The summand
\(F(\ell)=(2\ell+1)f_\ell P_\ell\) is sampled at even \(\ell\), with spacing
\(\Delta\ell=2\), and the Euler--Maclaurin formula~\cite{AbramowitzStegun}
gives
\begin{equation}
\adjustbox{max width=207.8pt}{$\displaystyle
\begin{aligned}
 \sum_{\substack{\ell>\ell_*\\ \ell\ {\rm even}}}F(\ell)
 &=\frac12\int_{\ell_*}^\infty\dd\ell\,F(\ell)\\
 &\quad+\frac12F(\ell_*)-\frac16F'(\ell_*)+\cdots .
\end{aligned}
$}\label{eq:app-eik-em}
\end{equation}
The integral replaces the sum when \(F\) changes little over
\(\Delta\ell=2\), that is over \(\Delta b=4/\sqrt s\).  In \(b\), \(F\)
oscillates through \(J_0(qb)\) on the scale \(1/q\) and through
\(e^{2\ii\delta(s,b)}\) on the scale \(|2\partial\delta/\partial b|^{-1}\),
which equals \(1/q\) at the stationary point~\eqref{eq:general-saddle}, where the
classical deflection equals \(q\).  Both phases therefore change by
\(O(4q/\sqrt s)\) per step in the region that dominates the stationary-phase
integral, Eq.~\eqref{eq:stationary-phase}.  The boundary term
\(\tfrac12F(\ell_*)\) is a single partial wave, bounded by
\(|f_\ell|\le2\), Eq.~\eqref{eq:pwu-unitarity}, and smaller by \(1/\ell_*\)
than the low-spin bound~\eqref{eq:method-low-spin-bound}, to which it is
assigned.

\paragraph*{Result.}  With \(f_\ell=\ii(1-e^{2\ii\delta_\ell})\),
\(\delta_\ell=\delta(s,b)\), Eq.~\eqref{eq:pwu-to-eikonal}, and
Eqs.~\eqref{eq:app-eik-b}--\eqref{eq:app-eik-em}, the prefactors combine as
\(16\pi\cdot\tfrac12\cdot\tfrac{\sqrt s}{2}\cdot\sqrt s=4\pi s\), which gives the
last line of Eq.~\eqref{eq:general-eikonal} with
\(b_*=(2\ell_*+1)/\sqrt s\) and relative corrections of order \(q/\sqrt s\).

\section{Renormalization of the one-loop \texorpdfstring{$(\partial\phi)^4$}{(∂φ)⁴} amplitude}
\label{app:P4}
\setcounter{equation}{0}
\gdef\theequation{E\arabic{equation}}

Each quartic vertex contains four derivatives, so the \(X\)-channel
bubble, \(X=s,t,u\), is
\begin{equation}
\begin{gathered}
 \ii\M^{\rm bubble,\,bare}_{X,(\partial\phi)^4}=\frac12\,\mu^{2\epsilon}\!\int\frac{\dd^d\ell}{(2\pi)^d}\\
 \times\frac{V_4(p_a,p_b,\ell,-\ell-P)\,V_4(p_c,p_d,-\ell,\ell+P)}
 {(\ell^2-\ii0)\bigl((\ell+P)^2-\ii0\bigr)},\\
 -P^2=X,
\end{gathered}\label{eq:app-bubble}
\end{equation}
in \(d=4-2\epsilon\), with \(V_4\) the quartic vertex, \(P\) the momentum
through the bubble and \(1/2\) the symmetry factor.  The numerator has degree
eight; after tensor reduction~\cite{PassarinoVeltman}, with massless tadpoles vanishing~\cite{tHooftVeltman}, the integral is a polynomial times the scalar bubble
\begin{equation}
\begin{aligned}
 &\mu^{2\epsilon}\!\int\frac{\dd^d\ell}{(2\pi)^d}\,
 \frac{1}{(\ell^2-\ii0)\bigl((\ell+P)^2-\ii0\bigr)}\\
 &\quad=\frac{\ii}{16\pi^2}\left[\frac1{\bar\epsilon}+2
 -\log\frac{-X-\ii0}{\mu^2}\right]+O(\epsilon),
\end{aligned}\label{eq:app-scalar-bubble}
\end{equation}
where \(1/\bar\epsilon=1/\epsilon-\gamma_E+\log4\pi\)~\cite{BBDM} and
\(\int_0^1\dd x\,\log[x(1-x)]=-2\) gives the constant.  The \(O(\epsilon)\)
parts of the \(d\)-dependent reduction coefficients, multiplying the pole,
add finite polynomial terms~\cite{PassarinoVeltman}.  Since pole and logarithm enter only through
Eq.~\eqref{eq:app-scalar-bubble}, they appear as
\(1/\bar\epsilon-\log(-X/\mu^2)\), so the logarithm coefficient fixed by
unitarity (Sec.~\ref{sec:quartic}) also fixes the pole.  With
\(N_X\equiv X^2(21X^2-YZ)\),
\begin{equation}
\begin{aligned}
 &\M^{\rm bubble,\,bare}_{(\partial\phi)^4}=\sum_{(X;Y,Z)}\left\{-\frac{g_4^2}{60\pi^2\Lambda^8} N_X\right.\\
 &\qquad\times\left.\biggl[\log\frac{-X-\ii0}{\mu^2}-\frac1{\bar\epsilon}\biggr]+R_X\right\},
\end{aligned}\label{eq:app-bare}
\end{equation}
where the degree-four polynomials \(R_X\) collect the finite rational
terms.  By Eq.~\eqref{eq:x2-sigma-sum} the pole multiplies
\(\sum_{(X;Y,Z)}N_X=42\,\sigma_2^2\),
a local contact term
\(\frac{42g_4^2}{60\pi^2\Lambda^8}\sigma_2^2/\bar\epsilon
=\frac{7g_4^2}{10\pi^2\Lambda^8}\sigma_2^2/\bar\epsilon\).  Since \(2a+3b=4\) has
the unique solution \((2,0)\), \(\sigma_2^2\) is the only crossing-symmetric
structure of degree four: the on-shell amplitude of the eight-derivative
operators \(\partial^8\phi^4\), with \((s^2+t^2+u^2)^2=4\sigma_2^2\); for the same reason
\(\sum_X R_X\propto\sigma_2^2\).  A counterterm with on-shell amplitude
\((g_4^2/\Lambda^8)\,\delta c\,\sigma_2^2\) cancels the pole for
\begin{equation}
 \delta c=-\frac{42}{60\pi^2}\,\frac1{\bar\epsilon}
 =-\frac{7}{10\pi^2}\,\frac1{\bar\epsilon}\label{eq:app-counterterm}
\end{equation}
in \(\overline{\rm MS}\), giving Eq.~\eqref{eq:x2-channel-decomposition} with the
counterterm~\eqref{eq:x2-P4}, where \(c_P(\mu)\) collects the finite counterterm and \(\sum_X R_X\).
Its value at a reference scale is fixed by matching to the completion~\cite{Weinberg},
equivalently by the coefficient of the eight-derivative operators, which
\(P_4\) renormalizes; only this boundary value is
ultraviolet sensitive.

The counterterm \(P_4\) enters only
the low-spin part, Eq.~\eqref{eq:x2-low-master}, which is bounded by unitarity
without the value of \(c_P\), and not the eikonal phase; the choice \(\mu^2=s\) moves only polynomials in
\(t\), again at \(b=0\), between the logarithms and \(P_4\).  The
ultraviolet-sensitive information is thus confined to local terms that do not
affect the high-spin analysis, while \(C(s)\), which does, is fixed by
unitarity; as discussed in Sec.~\ref{sec:coefficient}, its extrapolation to
arbitrary \(s\) is nevertheless limited to the energies at which the EFT partial waves are
small.

\section{RAMBO phase-space integration of the unitarity cuts}
\label{app:rambo}
\setcounter{equation}{0}
\gdef\theequation{F\arabic{equation}}

The absorptive parts used in this paper are phase-space integrals of products
of tree amplitudes, Eq.~\eqref{eq:pwu-cut}~\cite{Cutkosky}.  For the
\(s\)-channel banana (Fig.~\ref{fig:x3-cut}) the four internal lines are put
on shell, each propagator being replaced by its on-shell delta function, and
Eq.~\eqref{eq:pwu-cut} with \(n=4\) reads
\begin{equation}
\adjustbox{max width=\linewidth}{$\displaystyle
\begin{aligned}
 2\operatorname{Im}\M^{\rm banana}_{s,(\partial\phi)^6}
 &=\frac{1}{4!}\int\PhiPS_4(P;k_1,\dots,k_4)\\
 &\quad\times\M^{\rm tree}_{6,(\partial\phi)^6}(p_1,p_2,-k_1,\dots,-k_4)\\
 &\quad\times\M^{\rm tree}_{6,(\partial\phi)^6}(p_3,p_4,k_1,\dots,k_4),
\end{aligned}
$}\label{eq:banana-cut}
\end{equation}
\begin{figure*}[!t]
\edef\savedeqno{\the\value{equation}}\setcounter{equation}{17}
\hrule\vspace{0.1\baselineskip}
\begin{equation}
\widehat{\operatorname{Im}\M^{\rm bubble}_{s,(\partial\phi)^4}}
 =\frac{w_2}{2\cdot2!\,N}\sum_{r=1}^N
 \M^{\rm tree}_{(\partial\phi)^4}\bigl(p_1,p_2,-k_1^{(r)},-k_2^{(r)}\bigr)\,
 \M^{\rm tree}_{(\partial\phi)^4}\bigl(p_3,p_4,k_1^{(r)},k_2^{(r)}\bigr),
 \label{eq:rambo-two-body}
\end{equation}
\begin{equation}
\widehat{\operatorname{Im}\M^{\rm banana}_{s,(\partial\phi)^6}}
 =\frac{w_4(s)}{2\cdot4!\,N}\sum_{r=1}^N
 \M^{\rm tree}_{6,(\partial\phi)^6}\bigl(p_1,p_2,-k_1^{(r)},\dots,-k_4^{(r)}\bigr)\,
 \M^{\rm tree}_{6,(\partial\phi)^6}\bigl(p_3,p_4,k_1^{(r)},\dots,k_4^{(r)}\bigr),
 \label{eq:rambo-four-body}
\end{equation}
\vspace{0.1\baselineskip}\hrule
\setcounter{equation}{\savedeqno}
\end{figure*}
with \(P=p_1+p_2\) the total momentum, \(\M^{\rm tree}_{6,(\partial\phi)^6}\)
the sextic tree~\eqref{eq:x3-tree-permutation} and \(1/4!\) the symmetry factor
of the four identical intermediate particles, which the phase-space integral
counts \(4!\) times.  The momenta are incoming, as in
Sec.~\ref{sec:intro}, so the produced particles enter the first vertex with
\(-k_i\); the tree is real, so that \(\M^*=\M\) in Eq.~\eqref{eq:pwu-cut}.
Since every field of Eq.~\eqref{eq:lagrangian} carries one derivative, the
tree is linear in each external momentum, Eq.~\eqref{eq:x3-tree-permutation}.
After the integration over the \(k_i\), which can depend only on \(P\), the
product is therefore bilinear in \((p_1,p_2)\) and \((p_3,p_4)\), so that it
is at most quadratic in \(p_1\cdot p_3=t/2\); being of degree sixteen and
symmetric under \(t\leftrightarrow u\), it has the form \(s^6(a\,s^2+b\,tu)\) of
Eq.~\eqref{eq:banana-imaginary}.  The two-body cut of the quartic sector,
Eq.~\eqref{eq:pwu-cut} with \(n=2\), is elementary (Appendix~\ref{app:pw}),
but the four-body cut~\eqref{eq:banana-cut} has no elementary
parametrization.  Both are evaluated with the RAMBO
algorithm~\cite{RAMBO,RAMBOdiet}, which generates \(n\) massless momenta with
fixed total momentum and a constant phase-space weight; the construction, the
weight and the estimators follow, in the conventions of
Sec.~\ref{sec:intro}.

Following Ref.~\cite{RAMBO}, for each of the \(n\) momenta four independent
random numbers \(r_i^a\in(0,1)\), \(a=1,\dots,4\), uniformly distributed, are
combined into
\begin{equation}
\adjustbox{max width=\linewidth}{$\displaystyle
\tau_i=2r_i^1-1,\qquad \varphi_i=2\pi r_i^2,\qquad
 E_i=-\log\bigl(r_i^3r_i^4\bigr),
$}\label{eq:rambo-random}
\end{equation}
and
\begin{equation}
\adjustbox{max width=\linewidth}{$\displaystyle
q_i^\mu=E_i\Bigl(1,\ \sqrt{1-\tau_i^2}\cos\varphi_i,\
 \sqrt{1-\tau_i^2}\sin\varphi_i,\ \tau_i\Bigr).
$}\label{eq:rambo-q}
\end{equation}
Here \(\tau_i=\cos\theta_i\) and \(\varphi_i\) are uniform on
\((-1,1)\) and \((0,2\pi)\), and since the solid angle is
\(\dd\Omega=\dd\tau\,\dd\varphi\), the direction of \(\bm q_i\) is uniform on
the sphere; with \(|\bm q_i|=q_i^0=E_i\), each \(q_i\) is massless.  For
uniform \(r\), \(-\log r\) has the density \(e^{-x}\), and \(E_i\) is the sum of
two such independent variables, with density
\(\int_0^E\dd x\,e^{-x}e^{-(E-x)}=E\,e^{-E}\),
\begin{equation}
 \dd\mathcal P(E)=E\,e^{-E}\,\dd E .\label{eq:rambo-energy}
\end{equation}
The massless measure is
\(\dd^4q\,\delta(q^2)\theta(q^0)=\dd^3\bm q/2q^0=\frac12E\,\dd E\,\dd\Omega\), so
that Eqs.~\eqref{eq:rambo-random}--\eqref{eq:rambo-energy} generate each
\(q_i\) with the normalized distribution
\begin{equation}
\adjustbox{max width=\linewidth}{$\displaystyle
\frac{E\,e^{-E}\,\dd E\,\dd\Omega}{4\pi}
 =\frac{1}{2\pi}\,\dd^4q\,\delta(q^2)\theta(q^0)\,e^{-q^0},
$}\label{eq:rambo-q-density}
\end{equation}
which is Lorentz invariant up to the factor \(e^{-q^0}\); this is the property
from which the constant weight follows~\cite{RAMBO}.
The sum of the \(q_i\) is timelike, since future-directed null vectors that
are not all parallel add to a timelike vector, and in the mostly-plus metric
of Sec.~\ref{sec:intro}
\begin{equation}
 Q^\mu=\sum_{i=1}^nq_i^\mu,\qquad M=\sqrt{-Q^2}>0 .\label{eq:rambo-Q}
\end{equation}
A boost to the rest frame of \(Q\), followed by a common rescaling, maps the
\(q_i\) onto momenta with the required total momentum, which in the
centre-of-mass frame is
\begin{equation}
 P^\mu=\bigl(\sqrt s,\bm0\bigr).\label{eq:rambo-P}
\end{equation}
The boost and rescaling parameters are
\begin{equation}
\begin{gathered}\bm\nu=-\frac{\bm Q}{M},\qquad \gamma=\frac{Q^0}{M}=\sqrt{1+\bm\nu^2},\\
 \zeta=\frac{1}{1+\gamma},\qquad \alpha=\frac{\sqrt s}{M},\end{gathered}\label{eq:rambo-params}
\end{equation}
where \(\bm\nu=\gamma\bm\beta\) with \(\bm\beta=-\bm Q/Q^0\) the velocity of the
boost, and \(\gamma^2-\bm\nu^2=1\) follows from \((Q^0)^2-\bm Q^2=M^2\).  The
final momenta are
\begin{equation}
\begin{gathered}k_i^0=\alpha\bigl(\gamma q_i^0+\bm\nu\cdot\bm q_i\bigr),\\
 \bm k_i=\alpha\bigl(\bm q_i+q_i^0\,\bm\nu+\zeta\,(\bm\nu\cdot\bm q_i)\,\bm\nu\bigr),\end{gathered}\label{eq:rambo-boost}
\end{equation}
the Lorentz boost along \(\bm\nu\) written with
\(\zeta\bm\nu^2=\gamma-1\), multiplied by \(\alpha\).  Applied to \(Q\), with
\(\bm\nu\cdot\bm Q=-\bm Q^2/M\), Eq.~\eqref{eq:rambo-boost} gives
\begin{equation}
\adjustbox{max width=\linewidth}{$\displaystyle
\begin{aligned}
 \sum_ik_i^0&=\alpha\,\frac{(Q^0)^2-\bm Q^2}{M}=\alpha M=\sqrt s,\\
 \sum_i\bm k_i&=\alpha\,\bm Q\Bigl(1-\gamma+\zeta\,\frac{\bm Q^2}{M^2}\Bigr)=\bm 0,
\end{aligned}
$}\label{eq:rambo-check}
\end{equation}
since \(\zeta\bm Q^2/M^2=\zeta(\gamma^2-1)=\gamma-1\).  A Lorentz
transformation preserves \(q_i^2=0\), and a rescaling multiplies it by
\(\alpha^2\), so that
\begin{equation}
 k_i^2=0,\qquad \sum_{i=1}^nk_i^\mu=P^\mu .\label{eq:rambo-constraints}
\end{equation}
Reference~\cite{RAMBO} changes variables from the \(q_i\) to the \(k_i\),
\(\bm\nu\) and \(\alpha\), and shows that the
density~\eqref{eq:rambo-q-density} becomes the product of the Lorentz-invariant
phase-space measure of the \(k_i\) and a function of \(\bm\nu\) and \(\alpha\)
alone: the factor \(e^{-q^0}\) depends on the configuration only through
\(\sum_iq_i^0=Q^0\), which the map expresses through \(\bm\nu\) and
\(\alpha\).  The \(k_i\) are therefore distributed uniformly in phase space,
and every event has the same weight, equal to the total phase-space volume.
With the measure
\begin{equation}
\adjustbox{max width=\linewidth}{$\displaystyle
\dd\Phi_n(P)=(2\pi)^4\delta^{(4)}\Bigl(P-\sum_ik_i\Bigr)
 \prod_{i=1}^n\frac{\dd^3\bm k_i}{(2\pi)^32k_i^0},
$}\label{eq:rambo-measure}
\end{equation}
the factors of \(2\pi\) separate, \(\dd\Phi_n=(2\pi)^{4-3n}\,\dd\tilde\Phi_n\),
and the volume \(\tilde V_n(s)=\int\dd\tilde\Phi_n\) of the measure without them
obeys a recursion: inserting \(1=\int\dd m^2\,\dd^4K\,\delta^{(4)}(K-\sum_{i<n}k_i)\,\delta(K^2+m^2)\)
separates the first \(n-1\) momenta, of invariant mass \(m\), from \(k_n\),
\begin{equation}
\adjustbox{max width=\linewidth}{$\displaystyle
\begin{aligned}
 \tilde V_n(s)&=\int_0^s\dd m^2\,\tilde V_{n-1}(m^2)\,\tilde V_2(s;m^2),\\
 \tilde V_2(s;m^2)&=\int\frac{\dd^3\bm k}{2E_K\,2|\bm k|}\,\delta\bigl(\sqrt s-E_K-|\bm k|\bigr)
 =\frac\pi2\Bigl(1-\frac{m^2}{s}\Bigr),
\end{aligned}
$}\label{eq:rambo-recursion}
\end{equation}
where \(E_K=(\bm k^2+m^2)^{1/2}\) and the delta function fixes
\(|\bm k|=(s-m^2)/2\sqrt s\).  For \(m=0\), \(\tilde V_2=\pi/2\); if
\(\tilde V_{n-1}(m^2)=(\pi/2)^{n-2}m^{2(n-3)}/[\Gamma(n-1)\Gamma(n-2)]\), the
integral \(\int_0^s\dd m^2\,m^{2(n-3)}(1-m^2/s)=s^{n-2}/[(n-2)(n-1)]\) gives
the same form for \(n\), and by induction
\begin{equation}
\adjustbox{max width=\linewidth}{$\displaystyle
w_n(s)=\int\dd\Phi_n
 =(2\pi)^{4-3n}\Bigl(\frac\pi2\Bigr)^{n-1}\frac{s^{n-2}}{\Gamma(n)\Gamma(n-1)} ,
$}\label{eq:rambo-weight}
\end{equation}
the result of Ref.~\cite{RAMBO}.  For \(n=2\),
\begin{equation}
 w_2=\frac{1}{8\pi}=\int\frac{\dd\Omega}{32\pi^2},\label{eq:rambo-w2}
\end{equation}
the two-body measure used in Sec.~\ref{sec:coefficient}, and for \(n=4\), with
\(\Gamma(4)\Gamma(3)=12\) and \(8\cdot12\cdot256=24576\),
\begin{equation}
 w_4(s)=(2\pi)^{-8}\Bigl(\frac\pi2\Bigr)^3\frac{s^2}{\Gamma(4)\Gamma(3)}
 =\frac{s^2}{24576\,\pi^5}.\label{eq:rambo-w4}
\end{equation}

Since the weight is constant, a phase-space integral is \(w_n\) times the
mean of the integrand over the generated events, with an error that
decreases as \(N^{-1/2}\) by the central limit theorem~\cite{JamesMC}.  The external momenta
are taken in the centre-of-mass frame,
\begin{equation}
\begin{gathered}p_{1,2}=\frac{\sqrt s}{2}\bigl(1,0,0,\pm1\bigr), \\
p_{3,4}=\frac{\sqrt s}{2}\bigl(1,\pm\sqrt{1-x^2},0,\pm x\bigr), \\
x=1+\frac{2t}{s},\end{gathered}\label{eq:rambo-kinematics}
\end{equation}
where \(p_3\) and \(p_4\) are the physical outgoing momenta, at the scattering
angle \(\theta\) with \(\cos\theta=x\); then \(p_i^2=0\),
\(t=-(p_1-p_3)^2=-\frac s2(1-x)\) and \(u=-(p_1-p_4)^2=-\frac s2(1+x)\), so
that \(s+t+u=0\), and \(x=1\) is the forward direction \(t=0\).  For the
two-body cut of the quartic sector, Eq.~\eqref{eq:pwu-cut} with the
identical-particle factor \(1/2!\), divided by two, and for the banana,
Eq.~\eqref{eq:banana-cut} divided by two, the estimators are
Eqs.~\eqref{eq:rambo-two-body} and~\eqref{eq:rambo-four-body}.
\addtocounter{equation}{2}%
The first reproduces Eq.~\eqref{eq:x2-optical} within its statistical error.  For
two bodies the constant weight is elementary, since the direction of \(\bm k_1\)
in the rest frame of \(P\) is uniform on the sphere and
\(\int\dd\Phi_2=w_2\int\dd\Omega/4\pi\); an implementation that samples this
direction directly is an independent check of the code.  In the second, \(\M^{\rm tree}_{6,(\partial\phi)^6}\) is the fifteen-pairing contact
amplitude~\eqref{eq:x3-tree-permutation} evaluated with the invariants
\(s_{ij}\) of Sec.~\ref{sec:intro}.  This amplitude is a homogeneous polynomial
of degree six in the momenta, a sum of products of three \(s_{ij}\), each
quadratic, so that
\begin{equation}
\adjustbox{max width=\linewidth}{$\displaystyle
\M^{\rm tree}_{6,(\partial\phi)^6}(-p_1,\dots,-p_6)=\M^{\rm tree}_{6,(\partial\phi)^6}(p_1,\dots,p_6);
$}\label{eq:rambo-parity}
\end{equation}
the first factor of Eq.~\eqref{eq:rambo-four-body} may therefore be evaluated
with \(+k_i^{(r)}\), and both vertices use the same generated momenta.  At
\(t=0\), where \(p_3=p_1\) and \(p_4=p_2\), the summand is the square of one
tree and Eq.~\eqref{eq:rambo-four-body} is the optical theorem, positive event
by event.  Writing \(\omega_r\) for the summand of Eq.~\eqref{eq:rambo-four-body}
multiplied by \(w_4/(2\cdot4!)\), so that the estimator is the mean
\(\bar\omega=\frac1N\sum_r\omega_r\), the statistical error is the standard
error of the mean,
\begin{equation}
\adjustbox{max width=\linewidth}{$\displaystyle
\Delta=\biggl[\frac{1}{N}\Bigl(\frac1N\sum_{r=1}^N\omega_r^2
 -\bar\omega^{\,2}\Bigr)\biggr]^{1/2}
 \propto N^{-1/2},
$}\label{eq:rambo-error}
\end{equation}
the sample variance of the \(\omega_r\) divided by \(N\), since the events are
independent and identically distributed.

The closed form against which Eq.~\eqref{eq:rambo-four-body} is compared
is Eq.~\eqref{eq:banana-imaginary}.  With \(N=2\times10^6\) events per run the
relative statistical error is about \(5\times10^{-4}\),
Eq.~\eqref{eq:rambo-error}.  Table~\ref{tab:rambo} lists the result of
\(N=1.6\times10^7\) events, with the same generated \(k_i\) used at all angles,
in units of \(g_6^2s^8/(1\,857\,945\,600\,\pi^5\Lambda^{16})\), for which
Eq.~\eqref{eq:banana-imaginary} reads \(1210-31\,tu/s^2\).  All four angles agree
within one standard deviation.  The \(t\)-dependence is a \(0.6\%\) effect,
about thirty standard deviations of a single angle; because the four estimates
share the events, the difference of two angles is estimated by
Eq.~\eqref{eq:rambo-error} applied to the event-by-event difference
\(\omega_r(t)-\omega_r(t')\), whose fluctuations largely cancel.  The difference
between \(t=0\) and \(t=-s/2\), \(tu/s^2=1/4\), gives
\(b/4=7.694\pm0.046\), i.e.\ \(b=30.78\pm0.18\), \(1.2\) standard deviations
from \(31\).
\begin{table}[!t]
\centering
\caption{Four-particle cut of the \(s\)-channel banana,
Eq.~\eqref{eq:rambo-four-body}, from \(1.6\times10^7\) RAMBO events, in
units of \(g_6^2s^8/(1\,857\,945\,600\,\pi^5\Lambda^{16})\).}
\label{tab:rambo}
\footnotesize\setlength{\tabcolsep}{4pt}
\begin{tabular}{cccc}
\hline
\(t/s\) & \(tu/s^2\) & RAMBO & Eq.~\eqref{eq:banana-imaginary}\\
\hline
\(0\) & \(0\) & \(1209.80\pm0.23\) & \(1210\)\\
\(-0.3\) & \(0.21\) & \(1203.30\pm0.23\) & \(1203.49\)\\
\(-0.5\) & \(0.25\) & \(1202.11\pm0.23\) & \(1202.25\)\\
\(-0.6\) & \(0.24\) & \(1202.44\pm0.23\) & \(1202.56\)\\
\hline
\end{tabular}
\end{table}
Every generated event is also checked against the
constraints~\eqref{eq:rambo-constraints}, which hold to machine precision, and
the mean of a constant integrand against the volume~\eqref{eq:rambo-weight}.
Bose symmetry is checked by permuting the labels of the four final-state
momenta, which must leave each term of Eq.~\eqref{eq:rambo-four-body}
unchanged, since the tree~\eqref{eq:x3-tree-permutation} is symmetric under
all permutations of its legs.

The same events give the complete \(2\to4\) cross section in the forward
direction, with the tree~\eqref{eq:x6-exchange} in place of the contact term.
At \(t=0\) the summand is
\(\bigl(\M^{\rm tree}_{2\to4}\bigr)^2=\bigl(g_6A_6+g_4^2A_{44}\bigr)^2/\Lambda^{16}\),
a quadratic form in \(g_6\) and \(g_4^2\), and since \(A_6\) and \(A_{44}\) are
homogeneous of degree six, the mean of each product is proportional to
\(s^8\), \(w_4\propto s^2\) by Eq.~\eqref{eq:rambo-w4}.  With \(A_6\) and
\(A_{44}\) evaluated at every event,
\begin{equation}
\adjustbox{max width=\linewidth}{$\displaystyle
\begin{aligned}
 &\frac{w_4(s)}{2\cdot4!\,N\Lambda^{16}}\sum_{r=1}^N\bigl(g_6A_6+g_4^2A_{44}\bigr)^2\\
 &\quad=\frac{s^8}{\Lambda^{16}}\bigl(a_{66}\,g_6^2+a_{46}\,g_4^2g_6+a_{44}\,g_4^4\bigr),
\end{aligned}
$}\label{eq:rambo-2to4}
\end{equation}
where, with \(\langle\cdot\rangle\) the mean over events,
\begin{equation}
\adjustbox{max width=\linewidth}{$\displaystyle
\begin{gathered}
 a_{66}=\frac{w_4\langle A_6^2\rangle}{48\,s^8},\qquad
 a_{46}=\frac{2w_4\langle A_6A_{44}\rangle}{48\,s^8},\\
 a_{44}=\frac{w_4\langle A_{44}^2\rangle}{48\,s^8}.
\end{gathered}
$}\label{eq:rambo-a-coefficients}
\end{equation}
Eight independent runs of \(2.5\times10^7\) events each give
\(\pi^5a_{66}=6.5124(4)\times10^{-7}\), in agreement with
\(121/185\,794\,560=6.5126\times10^{-7}\) of Eq.~\eqref{eq:banana-imaginary}
within \(0.5\) standard deviations, \(\pi^5a_{46}=5.0110(3)\times10^{-6}\) and
\(\pi^5a_{44}=9.6653(6)\times10^{-6}\), with the statistical errors of
Eq.~\eqref{eq:rambo-error}.  The ratios, whose errors are smaller because the three terms share the
events, are \(a_{46}/a_{66}=7.69457(2)\) and \(a_{44}/a_{66}=14.84156(7)\).
Completing the square,
\begin{equation}
\adjustbox{max width=\linewidth}{$\displaystyle
\begin{aligned}
 &a_{66}g_6^2+a_{46}g_4^2g_6+a_{44}g_4^4\\
 &\quad=a_{66}\bigl[(g_6+c\,g_4^2)^2+d\,g_4^4\bigr],\\
 &c=\frac{a_{46}}{2a_{66}}=3.84728(1),\quad
 d=\frac{a_{44}}{a_{66}}-c^2=0.039968(4),
\end{aligned}
$}\label{eq:rambo-cd}
\end{equation}
the constants of Eq.~\eqref{eq:improved-positivity-im}.  By
Eq.~\eqref{eq:rambo-a-coefficients}, \(c=\langle A_6A_{44}\rangle/\langle
A_6^2\rangle\) is the coefficient of the least-squares fit
\(A_{44}\simeq c\,A_6\) over phase space, and
\begin{equation}
\adjustbox{max width=\linewidth}{$\displaystyle
 d=\frac{\langle(A_{44}-c\,A_6)^2\rangle}{\langle A_6^2\rangle}\ge0,\qquad
 \frac{\sqrt d}{c}=0.052,
$}\label{eq:rambo-residual}
\end{equation}
so that \(d\) is the mean square of the residual of that fit, non-negative as
required by the Cauchy--Schwarz inequality; the exchange is close to
proportional to the contact amplitude, with a root-mean-square residual of
\(5\%\) of \(c\,A_6\), and \(d/c^2=2.7\times10^{-3}\).

\onecolumn
\vspace*{0.6\baselineskip}\noindent\rule{\textwidth}{0.4pt}\par\vspace{0.9\baselineskip}
{\centering\normalfont\large\bfseries\MakeUppercase{\refname}\par}\vspace{0.8\baselineskip}
\begin{multicols}{2}
\makeatletter\patchcmd{\thebibliography}{\section*{\refname}\@mkboth{\MakeUppercase\refname}{\MakeUppercase\refname}}{}{}{\patchcmd{\thebibliography}{\section*{\refname}}{}{}{}}\makeatother
\end{multicols}

{\footnotesize
\begin{thebibliography}{99}
\setlength{\itemsep}{0pt plus 0.5pt}\setlength{\parsep}{0pt plus 0.5pt}

\bibitem{Weinberg}
S. Weinberg, \href{https://doi.org/10.1016/0378-4371(79)90223-1}{Physica A} \textbf{96}, 327 (1979)

\bibitem{Basile}
I. Basile, L. Buoninfante, F. Di Filippo \textit{et al}., \href{https://doi.org/10.21468/SciPostPhysLectNotes.98}{SciPost Phys. Lect. Notes} \textbf{98} (2025), arXiv:\href{https://arxiv.org/abs/2412.08690}{2412.08690[hep-th]}

\bibitem{PengShaoTokarevaXu}
G. Peng, L.-Q. Shao, A. Tokareva \textit{et al}., \href{https://doi.org/10.1007/JHEP03(2026)163}{JHEP} \textbf{03}, 163 (2026), arXiv:\href{https://arxiv.org/abs/2501.09717}{2501.09717[hep-th]}

\bibitem{PhamTruong}
T. N. Pham and T. N. Truong, \href{https://doi.org/10.1103/PhysRevD.31.3027}{Phys. Rev. D} \textbf{31}, 3027 (1985)

\bibitem{Adams}
A. Adams, N. Arkani-Hamed, S. Dubovsky \textit{et al}., \href{https://doi.org/10.1088/1126-6708/2006/10/014}{JHEP} \textbf{10}, 014 (2006), arXiv:\href{https://arxiv.org/abs/hep-th/0602178}{hep-th/0602178}

\bibitem{DeRham}
C. de Rham, S. Melville, A. J. Tolley \textit{et al}., \href{https://doi.org/10.1103/PhysRevD.96.081702}{Phys. Rev. D} \textbf{96}, 081702 (2017), arXiv:\href{https://arxiv.org/abs/1702.06134}{1702.06134[hep-th]}

\bibitem{BellazziniMoments}
B. Bellazzini, J. Elias Mir\'o, R. Rattazzi \textit{et al}., \href{https://doi.org/10.1103/PhysRevD.104.036006}{Phys. Rev. D} \textbf{104}, 036006 (2021), arXiv:\href{https://arxiv.org/abs/2011.00037}{2011.00037[hep-th]}

\bibitem{Tolley}
A. J. Tolley, Z.-Y. Wang, and S.-Y. Zhou, \href{https://doi.org/10.1007/JHEP05(2021)255}{JHEP} \textbf{05}, 255 (2021), arXiv:\href{https://arxiv.org/abs/2011.02400}{2011.02400[hep-th]}

\bibitem{EFThedron}
N. Arkani-Hamed, T.-C. Huang, and Y.-T. Huang, \href{https://doi.org/10.1007/JHEP05(2021)259}{JHEP} \textbf{05}, 259 (2021), arXiv:\href{https://arxiv.org/abs/2012.15849}{2012.15849[hep-th]}

\bibitem{CaronHuotVanDuong}
S. Caron-Huot and V. Van Duong, \href{https://doi.org/10.1007/JHEP05(2021)280}{JHEP} \textbf{05}, 280 (2021), arXiv:\href{https://arxiv.org/abs/2011.02957}{2011.02957[hep-th]}

\bibitem{CaronHuotSwampland}
S. Caron-Huot, D. Maz\'a\v{c}, L. Rastelli \textit{et al}., \href{https://doi.org/10.1007/JHEP07(2021)110}{JHEP} \textbf{07}, 110 (2021), arXiv:\href{https://arxiv.org/abs/2102.08951}{2102.08951[hep-th]}

\bibitem{Chiang}
L.-Y. Chiang, Y.-T. Huang, W. Li \textit{et al}., \href{https://doi.org/10.1007/JHEP03(2022)063}{JHEP} \textbf{03}, 063 (2022), arXiv:\href{https://arxiv.org/abs/2105.02862}{2105.02862[hep-th]}

\bibitem{Deproject}
L.-Y. Chiang, Y.-T. Huang, L. Rodina \textit{et al}., \href{https://doi.org/10.1007/JHEP05(2024)102}{JHEP} \textbf{05}, 102 (2024), arXiv:\href{https://arxiv.org/abs/2204.07140}{2204.07140[hep-th]}

\bibitem{NRTpositivity}
A. Nicolis, R. Rattazzi, and E. Trincherini, \href{https://doi.org/10.1007/JHEP05(2010)095}{JHEP} \textbf{05}, 095 (2010), arXiv:\href{https://arxiv.org/abs/0912.4258}{0912.4258[hep-th]}

\bibitem{Mandelstam}
S. Mandelstam, \href{https://doi.org/10.1103/PhysRev.112.1344}{Phys. Rev.} \textbf{112}, 1344 (1958)

\bibitem{Lehmann}
H. Lehmann, \href{https://doi.org/10.1007/BF02859794}{Nuovo Cimento} \textbf{10}, 579 (1958)

\bibitem{BEG}
J. Bros, H. Epstein, and V. Glaser, Nuovo Cimento \textbf{31}, 1265 (1964)

\bibitem{TokarevaXu}
A. Tokareva and Y. Xu, \href{https://doi.org/10.1103/86hv-m36b}{Phys. Rev. D} \textbf{112}, 085020 (2025), arXiv:\href{https://arxiv.org/abs/2502.10375}{2502.10375[hep-th]}

\bibitem{Distler}
J. Distler, B. Grinstein, R. A. Porto, and I. Z. Rothstein, \href{https://doi.org/10.1103/PhysRevLett.98.041601}{Phys. Rev. Lett.} \textbf{98}, 041601 (2007), arXiv:\href{https://arxiv.org/abs/hep-ph/0604255}{hep-ph/0604255}

\bibitem{ChandorkarRegge}
D. Chandorkar, S. D. Chowdhury, S. Kundu \textit{et al}., \href{https://doi.org/10.1007/JHEP05(2021)143}{JHEP} \textbf{05}, 143 (2021), arXiv:\href{https://arxiv.org/abs/2102.03122}{2102.03122[hep-th]}

\bibitem{HaringZhiboedov}
K. H\"aring and A. Zhiboedov, \href{https://doi.org/10.21468/SciPostPhys.16.1.034}{SciPost Phys.} \textbf{16}, 034 (2024), arXiv:\href{https://arxiv.org/abs/2202.08280}{2202.08280[hep-th]}

\bibitem{BellazziniRiembauRiva}
B. Bellazzini, M. Riembau, and F. Riva, \href{https://doi.org/10.1103/PhysRevD.106.105008}{Phys. Rev. D} \textbf{106}, 105008 (2022), arXiv:\href{https://arxiv.org/abs/2112.12561}{2112.12561[hep-th]}

\bibitem{YeHeGu}
Y. Ye, B. He, and J. Gu, \href{https://doi.org/10.1007/JHEP12(2024)046}{JHEP} \textbf{12}, 046 (2024), arXiv:\href{https://arxiv.org/abs/2408.10318}{2408.10318[hep-ph]}

\bibitem{Regge}
T. Regge, \href{https://doi.org/10.1007/BF02728177}{Nuovo Cimento} \textbf{14}, 951 (1959)

\bibitem{Collins}
P. D. B. Collins, \textit{An Introduction to Regge Theory and High-Energy Physics} (Cambridge: Cambridge University Press, 1977)

\bibitem{AlberteQED}
L. Alberte, C. de Rham, S. Jaitly \textit{et al}., \href{https://doi.org/10.1103/PhysRevD.103.125020}{Phys. Rev. D} \textbf{103}, 125020 (2021), arXiv:\href{https://arxiv.org/abs/2012.05798}{2012.05798[hep-th]}

\bibitem{HerreroValea}
M. Herrero-Valea, R. Santos-Garcia, and A. Tokareva, \href{https://doi.org/10.1103/PhysRevD.104.085022}{Phys. Rev. D} \textbf{104}, 085022 (2021), arXiv:\href{https://arxiv.org/abs/2011.11652}{2011.11652[hep-th]}

\bibitem{Cutkosky}
R. E. Cutkosky, \href{https://doi.org/10.1063/1.1703676}{J. Math. Phys.} \textbf{1}, 429 (1960)

\bibitem{Szego}
G. Szeg\H{o}, \textit{Orthogonal Polynomials}, 4th ed. (Providence: American Mathematical Society, 1975)

\bibitem{Titchmarsh}
E. C. Titchmarsh, \textit{The Theory of Functions}, 2nd ed. (Oxford: Oxford University Press, 1939)

\bibitem{ChewFrautschi}
G. F. Chew and S. C. Frautschi, \href{https://doi.org/10.1103/PhysRevLett.7.394}{Phys. Rev. Lett.} \textbf{7}, 394 (1961)

\bibitem{LevySucher}
M. L\'evy and J. Sucher, \href{https://doi.org/10.1103/PhysRev.186.1656}{Phys. Rev.} \textbf{186}, 1656 (1969)

\bibitem{DiVecchiaEikonal}
P. Di Vecchia, C. Heissenberg, R. Russo \textit{et al}., \href{https://doi.org/10.1016/j.physrep.2024.06.002}{Phys. Rept.} \textbf{1083}, 1 (2024), arXiv:\href{https://arxiv.org/abs/2306.16488}{2306.16488[hep-th]}

\bibitem{AbramowitzStegun}
M. Abramowitz and I. A. Stegun, \textit{Handbook of Mathematical Functions} (New York: Dover, 1972)

\bibitem{tHooft}
G. 't Hooft, Phys. Lett. B \textbf{198}, 61 (1987)

\bibitem{ACV}
D. Amati, M. Ciafaloni, and G. Veneziano, Phys. Lett. B \textbf{197}, 81 (1987)

\bibitem{KabatOrtiz}
D. N. Kabat and M. Ortiz, \href{https://doi.org/10.1016/0550-3213(92)90627-N}{Nucl. Phys. B} \textbf{388}, 570 (1992), arXiv:\href{https://arxiv.org/abs/hep-th/9203082}{hep-th/9203082}

\bibitem{Giddings}
S. B. Giddings, M. Schmidt-Sommerfeld, and J. R. Andersen, \href{https://doi.org/10.1103/PhysRevD.82.104022}{Phys. Rev. D} \textbf{82}, 104022 (2010), arXiv:\href{https://arxiv.org/abs/1005.5408}{1005.5408[hep-th]}

\bibitem{BenderOrszag}
C. M. Bender and S. A. Orszag, \textit{Advanced Mathematical Methods for Scientists and Engineers} (New York: McGraw-Hill, 1978)

\bibitem{Macdonald}
I. G. Macdonald, \textit{Symmetric Functions and Hall Polynomials}, 2nd ed. (Oxford: Oxford University Press, 1995)

\bibitem{tHooftVeltman}
G. 't Hooft and M. Veltman, \href{https://doi.org/10.1016/0550-3213(72)90279-9}{Nucl. Phys. B} \textbf{44}, 189 (1972)

\bibitem{RAMBO}
R. Kleiss, W. J. Stirling, and S. D. Ellis, \href{https://doi.org/10.1016/0010-4655(86)90119-0}{Comput. Phys. Commun.} \textbf{40}, 359 (1986)

\bibitem{RAMBOdiet}
S. Pl\"atzer, arXiv:\href{https://arxiv.org/abs/1308.2922}{1308.2922[hep-ph]}

\bibitem{Froissart}
M. Froissart, \href{https://doi.org/10.1103/PhysRev.123.1053}{Phys. Rev.} \textbf{123}, 1053 (1961)

\bibitem{Martin}
A. Martin, \href{https://doi.org/10.1103/PhysRev.129.1432}{Phys. Rev.} \textbf{129}, 1432 (1963)

\bibitem{JinMartin}
Y. S. Jin and A. Martin, \href{https://doi.org/10.1103/PhysRev.135.B1375}{Phys. Rev.} \textbf{135}, B1375 (1964)

\bibitem{Dvali}
G. Dvali, G. F. Giudice, C. Gomez \textit{et al}., \href{https://doi.org/10.1007/JHEP08(2011)108}{JHEP} \textbf{08}, 108 (2011), arXiv:\href{https://arxiv.org/abs/1010.1415}{1010.1415[hep-ph]}

\bibitem{PassarinoVeltman}
G. Passarino and M. Veltman, \href{https://doi.org/10.1016/0550-3213(79)90234-7}{Nucl. Phys. B} \textbf{160}, 151 (1979)

\bibitem{BBDM}
W. A. Bardeen, A. J. Buras, D. W. Duke \textit{et al}., \href{https://doi.org/10.1103/PhysRevD.18.3998}{Phys. Rev. D} \textbf{18}, 3998 (1978)

\bibitem{JamesMC}
F. James, Rep. Prog. Phys. \textbf{43}, 1145 (1980)

\end{thebibliography}
}
\end{document}